\documentclass[preprint]{aastex631}
\usepackage{mathtools}
\usepackage{upgreek}
\usepackage{rotating}
\usepackage{graphicx}
\usepackage{gensymb}

\received{...}
\revised{...}
\accepted{...}
\submitjournal{ApJ}

\shorttitle{Dissecting the Extended X-ray Emission in the Merging Pair NGC 6240}
\shortauthors{Paggi et al.}

\begin{document}
	
\title{Dissecting the Extended X-ray Emission in the Merging Pair NGC 6240:\\Photo-ionization and Winds}

\correspondingauthor{Alessandro Paggi}
\email{alessandro.paggi@unito.it}

\author[0000-0002-5646-2410]{A. Paggi}
\affiliation{Dipartimento di Fisica, Universit\`{a} degli Studi di Torino, via Pietro Giuria 1, I-10125 Torino, Italy}
\affiliation{INAF-Osservatorio Astrofisico di Torino, via Osservatorio 20, I-10025 Pino Torinese, Italy}
\affiliation{Istituto Nazionale di Fisica Nucleare, Sezione di Torino, via Pietro Giuria 1, I-10125 Torino, Italy}

\author[0000-0002-3554-3318]{G. Fabbiano}
\affiliation{Center for Astrophysics \(|\) Harvard \& Smithsonian, 60 Garden St. Cambridge MA 02138, USA}

\author[0000-0001-9226-8992]{E. Nardini}
\affiliation{Dipartimento di Fisica e Astronomia, Universit\`{a} di Firenze, via G. Sansone 1, I-50019 Sesto Fiorentino, Italy}
\affiliation{INAF-Osservatorio Astrofisico di Arcetri, Largo Enrico Fermi 5, I-50125 Firenze, Italy}

\author[0000-0003-1769-9201]{M. Karovska}
\affiliation{Center for Astrophysics \(|\) Harvard \& Smithsonian, 60 Garden St. Cambridge MA 02138, USA}

\author[0000-0001-5060-1398]{M. Elvis}
\affiliation{Center for Astrophysics \(|\) Harvard \& Smithsonian, 60 Garden St. Cambridge MA 02138, USA}

\author[0000-0003-4874-0369]{J. Wang}
\affiliation{Department of Astronomy, Physics Building, Xiamen University, Xiamen, Fujian, 361005, China}

\begin{abstract}
We present  a  detailed  spectral  and  imaging  analysis  of  the  central  \(15''\) radius (\(\sim 7.5 \text{ kpc}\)) region of the merger galaxy NGC 6240 that makes use of all the available \textit{Chandra}-ACIS data (\(0.3 - 3 \text{ keV}\) effective exposure of \(\sim 190 \text{ ks}\)). This region shows extended X-ray structures with lower energy counterparts imaged in CO, [O III] and H\(\alpha\) line emission. We find both photo-ionized phases of possible nuclear excitation and thermal shock-excited emission in the different large-scale components: the north-west ``loop" detected in H\(\alpha\), the {region surrounding the two nuclei}, the large outflow region to the north-east detected in [O III], and the southern X-ray extensions. The latter could be the ionization cone of the {northern} nucleus{, with the N counterpart being obscured by the galaxy disk}. The radial distribution of the X-ray surface brightness suggests a confined hot interstellar medium at \(r  < 2.5 \text{ kpc}\),  with a free-flowing wind at larger radii;  if the confinement is magnetic, we estimate B-field values of \(\sim 100\,\upmu \text{G}\) , similar to those measured in the halo of M82. The thermal gas {of the extended halo} at \(kT \sim 1 \text{ keV}\) absorbs soft X-rays from the AGN, but not the extreme ultraviolet radiation leading to a rapid increase in \(F_{\text{[O III]}}/F_{\text{X}}\) beyond \(\sim 3 \text{ kpc}\). The \(\alpha\) element to Fe abundance ratios of the thermal components in the different regions of the extended X-ray emission are generally compatible with SNe II yields, confirming the importance of the active star formation in NGC 6240.
\end{abstract}

\keywords{X-ray active galactic nuclei, Interacting galaxies, Active galaxies}

\section{Introduction}\label{sec:intro}

{NGC 6240 is a} highly disturbed merger galaxy at a redshift of \(0.0245\) \citep[{\(D= 108 \text{ Mpc}\), \(1\arcsec = 500 \text{ pc}\)}][]{1993ApJ...414L..13D}, {that} provides a relatively nearby case study of the complex {evolutionary} phenomena connected with the {merging of} galaxies and their nuclear massive black holes. In the optical band, NGC 6240 presents long tidal tails and giant butterfly{-}shaped H\(\alpha\){,} and X-ray{,} emitting loops in the central \(\sim 15\) kpc as well as a disturbed disk crossed by broad dust lanes \citep{1979MNRAS.189...79F, 1987AJ.....93..276H, 2002MNRAS.333..709L, 2004AJ....127...75G}. Its strong infrared (IR) emission indicates active merger-induced star formation \citep{1998Natur.395..859G}. NGC 6240 contains two well-defined nuclear regions detected at both optical \citep{1983A&A...118..166F} and IR wavelengths \citep{2000AJ....119..991S}. {Two} highly obscured active galactic nuclei (AGNs) were discovered {in this pair of nuclear regions, North (N) and South (S),} with \textit{Chandra} \citep{2003ApJ...582L..15K} and {were} subsequently located with high accuracy in radio VLBA images \citep{2004AJ....127..239G}. The region near and in between the nuclei is rich in molecular gas \citep[e.g.,][]{1999Ap&SS.266..157T, 2020ApJ...890..149T}. Complex kinematics have been reported in both CO and [O III] line emission, suggesting both nuclear outflows and starburst-induced winds \citep{2013A&A...549A..51F, 2013A&A...558A..87F, 2018Natur.556..345M}.

Deep \textit{Chandra} observations have revealed a rich and complex X-ray morphology, both in the nuclear regions and at larger radii{, which we summarize below. }

{The larger-scale X-ray properties of the merger NGC 6240 are discussed in \citet{2013ApJ...765..141N}. These authors find that the merger resides within a luminous soft X-ray halo (\(0.4-2.5 \text{ keV}\) luminosity of \(4 \times {10}^{41} \text{erg} \text{ s}^{-1}\)), extending outwards from \(\sim 7.5 \text{ kpc}\) to a projected physical size of \(\sim 110 \times 80 \text{ kpc}\). The halo has a fairly flat radial surface brightness distribution, radially uniform hot gas temperature of \(\sim 7.5\) million K (\(0.65 \text{ keV}\)), and a total mass of a \(\sim {10}^{10}\,M_{\sun}\). Since most of the detected photons have energies \(< 2.5 \text{ keV}\), \citeauthor{2013ApJ...765..141N} restricted their spectral analysis to this energy range. Modeling the spectral emission of the soft halo with thermal optically thin models, they found that the relative abundances of the main \(\upalpha\)-elements (O, Ne, Mg, Si) with respect to iron are several times the solar value, with no significant radial variations. \citeauthor{2013ApJ...765..141N} note that the lack of strong radial abundance gradients implies a uniform enrichment by type II supernovae out to the largest scales. Moreover, the temperature of the halo is significantly higher than expected from a gaseous halo thermalized in the gravitational potential of the system. It is also higher than suggested by merger simulations (expected \(kT \sim 0.25 \text{ keV}\)). These results led \citet{2013ApJ...765..141N} to suggest that a widespread, enhanced star formation, proceeding at a steady rate over the entire dynamical timescale (\(\sim 200 \text{ Myr}\)), contributes to both the energy and metal enrichment of the halo.}

{Within the inner \(\sim 7.5 \text{ kpc}\) of NGC 6240, \citet{2013ApJ...765..141N} observe instead a significantly greater surface brightness and higher temperatures (rising to \(\sim 1.2 \text{ keV}\) at \(\sim 2 \text{ kpc}\)), both with steep radial profiles consistent with an expanding adiabatic wind. This is the region with the most disturbed morphology, where tidal effects, the starburst, and the AGN emission are the strongest. Using the full spectral range of \textit{Chandra} (up to \(\sim 7 \text{ keV}\)), \citet{2013A&A...558A..87F} reported the detection of a hard X-ray emission component in the spectrum from this region. This emission would be  consistent with the shock-ionized gas from a nuclear AGN wind matching the terminal velocities of \(\sim 400 \text{ km} \text{ s}^{-1}\) that they infer from the CO(1-0) line, which they ascribe to nuclear winds from the dual AGN. \citet{2014ApJ...781...55W} also reported extended hard X-ray emission from a \(kT \sim 6 \text{ keV}\) (\(\sim 70 \text{ MK}\)) hot gas over a spatial scale of \(5 \text{ kpc}\), indicating the presence of fast shocks with a velocity of \(\sim 2200 \text{ km} \text{ s}^{-1}\). Moreover, they mapped the spatial distribution of this highly ionized gas using the \(6.7 \text{ keV}\) Fe XXV line, which shows a remarkable correspondence to the large-scale morphology of H2(1-0) S(1) line emission and H\(\upalpha\) filaments. While not excluding nuclear winds, they note that the propagation of fast shocks originating in the starburst-driven wind into the ambient dense gas can account for this morphological correspondence.}

{A re-analysis of the innermost \(\sim 2''\) (\(\sim 2 \text{ kpc}\)) radius region in the hard X-ray, at the highest resolution allowed by the \textit{Chandra} point response function, was recently reported by \citet{2020ApJ...902...49F}. Focusing on the hard energy bands containing the hard spectral continuum (\(5.5-5.9 \text{ keV}\)), the redshifted neutral Fe K\(\upalpha\) line (\(6.0-6.4 \text{ keV}\)), and the redshifted thermal Fe XXV line (\(6.4-6.7 \text{ keV}\)), these authors were able to resolve structures with sizes from \(\sim 1 \text{ kpc}\) to \(< 200 \text{ pc}\). They found significant extended emission in both continuum and Fe lines in the \(\sim 2\arcsec\) (\(\sim 1 \text{ kpc}\)) region surrounding the nuclei, in the region between the N and S AGN, and in a sector of position angle \(120\degree-210\degree\) extending to the SE from the centroid of the S AGN surface brightness. The extended neutral Fe K\(\upalpha\) emission is likely to originate from the fluorescence of X-ray photons interacting with dense molecular clouds, providing a complementary view to recent high-resolution Atacama Large Millimeter/submillimeter Array (ALMA) studies. The nonthermal emission (i.e. neutral Fe K\(\upalpha\)) is more prevalent in the region in between the two active X-ray nuclei and in the N AGN.}

In this paper, we use the deep (\(\sim 300 \text{ ks}\)) co-added \textit{Chandra} ACIS dataset on NGC 6240, to probe in detail {the spatial and spectral properties of} the inner \(\sim 7.5 \text{ kpc}\) region. {This study is complementary to the work of \citep{2013ApJ...765..141N}, in that it explores this region with the ultimate \textit{Chandra} spatial resolution. It is also complementary to the work by \citet{2014ApJ...781...55W} and \citet{2020ApJ...902...49F}, in that the spectral analysis concentrates on the energy range \(< 3 \text{ keV}\), where the emission from the extended loops and filaments is particularly strong. This spectral range is sensitive to both thermal emission from the ISM (with the exclusion of the strong shocks explored by \citealt{2013A&A...558A..87F}, \citealt{2014ApJ...781...55W} and \citealt{2020ApJ...902...49F}), and to the non-thermal photo-ionization emission from the dual AGN (see e.g. \citealt{2018ApJ...865...83F} in the case of ESO 428-G014).}

After producing the merged data cube (Sect. \ref{sec:analysis}), we use sub-pixel imaging{,} image enhancement and reconstruction techniques to provide the most detailed spatially resolved {X-ray} images of NGC 6240 (Sect. \ref{sec:imaging}). Based on this image, we perform spectral analysis from selected regions, to constrain the emission parameters (Sect. \ref{sec:spectra}). We discuss the implications of our results in Sect. \ref{sec:discussion} and summarize our findings in Sect. \ref{sec:summary}.

In this paper we assume a flat \(\Lambda\)CDM cosmology with \(H_0 = 69.6 \text{ km} \text{ s}^{-1} \text{ Mpc}^{-1}\), \(\Omega_M=0.286\) and \(\Omega_\Lambda=0.714\) \citep{2014ApJ...794..135B}\footnote{With this cosmology, the redshift of \(0.0245\) of NGC 6240 corresponds to a luminosity distance of \(107.5 \text{ Mpc}\) and an angular scale of \(496 \text{ pc}/\arcsec\).}.

\section{Data Preparation and Analysis Methods}\label{sec:analysis}

We have used the same data set as in \citet{2013ApJ...765..141N}, \citet{2014ApJ...781...55W} and \citet{2020ApJ...902...49F}, which includes all the available ACIS-S observations with NGC 6240 at the aimpoint.  It consists of two imaging ACIS-S observations (ObsID 1590, 12713) with a total exposure time of \(183 \text{ ks}\), and of two ACIS-S HETG grating observations (ObsID 6908, 6909) with a total exposure time of \(302 \text{ ks}\). The observations were retrieved from \textit{Chandra} Data Archive through ChaSeR service\footnote{\href{http://cda.harvard.edu/chaser}{http://cda.harvard.edu/chaser}}. We use the 0th order images of the grating observations. Given the response of the gratings, the combined effective ACIS-S imaging exposure times are \(\sim 190 \text{ ks}\) (\(0.3-3 \text{ keV}\)), \(\sim 210 \text{ ks}\) (\(3-6 \text{ keV}\)), and \(\sim 360 \text{ ks}\), at higher energies.

As in \citet{2020ApJ...902...49F}, data have been analyzed with {the} \textsc{CIAO} \citep{2006SPIE.6270E..1VF} data analysis system version 4.12 and {the} \textit{Chandra} calibration database CALDB version 4.9.1, adopting standard procedures.

\begin{figure}
	\centering
	\includegraphics[scale=1]{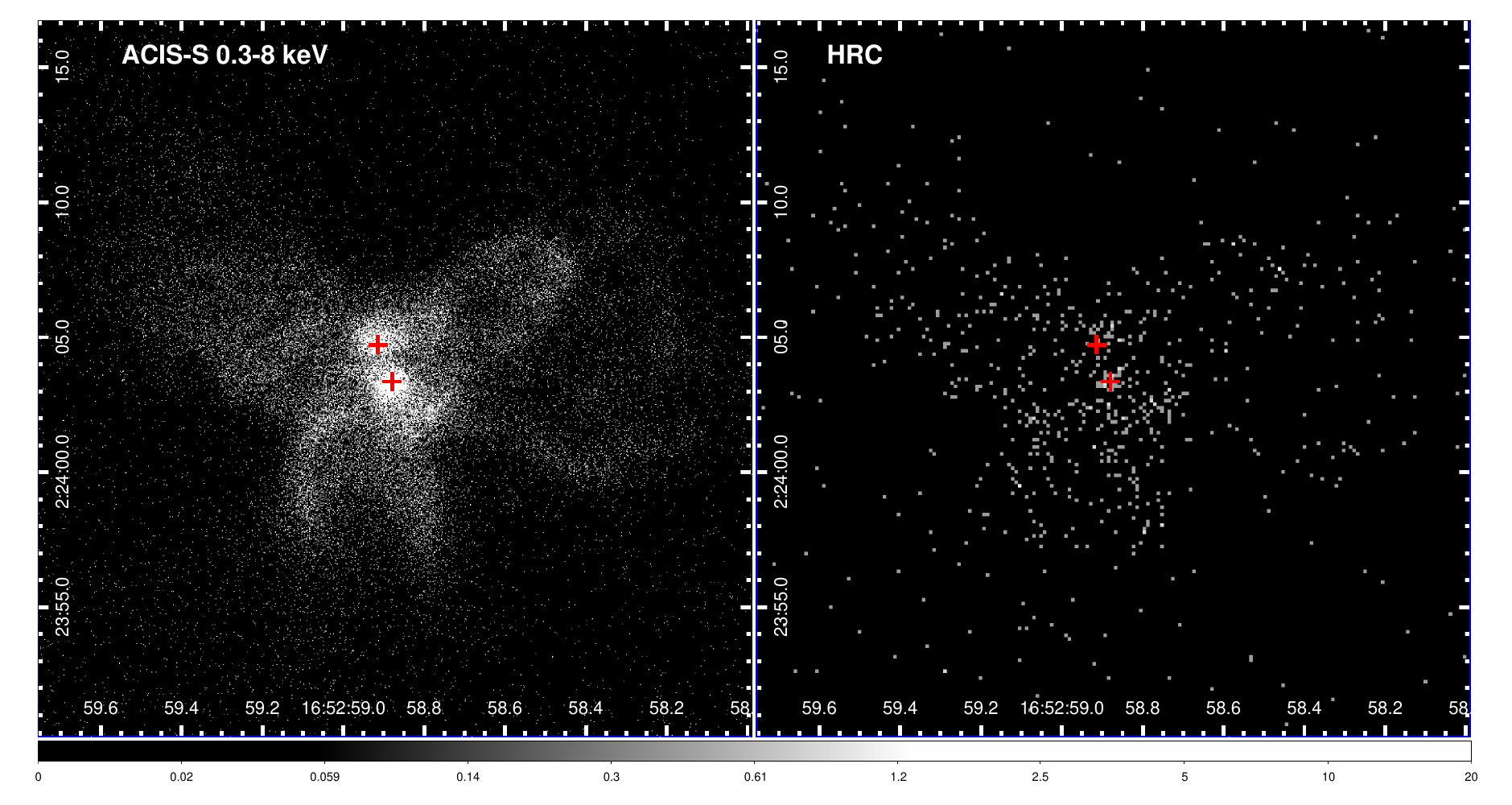}
	\caption{Left: ACIS merged image (\(0.3 – 8\) keV), with pixel size \(1/16\) ACIS instrument pixel, and no image processing. Right: HRC image (ObsID 438). The gray scale gives the logarithmic intensity scale in counts per image pixel. The positions of the nuclear centroids in {both images} are marked by red crosses.}\label{fig:hrc}
\end{figure}

To optimize the spatial resolution of the data, we use the same merged data set and merged PSF as in \citet{2020ApJ...902...49F}. {The data were merged using the peak emission of the strong hard-band dual AGN sources (see \citealt{2020ApJ...902...49F} for details on the merging procedure). We} used subpixel binning of \(1/16\) of the \(0\farcs492\) ACIS instrumental pixel\footnote{\href{https://cxc.harvard.edu/proposer/POG/html/chap6.html\#tab:acis\_char}{https://cxc.harvard.edu/proposer/POG/html/chap6.html\#tab:acis\_char}}. This method has been validated by several works, including our own \citep[see e.g. ][]{2004ApJ...615..161H, 2007ApJ...657..145S, 2010ApJ...710L.132K, 2011ApJ...729...75W, 2011ApJ...736...62W, 2011ApJ...742...23W, 2012ApJ...756...39P, 2018ApJ...865...83F}. It is conceptually comparable to the HST drizzle imaging \citep{2002PASP..114..144F} and exploits the sharp central peak of the \textit{Chandra} PSF and the well-characterized \textit{Chandra} dither motion\footnote{\href{https://cxc.harvard.edu/proposer/POG/html/chap5.html\#tth\_sEc5.3}{https://cxc.harvard.edu/proposer/POG/html/chap5.html\#tth\_sEc5.3}} to retrieve the full \textit{Chandra} mirror resolution. Fig. \ref{fig:hrc} compares the \(1/16\) pixel full band (\(0.3-8 \text{ keV}\)) ACIS merged image (left panel) with the \(10 \text{ ks}\) HRC image (ObsID 438, PI Murray; \citealt{2002MNRAS.333..709L}). The HRC instrumental readout pixel is \(0\farcs1318\), so that the full resolution of the PSF (\(FWHM \sim 0\farcs 2\)) is exploited\footnote{\href{https://cxc.harvard.edu/proposer/POG/html/chap4.html\#tth\_sEc4.2.3}{https://cxc.harvard.edu/proposer/POG/html/chap4.html\#tth\_sEc4.2.3}}. Although the signal to noise of the HRC image is much inferior to that of the long ACIS exposure, the good agreement between the main features of the extended emission demonstrate the validity and power of the \(1/16\) pixel imaging. Note that given the softer energy response of the HRC, the two nuclear sources are less prominent than in the ACIS image. These nuclei have large intrinsic absorbing columns \citep{2003ApJ...582L..15K, 2017MNRAS.471.3483N}.

\section{Broad-band Imaging}\label{sec:imaging}

We produced images using two different methods: (1) adaptive smoothing, with the CIAO tool \textsc{dmimgadapt}, which uses a range of Gaussian kernels to smooth the data; and (2) image reconstruction, with the image restoration algorithm Expectation through Markov Chain Monte Carlo \citep[EMC2,][sometimes referred to as PSF deconvolution]{2004ApJ...610.1213E, 2005ApJ...623L.137K, 2007ApJ...661.1048K, 2010ApJ...710L.132K, 2014ApJ...781...55W} with the PSF models derived in \citet{2020ApJ...902...49F} {from the nuclear sources}.

To better visualize the extended emission, we have produced adaptively smoothed maps with the CIAO tool \textsc{dmimgadapt}. For {the full-scale image in} the \(0.3-3 \text{ keV}\) energy band, which contains the line-dominated soft emission (see Sect. \ref{sec:spectra}), we used {\(1/8\)} binned data smoothed with Gaussians, with {\(16\)} counts under the kernel and sizes ranging from \(1/2\) to \(30\) image pixels, in \(30\) logarithmic steps, and we also further smoothed the resulting image with {a 2} pixels Gaussian. For the \(4-7 \text{ keV}\) band, where the extended emission has a lower count rate, we followed the same procedure, but used \(10\) counts under the kernel {and a \(1/16\) pixel, to better isolate the bright nuclear sources (see \citealt{2020ApJ...902...49F}).}

The left panel of Fig. \ref{fig:adapsmooth} shows the adaptively smoothed image in the \(0.3-3 \text{ keV}\) band, where the emission spectrum displays clear emission line features (Sect. \ref{sec:spectra}). The \(0.3-3 \text{ keV}\) image reconstructed with the EMC2 algorithm ({\(1/8\) pixel}, \(100\) iterations) is shown in the right panel of Fig. \ref{fig:adapsmooth}. It shows more vividly all the features suggested by the adaptively smoothed image.

Fig. \ref{fig:adapsmooth_4-7} shows the adaptively smoothed image in the \(4-7 \text{ keV}\) band, where the emission is dominated by a featureless spectral continuum and the \(6.4 \text{ keV}\) (neutral Fe-K\(\alpha\)) and \(6.7 \text{ keV}\) (Fe XXV) emission lines \citep{2014ApJ...781...55W, 2020ApJ...902...49F}. These images are displayed in logarithmic scale, to enhance the large-scale low-surface-brightness emission.

The soft \(0.3-3 \text{ keV}\) emission is dominated by large-scale extended features. A prominent loop of emission extends to the NW out to \(\sim 10 \text{ kpc}\) from the nuclei, and a long filament is visible to the S of this loop. Two similar sized prominent protrusions are visible to the S of the nuclei and a large extended feature with a cross-ridge of enhanced emission is also visible to the NE. These features are highly statistically significant. For example, even the faint ``yellow" clumps in the NW quadrant {(see Fig. \ref{fig:adapsmooth})} contain \(\sim 100\) counts in \(\sim 0\farcs 6\) radius circles, compared with the average \(\sim 50\) counts in similar size regions in the ``green" plateau in the same area, resulting {in} \(>4\,\sigma\)  excesses. In the same \(0\farcs 6\) radius circle, the ``dark blue" regions surrounding the ``green" plateau yield \(\sim 8\) counts. 

The \(4-7 \text{ keV}\) image is instead dominated by the two highly absorbed Compton-thick active galactic nuclei (CT AGNs) and the circum-nuclear emission \citep[see][]{2003ApJ...582L..15K, 2014ApJ...781...55W, 2020ApJ...902...49F}. The nuclei are not prominent in the \(0.3-3 \text{ keV}\) map, where their positions from the \(4-7 \text{ keV}\) image are shown as crosses. Extended emission is also visible out to \(\sim 3 \text{ kpc}\) radius. This hard emission follows the general footprint of the softer emission (see Fig. \ref{fig:adapsmooth}), as has been recently observed in several nearby CT AGNs \citep[e.g.,][]{2017ApJ...842L...4F, 2018ApJ...865...83F, 2020ApJ...900..164M, 2021ApJ...910...19J, travascio2021}.

A zoom-in of the central region is shown in Fig. \ref{fig:adapsmooth_zoom}, where the data are {binned in 1/16 pixel and} plotted in a linear scale to better highlight the bright circumnuclear filamentary features, which are especially prominent to the SW of the southern AGN. Again, these features are highly significant, with \(\sim 400\) (the {northern} filaments) and \(\sim 1500\) (the {southern} filaments) excess counts over the local intense average extended emission.

\section{Spectral Analysis}\label{sec:spectra}

{We performed spectral analysis on several emission regions, using the morphology of the emission as guidance. 
Given that the morphology of the extended emission is complex (see Fig. \ref{fig:adapsmooth}), we do expect that the spectral characteristics may also vary in different regions. We therefore singled out 10 individual regions containing significant morphological features, for detailed spectral analysis. These extraction regions all contain a large number of counts, as to get good spectral constraints. They are shown in Fig. \ref{fig:regions} and include:
\begin{itemize}
	\item Two circular regions which in the hard band are dominated by the point-like AGN emission (see \citealt{2020ApJ...902...49F}; also, Fig. \ref{fig:adapsmooth_4-7}), denominated as ‘N Nuclear’ (2606 net counts in the \(0.3-3 \text{ keV}\) band) and ‘S Nuclear’(4762 net counts);
	\item The surrounding central emission region (nuclear regions excluded, 3924 net counts);
	\item Two regions in the area where the [OIII] emission suggests a nuclear outflow \citep{2018Natur.556..345M}. These are the ‘Outer Outflow’ region (2844 net counts) and the region at smaller radii where the X-ray image shows a luminous ridge  perpendicular to the outflow direction, which could indicate localized shocks (‘Outflow ridge’, 2096 net counts).
	\item Regions where there are spatially coincident soft X-ray and H\(\upalpha\) features (see Fig. \ref{fig:adapsmooth}; also \citealt{2002MNRAS.333..709L, 2016ApJ...820...48Y}). These are the two Southern extensions, one to the East (‘ES Extension’ – 3038 net counts), and the other to the West (‘WS Extension’- 2997 counts); a ‘NW  Loop’ (3875 counts), and a ‘W  Filament’ (1237 counts). We also considered separately the northern continuation of the SW Extension, which partially overlaps the Central region and presents a strong surface brightness enhancement (‘NES Extension’ – 1736 net counts).
\end{itemize}
We also extracted the spectrum of the entire emission within a circle of radius 15’’ (7.5 kpc, encompassing all the strong extended emission visible in Fig. \ref{fig:adapsmooth}), to obtain the overall spectral characterization of the extended emission and in particular to determine the narrow band regions containing the most prominent lines. These lines will then be used for our narrow-band mapping of the extended emission (see Section \ref{sec:narrowband}). This region (named ‘All’) contains 35869 net counts in the 0.3-3.0 keV energy band (see below for background subtraction).}
	
{For each spectral extraction region,} we produced spectral response matrices weighted by the count distribution within the aperture (as appropriate for extended sources). Background spectra were extracted in large (\(\sim 80\arcsec\)), source-free regions in ACIS-S chip 7, and subtracted from source spectra. We made use of the \(\chi^2\) fit statistic, binning the spectra to obtain a minimum of \(20\) counts per bin. Spectral fitting was performed in the \(0.3-3\text{ keV}\) energy range with {the} \textsc{Sherpa} application \citep{2001SPIE.4477...76F}.

To {describe} the extracted spectra we adopted two classes of models.
\begin{enumerate}
	\item A \textit{phenomenological model} comprising, in addition to the photo-electric absorption by the Galactic column density along the line of sight \(N_H = 5.47\times {10}^{20}\text{ cm}^{-2}\) \citep{2016A&A...594A.116H}, a power-law with photon index fixed to \(1.8\) and several red-shifted Gaussian lines with widths fixed to \(10 \text{ eV}\). {The \(1.8\) power-law gives a good approximation of the hard AGN emission that may be present in the central regions given to PSF wing spillover (e.g. see \citealt{2006ApJ...648..111L, 2018ApJ...865...83F}).  The redshift assumed for the emission lines is systemic only (0.0245 for D = 108 Mpc)}. {Were removed from the model} all the lines whose normalization was constrained only with an upper limit. In addition, we included in this model an intrinsic photo-electric absorption at the source redshift.
	\item A \textit{physical model} comprising, in addition to the Galactic and intrinsic photo-electric absorptions, up to two thermal plasma components and/or up to two photo-ionization components. The thermal plasma is represented {by} an \textsc{vapec}\footnote{\href{https://heasarc.gsfc.nasa.gov/xanadu/xspec/manual/XSmodelApec.html}{https://heasarc.gsfc.nasa.gov/xanadu/xspec/manual/XSmodelApec.html}} model. {In this model, the Fe abundance} is left free to vary, while the abundances of the \(\alpha\) elements O, Ne, Mg and Si are linked {(but can vary together)} during the fit. Since we are interested in the abundance ratio of the \(\alpha\) elements to Fe {(which is a diagnostic of the age of the stellar population)}, we slightly modified the model in order to evaluate this ratio, to allow its errors to be directly determined during the fit \citep{2006ApJ...639..136H}. In addition, when two thermal plasma components were included in the model, we {linked the} Fe abundances and the \(\alpha\)/Fe abundance ratios of the two components.
	
	For the photo-ionization components, we produced grid
	models with the Cloudy\footnote{\href{http://www.nublado.org/}{http://www.nublado.org/}} c08.01 package \citep{1998PASP..110..761F}. We assumed the ionization source to be {a typical} AGN continuum (with a ``big bump" temperature \(T = {10}^6 \text{ K}\), {an} X-ray to UV ratio \(\alpha_{ox} = -1.30\) and an X-ray power-law component of spectral energy index \(\alpha_x = -0.8\)) illuminating a cloud with plane-parallel geometry and constant electron density \(n_e = {10}^5 \text{ cm}^{-3}\). The grid of models so obtained are parameterized in terms of the ionization parameter \(U\) (varying in the range \(\log{U} = [-3.00 : 2.00]\) in steps of \(0.25\)) and the hydrogen column density \(N_H\) (expressed in \(\text{cm}^{-2}\) varying in the range \(\log{N_H} = [19.0 : 23.5]\) in steps of \(0.1\)), taking into account only the reflected spectrum from the illuminated face of the cloud \citep{2010MNRAS.405..553B, 2011A&A...526A..36M}.
\end{enumerate}

{The results of the spectral fits for the phenomenological model, which includes the best fit rest-frame energies of the emission lines, are summarized in Table \ref{tab:line_spectra}. These results are useful for pinpointing the spectral regions of line emission, but should not be used to infer ‘physical’ line fluxes, especially in the range \(< 1.5 \text{ keV}\) where several emission lines could contribute that cannot be entirely spectrally resolved at the resolution of ACIS (\(\sim 100 \text{ eV}\)).}
	
{Table \ref{tab:line_phys} summarizes the results of the best fits to each single component and multiple-component physical models for each region, and highlights those we suggest as best-fit for each region, as described below.  The full set of results for all the models considered is reported in Table \ref{tab:line_phys_long}. We start with single component thermal (\(kT\)) and photoionization (\(U\)) models, and then add additional components in each case as required to get an acceptable fit. In some cases, it is not possible to choose a best-fit model based on \(\chi^2\) values only. In these cases, we followed an additional criterion (see \citealt{2018ApJ...865...83F} for a detailed discussion), which also considers the behavior of the fit residuals in the various spectral sub-ranges. Whenever a given model resulted in a spectral range of correlated residuals, we took this as an indication of the model not dealing adequately with that particular region of the spectrum and we added spectral components to obtain a more random run of residuals overall. For the ‘All’ region we were not able to get acceptable fits to physical models, likely due to the different plasma components that are responsible for the emission in the different regions that end up mixed together in this large extraction region.  The elemental abundances derived from the APEC model (see Table \ref{tab:line_phys}) are physically motivated, within the full set of elements included in the model. The only caution here is that these estimates may underestimate the abundances if large areas of the image including a range of spectral emissions are averaged together. This is because, given the ACIS spectral resolution, a mix of spectra may result in an apparent increase of the continuum emission. This effect was clearly demonstrated in the case of the Antennae galaxy \citep{2006ApJS..162..113B}.}

{Fig. \ref{fig:spectrum_all} shows the spectrum of the ‘All’ region with the best-fit phenomenological model and fit residuals. Based on this figure and on the phenomenological model results (Table \ref{tab:line_spectra}), we have selected eight narrow spectral bands that we will use for narrow band imaging. These energy bands and the contributing emission lines are listed in Table \ref{tab:line_energy}. The spectra, best-fit models and residuals for the 10 individual regions are shown in Fig. \ref{fig:line_regions} (phenomenological model) and Fig. \ref{fig:phys_regions} (physical models). Figs. \ref{fig:line_regions} and \ref{fig:phys_regions} clearly show that spectra from different regions may be substantially different. In particular, the N Nuclear region has significantly less emission in the softer band (\(< 1.5 \text{ keV}\)) than the S Nuclear region, suggesting a localized higher absorption in this region; the spectral fits also suggest a larger \(N_H\) in the N Nuclear region (see Tables \ref{tab:line_spectra}, \ref{tab:line_phys}). Note that the AGN emission itself is not contributing directly to the soft band under analysis, given the large intrinsic nuclear absorption columns of these two Compton-Thick AGNs \citep{2003ApJ...582L..15K}, which instead dominate the emission at energies \(>3 \text{ keV}\) (Fig. \ref{fig:adapsmooth_4-7}; also \citealt{2020ApJ...902...49F}). The spectra of the higher surface brightness regions (Central, NES Extension, Outflow Ridge, NW Loop) are relatively harder than those from the regions at larger radii (Outflow, ES and WS Extensions, W Filament).}

{Below we discuss our results and their implications.}

\section{Discussion}\label{sec:discussion}

\subsection{{Physical} properties of the Hot ISM of NGC 6240}\label{sec:spectral_results}

The spectral analysis of the \(0.3 – 3 \text{ keV}\) emission shows that the continuum level (represented by the power-law component normalization in the phenomenological model, see Table \ref{tab:line_spectra}) is comparable in the two nuclear regions. The strongest emission lines in these regions are the Fe XX 3d2p and Ne X Ly\(\alpha\), with the N nucleus showing additional lines at lower-energies, in particular strong Fe XVIII and O VIII RRC lines. The spectra extracted from the regions closer to the two nuclei - namely the Central region and the NES extension - are similar to the nuclear ones, while the spectra from the outermost regions are significantly softer, characterized by lower-energy lines like the O VII triplet, Fe XVIII, and Fe XVII 3d2p. This is consistent with the results from the parameters of the thermal components in the physical model fits (Table \ref{tab:line_phys}), which show the presence of a lower temperature, \(kT \sim 1 \text{ keV}\), gas in the outer regions and a hotter \(kT \sim 2 \text{ keV}\) phase {closer} to the nuclei.

If collisional ionization is the predominant excitation mechanism, these spectral differences suggest stronger shocks in the innermost circumnuclear regions. This would be in agreement with the presence of {previously reported} strong Fe XXV line emission in these regions, at \(6.7 \text{ keV}\) in the rest frame of NGC 6240, which has been connected with strong nuclear winds (velocities of \(600\) to over \(800 \text{  km} \text{ s}^{-1}\), \citealt{2013A&A...549A..51F, 2013A&A...558A..87F}) or also with strong SN shocks in the central active star formation regions of the merger \citep{2014ApJ...781...55W}. The gas masses emitting the X-rays in the major {morphological} features (e.g., NW loop, Outer Outflow), estimated from the thermal spectral component emission measure and assuming cylindrical volumes with filling factor \(1\), are \(\sim 1 \times {10}^{8} M_{\sun}\), with cooling times \((30-120) \text{ Myr}\). These are clearly transient events in the lifetime of NGC 6240 {as they} are comparable to merger timescales \citep{2010A&A...524A..56E, 2020ApJ...890..149T}. 

Figure \ref{fig:components} shows the decomposition of the best fit spectrum extracted in the Central region of NGC 6240 (see Fig. \ref{fig:regions}), with the thermal and photo-ionization components shown with red and blue lines, respectively. From this figure it appears evident that thermal emission is the dominant process near to the double AGN. However, the photo-ionization component is a non{-}negligible contributor to the total emission, accounting {for} \(\sim 20\%\) of the total intrinsic flux. 

We have revisited the radial distribution of the X-ray surface brightness using our higher spatial resolution images {(see the previous work by \citealt{2013ApJ...765..141N} for the full scale surface brightness profiles at large radii)}. Figure \ref{fig:sur_bri} shows detailed radial profiles of the inner \(20''\) region of NGC 6240, extracted in the azimuthal bins presented in the upper-left panel. The {profiles} extracted {from} the NW cone (containing the H\(\alpha\) loop), {the} NE cone (containing the [O III] outflow), {the} SE cone (containing the southern X-ray extension), and {the} SW cone (containing the W filament) are presented in the upper-central, upper-right, lower-left and lower-central panels, respectively, {while the profile extracted from the full \({360}^{\degree}\) sector is presented in the lower-right panel.} These profiles all have a {broken power-law} form, flatter {in} the interior, and steeper in the outer regions, with the break occurring between \(\sim 1''\) and \(\sim 5''\). Depending on the azimuthal bins, we find slopes {\(\sim {r}^{-\alpha}\)} outside \(5''\) {\(\alpha_{>5}= 3 - 4\), while within  \(5''\) we find \(\alpha_{<5} = 0.4 - 0.9\)}.

A radial dependence of the surface brightness of \(r^{-3}\) would {result} from a radial density dependence of \(r^{-2}\), i.e. freely expanding wind (as discussed e.g. in the case of M82, \citealt{1988ApJ...330..672F}; and NGC 6240, \citealt{2013ApJ...765..141N}). Possibly steeper slopes, as allowed by the surface brightness fits for \(r>5''\), may indicate adiabatic cooling of the expanding halo at the larger radii. We note that adiabatic cooling times for the thermal components, despite the large uncertainties, are estimated as \(2-10\text{ Myr}\), about \(5-30\) {times} shorter than the radiative cooling times of \(30-120\text{ Myr}\), estimated from the flux and thermal energy content in the different spectral extractions regions. {Therefore adiabatic cooling is a viable explanation.}

The significantly flatter radial dependence of the surface brightness at smaller radii \(r<5''\) shows that the extended X-ray emission in this region is not dominated by a free-flowing wind, {suggesting some confinement of the} hot plasma in the inner region. By balancing the thermal pressure \(P_{\text{th}} = n_{\text{H}}\, kT / \mu\) (where \(n_{\text{H}}\) is the thermal gas particle density estimated from the thermal spectral component emission measure, and \(\mu\) is the mean molecular weight) with the magnetic pressure \(P_B=B^2/8\pi\) in this inner region, we find that magnetic confinement would require equipartition magnetic fields of  the order of \(\sim 100\,\upmu\text{G}\). These fields would be \(\sim 1-2\) orders of magnitude larger than magnetic fields measured in the large-scale (\(\sim 100 \text{ kpc}\)) X-ray emitting gas around radio galaxies \citep[e.g.,][]{2005ApJ...626..733C, 2008A&A...482...97S}, but {lies within} the range of the magnetic fields measured in the higher density \(\sim \text{ kpc}\) scale outflow of the starburst galaxy M82 (a few \(100\,\upmu\text{G}\), \citealt{2021ApJ...914...24L}), a better analogy to the highly disturbed merger NGC 6240.

\subsection{{Photo-ionized} emission}\label{photoion}

We find different ionization parameters in different regions (see Table \ref{tab:line_phys}). A mildly photo-ionized component compatible with \(\log{(U)}\sim 0\) is found both in the {N Nuclear} region and in the {ES and WS Extension} regions {(see Section \ref{sec:spectra} and Fig. \ref{fig:regions})}. A {more highly ionized} component with \(\log{(U)}\sim 1\) is found in the NW Loop region, while both photo-ionization components are found in the {S Nuclear and in the Outer Outflow regions}. For a similar nuclear photon flux, these differences suggest a range of cloud density in {the} different regions. This may be the case, since the absorbing intrinsic hydrogen column densities estimated with these models are comparable with those estimated {from} the phenomenological models, being significantly larger in the northern (\({1.13}_{-0.07}^{+0.10}\times{10}^{22}\text{ cm}^{-2}\)) and southern (\({0.69}_{-0.04}^{+0.16}\times{10}^{22}\text{ cm}^{-2}\)) nuclei with respect to the outer regions, {which have} \({0.39}_{-0.06}^{+0.13}\times{10}^{22}\text{ cm}^{-2}\) in the outflow ridge. The presence of high-density absorbing clouds in the inner nuclear regions is also demonstrated by the presence of fluorescent {neutral} Iron emission (\(6.4 \text{ keV}\) line) in the high-resolution narrow band \textit{Chandra} images of the circumnuclear regions, and their overall spatial correspondence with molecular line regions imaged with ALMA \citep{2020ApJ...902...49F}.

To shed additional light on the properties of the photoionized medium, we have investigated the properties of the [O III] to soft X-ray (\(0.5-2 \text{ keV}\)) flux ratio, \(F_{\text{[O III]}}/F_{0.5-2}\). For a single photoionized medium, this is expected to have an approximately power-law dependence on the radius {(assuming constant velocity and mass flux)}, depending on the radial density profile \citep{2006A&A...448..499B}. In the case of an outflowing nuclear wind, this ratio is expected to be constant (\citealt{2011ApJ...736...62W} in NGC 4151).

The upper-left panel of Fig. \ref{fig:OIII_X} shows the map of \(F_{\text{[O III]}}/F_{0.5-2}\). We estimated the [O III] flux from the HST-WFC3 with FQ508N filter image (see Sect. \ref{inner}) and the \(0.5-2 \text{ keV}\) flux from the observed \textit{Chandra} count rates assuming an exposure weighted average conversion factor of \(3.265\times{10}^{-11}\text{ erg} \text{ cm}^{-2}\) (corrected for {Galactic} absorption). This map shows that the regions associated with the giant loops and outflow regions have \(F_{\text{[O III]}}/F_{0.5-2}\) in the range \(\sim 15-60\) (green, see the color scale in {Fig.} \ref{fig:OIII_X}). These values are comparable to those reported in Seyfert galaxies \citep{2006A&A...448..499B} and some of the photoionized clouds of NGC 4151 \citep{2009ApJ...704.1195W, 2011ApJ...736...62W, 2011ApJ...742...23W}. The region within the central \(\sim 2.5 \text{ kpc}\) (\(r \sim 5''\)) shows lower values of the ratio (\(\sim 1-8\), see the blue bins in the figure). These lower values are consistent with those measured in the jet termination regions of NGC 4151, where shock-heated emission is present \citep{2009ApJ...704.1195W, 2011ApJ...736...62W, 2011ApJ...742...23W}, and {would be consistent} with the presence of strong thermal emission in these inner regions of NGC 6240 ({Section} \ref{sec:spectral_results}).

We have superimposed the same azimuthal bins used for the surface brightness profile extraction in Fig. \ref{fig:sur_bri} on the \(F_{\text{[O III]}}/F_{0.5-2}\) map in Fig. \ref{fig:OIII_X}. From these regions we have derived the radial \(F_{\text{[O III]}}/F_{0.5-2}\) profiles, also shown in  Fig. \ref{fig:OIII_X}. On these profiles, we also plot the values measured at the same physical radii for individual clouds of NGC 4151 by \citep{2011ApJ...736...62W}.

{These variations of  \(F_{\text{[O III]}}/F_{0.5-2}\) within NGC 6240 caution} against simple interpretations of the physical state of AGN emitting regions based only on ``average" \(F_{\text{[O III]}}/F_{0.5-2}\) measurements. In the case of NGC 4151, \citet{2011ApJ...736...62W} argued that the similarity of the \(F_{\text{[O III]}}/F_{0.5-2}\) ratios measured from individual clouds over a range of radii from \(\sim 100 \text{ pc}\) to \(\sim 1 \text{ kpc}\) demonstrated the presence of a nuclear wind (excluding the two uncertain measurements, and the clouds {interacting with the jet}, where the \(F_{\text{[O III]}}/F_{0.5-2}\) value is lower, because of the additional thermal emission). While we obtain similar values, and ``flat" [O III]/X-ray profiles within \(r\sim 5''\) in NGC 6240, both the spectral analysis and the X-ray surface brightness radial profiles in the same regions indicate a prevalent, confined, thermally emitting hot gas (Sec. \ref{sec:spectral_results}). Moreover, significant {structures are evident in} the 2D distribution of the \(F_{\text{[O III]}}/F_{0.5-2}\) ratio (Fig. \ref{fig:OIII_X}).

At radii \(r > 5''\), Fig. \ref{fig:OIII_X} shows that the \(F_{\text{[O III]}}/F_{0.5-2}\) ratio increases almost  monotonically as the X-ray flux decreases (Fig. \ref{fig:sur_bri}), reaching a value of \(\sim 100\) at a radius of \(\sim 20''\) (\(\sim 10 \text{ kpc}\)). At these large radii the X-ray emission is dominated by the thermal emission of an expanding halo (Sec. \ref{sec:spectral_results}; \citealt{2013ApJ...765..141N}).

{The observed \(F_{\text{[O III]}}/F_{0.5-2}\) ratio should then decrease as \(F_{0.5-2}\) is  being overestimated, and the photoionized \(F_{\text{[O III]}}/F_{0.5-2}\) should be constant \citep{2009ApJ...704.1195W, 2011ApJ...736...62W}, contrary to our results. The observed increase could be due to an intervening  warm, partially ionized, column of gas that gradually absorbs the AGN photoionizing X-rays while letting through the UV that excite the [O III] emission \citep{1984ApJ...281...90H}. Similar ratios were seen in NGC 4151 at the edges of a bicone where additional absorption is plausible (\citealt{2009ApJ...704.1195W}, see the two high value points in Fig. \ref{fig:OIII_X}). Note that this absorber is different from the cold absorber seen in the nuclei. To produce the observed gradual increase in \(F_{\text{[O III]}}/F_{0.5-2}\) this absorber must  be distributed over the \(\sim 5''-20''\) region. Once the thermal emission dominates, however, this explanation is insufficient. A candidate for this putative warm absorbing medium is the thermally emitting hot gas at \(kT \sim 1 \text{ keV}\) seen in the region.} Its density, evaluated from the emission measure of the spectral thermal components, is in fact \(\sim 0.4 \text{ cm}^{-3}\), corresponding to a column density \(\sim 3\times{10}^{21} \text{ cm}^{-2}\) integrated {radially} along the inner \(5''\), {which is in the right range to absorb \(\sim 1 \text{ keV}\) X-rays.}

\subsection{Narrow-Band Imaging}\label{sec:narrowband}

Our spectral analysis (Sect. \ref{sec:spectra}) has shown differences in the emission line properties from different regions of NGC 6240. Using the spectral data as a guide, we have produced images to study the {finer-scale} morphology of the X-ray {line} emission {using several narrow} spectral bands. Given the {limited energy resolution} of ACIS {(\(\sim 100 \text{ eV}\))} we cannot image the emission in individual lines, which are often spectrally blended. However, we can use the integrated spectrum of Fig. \ref{fig:spectrum_all} to select different emission features for imaging. The {bands} chosen for imaging, together with the emission lines that contribute to the emission, are listed in Table \ref{tab:line_energy}, while the resulting narrow-band images are presented in {the various panels of Fig.} \ref{fig:narrow-bands}.

Comparison of the images in the different spectral bands shows that:
\begin{itemize}
\item[a)] The emission features are more prominent at larger radii (outside the central \(\sim 2 \text{ kpc}\)) for the lower energies (\(<0.95 \text{ keV}\)). This could be due {to both} the lower AGN excitation of the ISM because of the Compton thick nuclear obscuration, and also an effect of larger line of sight \(N_H\) in the dustier central regions, as suggested by the large \(\sim 3\times {10}^{10} M_{\sun}\) molecular mass gas estimated in these regions with ALMA data \citep{2020ApJ...890..149T}.

\item[b)] The Outer Outflow [O III] region \citep{2018Natur.556..345M} appears to be more smoothly elongated at energies \(< 0.95 \text{ keV}\), where the O VII, O VIII and Ne IX lines contribute to the emission, than at higher energies. In particular the Outflow Ridge perpendicular to the outflow axis is particularly prominent at \(\sim 1 \text{ keV}\), and this region is the principal contributor to the emission at higher energies, consistent with the spectral results (Sect. \ref{sec:spectra}, compare the Outer Outflow and the Outflow Ridge spectra). The relatively strong ridge emission in the \(0.95 - 1.15 \text{ keV}\) band is interesting, because this is the spectral band to which the Ne X line contributes. O VII, O VIII, Ne IX and Ne X line emission has been related to strong shock excitation in nearby Seyfert galaxies \citep{2010ApJ...719L.208W, 2012ApJ...756...39P, 2019ApJ...870...69F, 2019ApJ...872...94M}, from interaction with radio jets. In the case of NGC 6240 we speculate that this emission may be generated by the interaction of the fast nuclear wind (\(\sim 800 \text{ km} \text{ s}^{-1}\), \citealt{2013A&A...549A..51F, 2013A&A...558A..87F}), or star formation energized outflow \citep{2014ApJ...781...55W} with local dense {molecular} clouds.

\item[c)] The large NW Loop traced by H\(\alpha\) emission \citep{2013A&A...549A..51F, 2018Natur.556..345M} has its X-ray emission peaking between \(0.7\) and \(1.15 \text{ keV}\), dominated by O VIII and Ne X emission lines. This structure is also detected at higher energies, with the contribution of Mg XI, Si XIII and S XV lines, usually associated with starburst activity \citep{2002MNRAS.335..241S, 2002A&A...382..843P}. The W filament is mainly detected between \(0.7\) and \(1.15 \text{ keV}\), and it appears dominated by iron lines (Fe XVIII, Fe XVII and Fe XX).

\item[d)] The two arms protruding southwards - namely the ES Extension and the WS Extension - both peak between \(0.7\) and \(1.15 \text{ keV}\), being dominated by O VIII, Fe XX and Ne X emission. The physical model suggests for these two regions the presence of a thermal gas component (with \(kT \sim 0.8 \text{ keV}\) and \(kT \sim 1 \text{ keV}\) for the ES and WS extension, respectively) with an additional mildly photo-ionized component in {both} regions, possibly related with the activity of the northern nucleus \citep{2018Natur.556..345M}.
\end{itemize}

The picture that results from these {narrow-band images is} the following:
\begin{enumerate}
\item The northern nucleus is characterized by a mildly photo-ionized component with \(\log{U}\sim 0\), and the same component is found in the ES and WS extension, dominated by O VIII and Ne X lines. These two southern extensions may {delineate} the edges of the ionization cone originating from the {northern} nucleus, where the nuclear wind interacts with the local dense ISM. {There is no counterpart to this half-bi-cone to the N, which is unusual in CT-AGNs \citep{2019ApJ...884..163F}.}
\item The southern nucleus shows an additional photo-ionization component with \(\log{(U)}\sim 1\), {that is} also found in the outer outflow region and in the outflow ridge, with these regions being dominated by O VII, O VIII, Ne IX and Ne X line emission, {that is associated with} strong shock excitation. This suggests a photoionized outflow from the southern nucleus, where the fast, outflowing nuclear wind \citep{2013A&A...549A..51F, 2013A&A...558A..87F, 2014ApJ...781...55W} is pushing out pre-existing ISM clouds {and} giving rise to shock ionization in denser {regions}. {As for the northern nucleus, there is no corresponding W cone.}
\item The H\(\alpha\) NW loop emission is also modeled with a \(\log{(U)}\sim 1\) photo-ionization component, however the strong Mg XI, Si XII and S XV lines detected here suggest {that  starburst activity is dominant in this region}.
\end{enumerate}

\subsection{{Inner} \(2 \text{ kpc}\) emission}\label{inner}
Fig. \ref{fig:optical-x} shows the \(\sim 2 \text{ kpc}\) central region of {NGC 6240} as imaged by HST-WFC3 with FQ508N filter (left panel), HST-WFC3 with F673N filter (center panel), and by \textit{Chandra}/ACIS-S in the \(0.3-3 \text{ keV}\) band (right panel; see also Fig. \ref{fig:adapsmooth_zoom} for a more {detailed} representation of the \(0.3-3 \text{ keV}\) ACIS data in the inner region). The regions around the two nuclei show a close resemblance of the morphology of the optical line emission {to} the overall soft X-ray emission, including the arc to the S and SW of the southern nucleus. While these morphological similarities may suggest a similar emission mechanism, there are subtle differences in both the optical and X-ray emission that support a more complex picture.

Fig. \ref{fig:narrow-bands_zoom} {shows} a different morphology of the X-ray emission as a function of energy.
The northern nucleus becomes visible at higher X-ray energies with respect to the southern one (see also Fig. 5 in \citealt{2017MNRAS.471.3483N}), consistent with the higher absorbing column estimated for the former (see Sect. \ref{sec:spectral_results}). Circumnuclear emission for the northern nucleus appears at energies \(>1.6 \text{ keV}\), while for the southern nucleus this is already visible above \(0.95 \text{ keV}\). The X-ray and H\(\alpha\) region to the SE does not have comparably strong [O III], suggesting a star formation origin. {There is an} X-ray arc {\(\sim 1''\)} south of the southern AGN {that} appears different at different energies, suggesting either local differences in cloud densities or different local conditions of interstellar shocks. {The arc may also result from the counterpart to the outflow in the northeast that, instead of escaping to large radii, interacts with molecular clouds in the region preventing a large-scale bi-cone from forming on this side. There is no similar feature corresponding to the counterpart of the northern nucleus outflow that complements the ES+WS outflow to the north. This could be because of obscuration from the galactic disk (see Fig. 1 of \citealt{2018Natur.556..345M}).}

\subsection{{Metal} enrichment of the hot ISM}

\textit{Chandra} ACIS spectra of galaxies have been used to constrain the metal abundances in the hot ISM. In particular, in merging and interacting galaxies the metal abundance of \(\alpha\) elements (O, Ne, Mg, Si) has been found to exceed the solar values, with ratios relative to Fe typical of the yield of SN II (see the Antennae, \citealt{2006ApJS..162..113B, 2006ApJ...636..158B}; NGC 4490, \citealt{2010ApJ...723.1375R}). These results are consistent with active star formation occurring in these galaxies, leading to fast evolution of the most massive stars and the consequent {SNe II} explosions. The extended halo of NGC 6240 was found to be enriched in \(\alpha\) elements \citep{2013ApJ...765..141N}, leading to a picture of \(\alpha\) element enriched winds propagating from the internal active region to the surrounding gaseous halo of this system. 

In this paper we studied the region where star formation activity is {currently} occurring as the result of the merging interaction. Table \ref{tab:line_phys} shows that the iron abundance of {the} thermal components is found to be about solar in the nuclear regions, while  it is lower in the NES Extension, WS Extension and NW Loop regions (\(\sim 0.3\) solar) and especially so in the Outflow Ridge, Outer Outflow, ES Extension and W Filament regions (\(\sim 0.1\) solar), consistent with the values reported by \citet{2013ApJ...765..141N}. The ratios of the abundance of \(\alpha\) elements to iron is found to be \(\gtrsim 2\) in all the regions considered in the present analysis. In Fig. \ref{fig:abundances} we plot these ratios and compare them with values obtained by \citet{2013ApJ...765..141N}, that appear to be consistent. We also compare our results with the expected yields by SNe Ia and SNe II \citep[][and reference therein]{2006ApJ...636..158B}. This figure shows that the ratios are consistent with {SNe II} {- and strongly inconsistent with SNe Ia -} yields for all the regions we have studied spectrally in NGC 6240, in agreement with the results of \citet{2013ApJ...765..141N} for the outer halo.

\section{Summary and Conclusions}\label{sec:summary}

\begin{figure}
	\centering
	\includegraphics[scale=0.17]{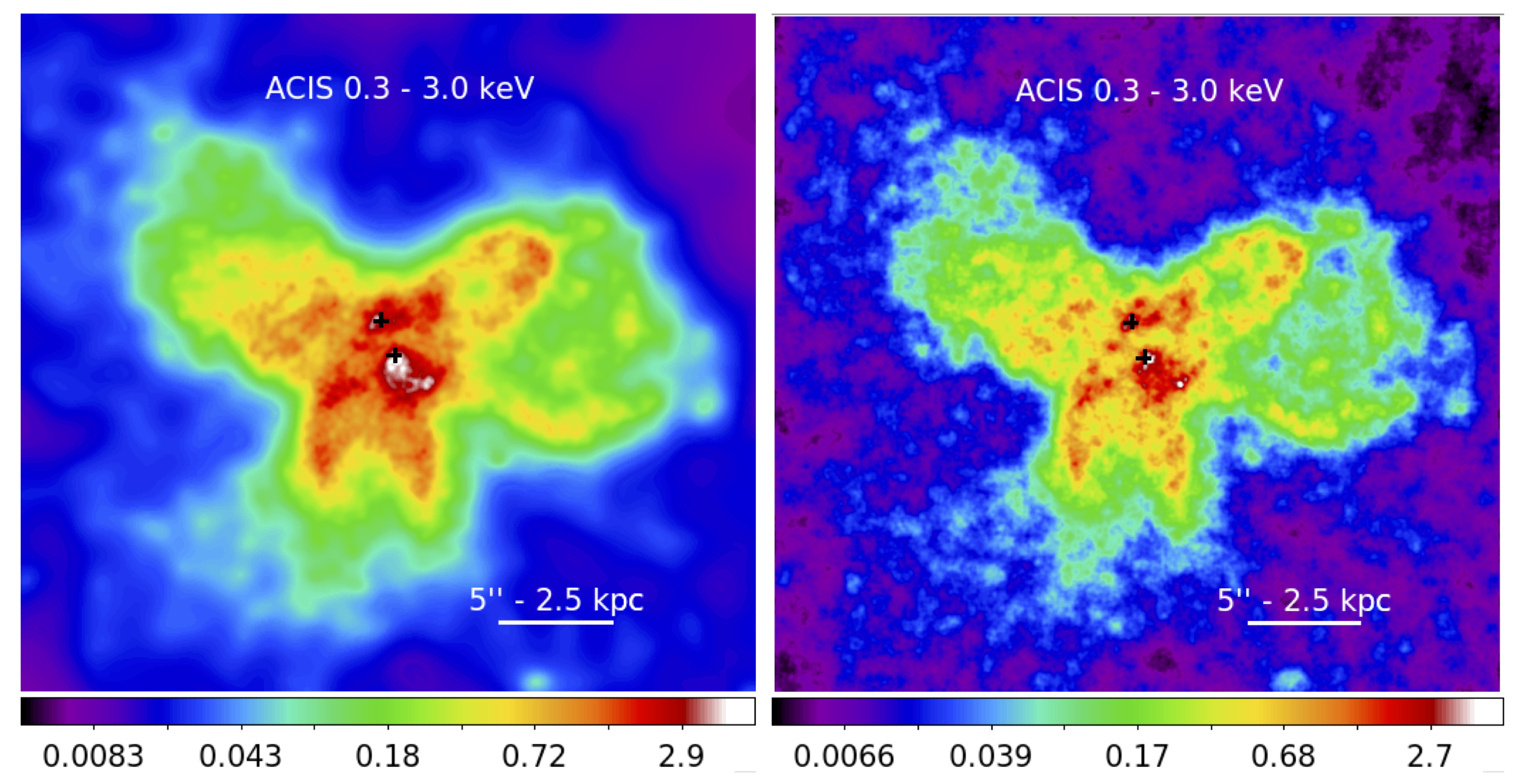}
	\caption{(Left panel) Adaptively smoothed image of the \textit{Chandra} ACIS merged data of NGC 6240 in the \(0.3 - 3 \text{ keV}\) energy band, with logarithmic color scale in counts per image pixel ({\(1/8\)} of ACIS pixel).  (Right panel) EMC2 reconstruction of the same data. N is to the top and E to the left of each image. The crosses are at the peak emission position of the two nuclear sources in the hard band image (see Fig. \ref{fig:adapsmooth_4-7}).}\label{fig:adapsmooth}
\end{figure}

\begin{figure}
	\centering
	\includegraphics[scale=0.5]{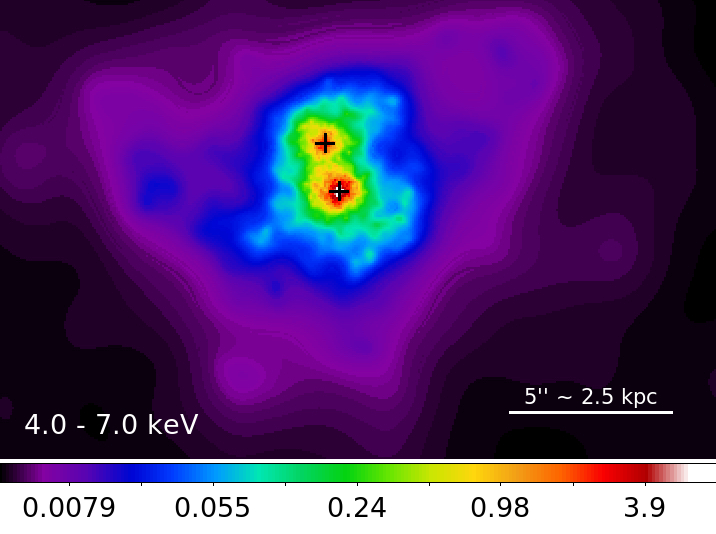}
	\caption{Adaptively smoothed image of the \textit{Chandra} ACIS merged data of NGC 6240, in the \(4-7 \text{ keV}\) energy {band}. N is to the top and E to the {left}. The color scale is logarithmic and the indicated color values at the {bottom} are in counts per image pixel (\(1/16\) of ACIS pixel). The crosses are the peak emission position of the two nuclear {sources}.}\label{fig:adapsmooth_4-7}
\end{figure}

\begin{figure}
	\centering
	\includegraphics[scale=0.39]{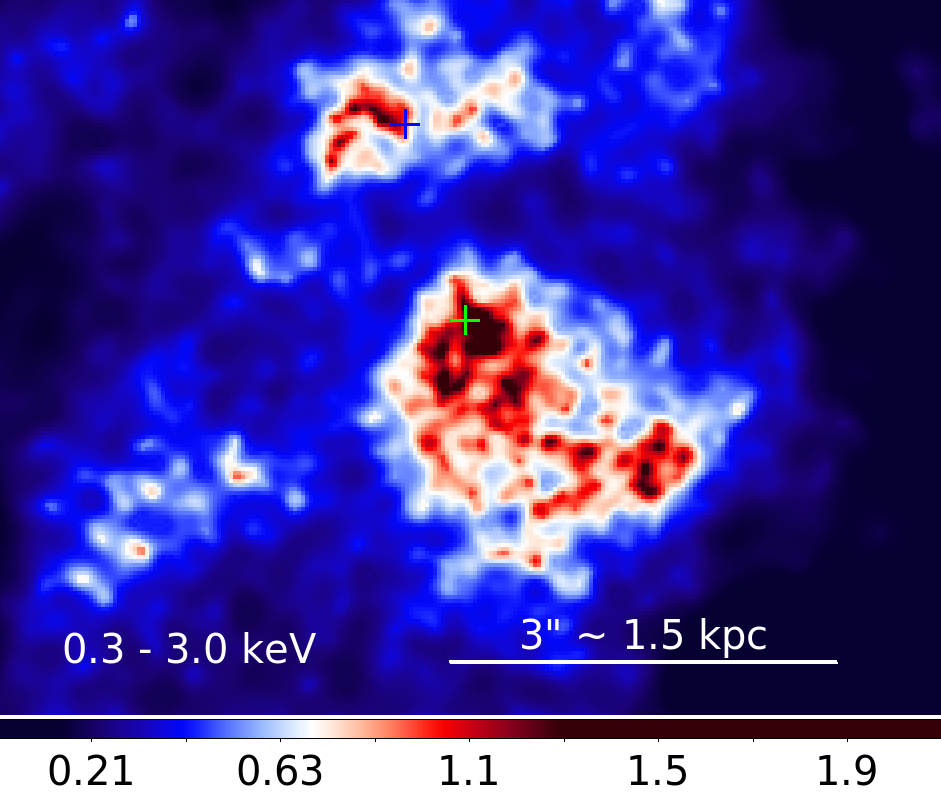}
	\caption{Adaptively smoothed image of the central region of the \(0.3-3 \text{ keV}\) \textit{Chandra} ACIS merged image of NGC 6240 (see main text). N is to the top and E to the left. The color scale is linear and has been chosen to highlight the filamentary emission. The indicated color values at the bottom of the image are in counts per image pixel (\(1/16\) of ACIS pixel). The crosses are the peak emission position of the two nuclear sources in the hard band image (see Fig. \ref{fig:adapsmooth_4-7}).}\label{fig:adapsmooth_zoom}
\end{figure}

\begin{figure}
	\centering
	\includegraphics[scale=1]{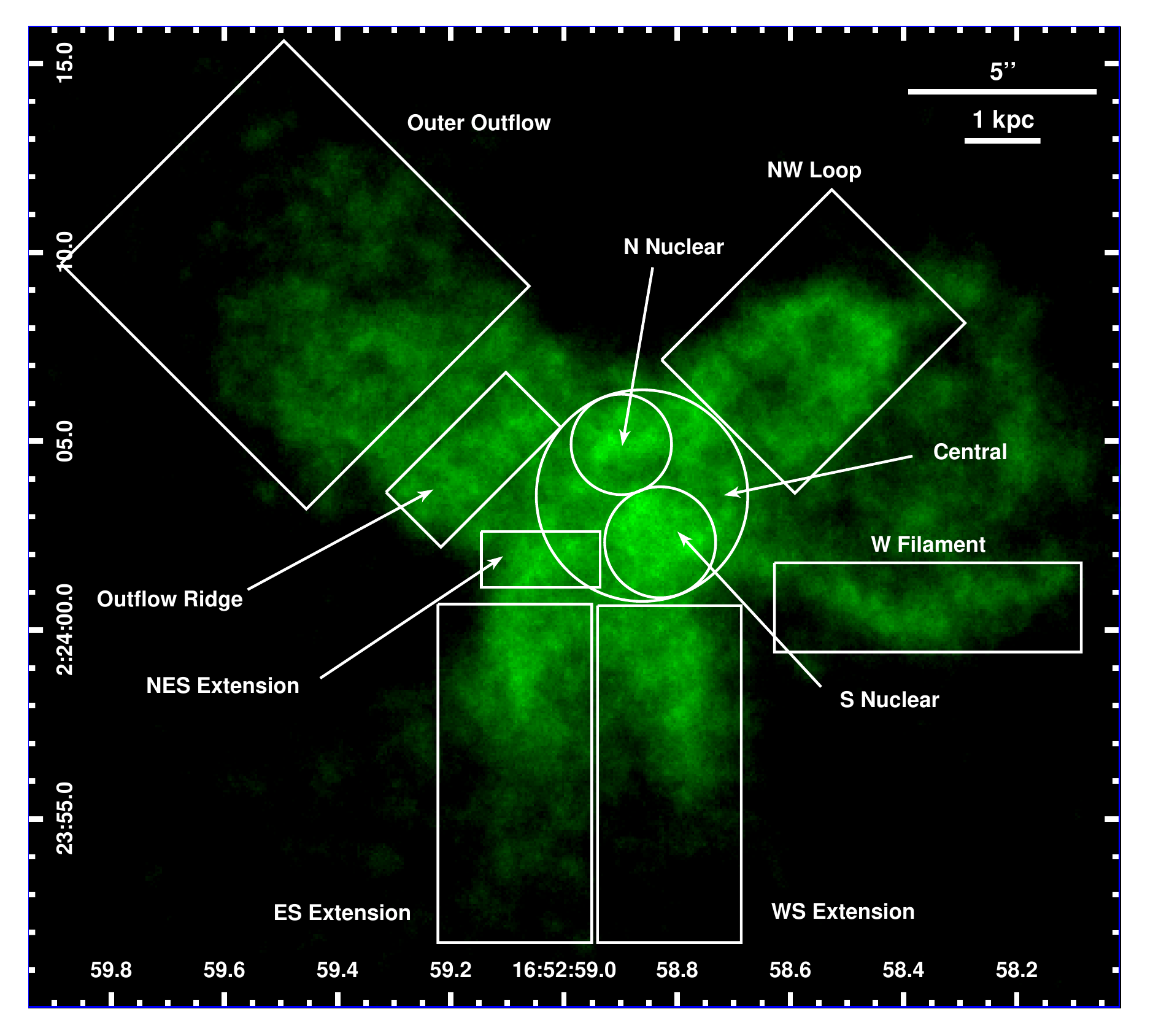}
	\caption{EMC2 reconstructed image of the \(1/16\) sub-pixel \textit{Chandra} ACIS merged data of NGC 6240 in the \(0.3 - 3 \text{ keV}\) energy band, with overlaid in white the spectral extraction regions discussed in Sects. \ref{sec:spectra} and \ref{sec:spectral_results}. Note that the Central region does not include the N Nucleus and S Nucleus regions.}\label{fig:regions}
\end{figure}

\begin{figure}
	\centering
	\includegraphics[scale=0.4]{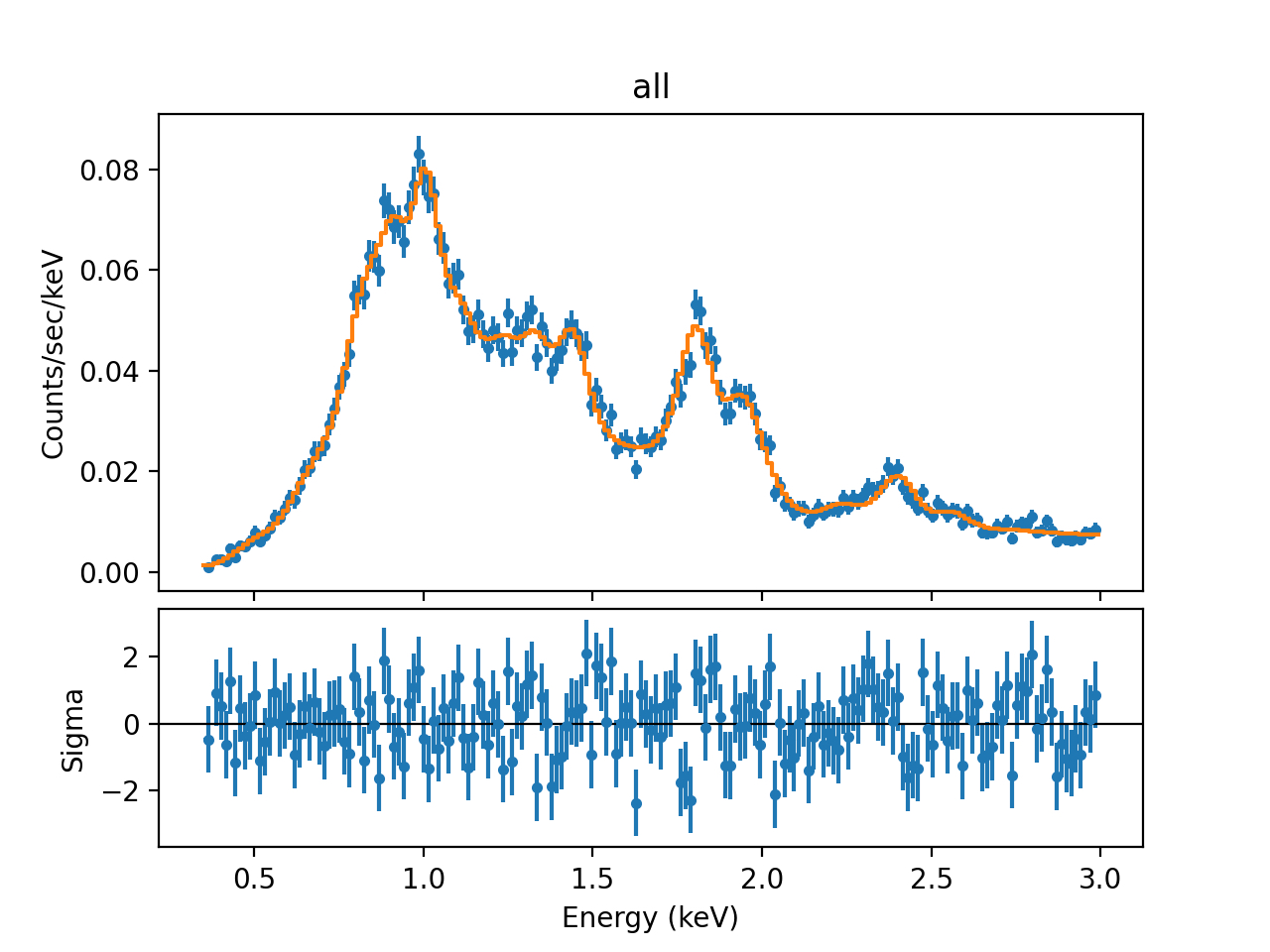}
	\caption{Best-fit phenomenological model (power-law plus Gaussian lines) for the spectrum extracted {from} a central \(15''\) region (upper panel) and corresponding residuals (lower panel).}\label{fig:spectrum_all}
\end{figure}

\begin{figure}
	\centering
	\includegraphics[scale=0.6]{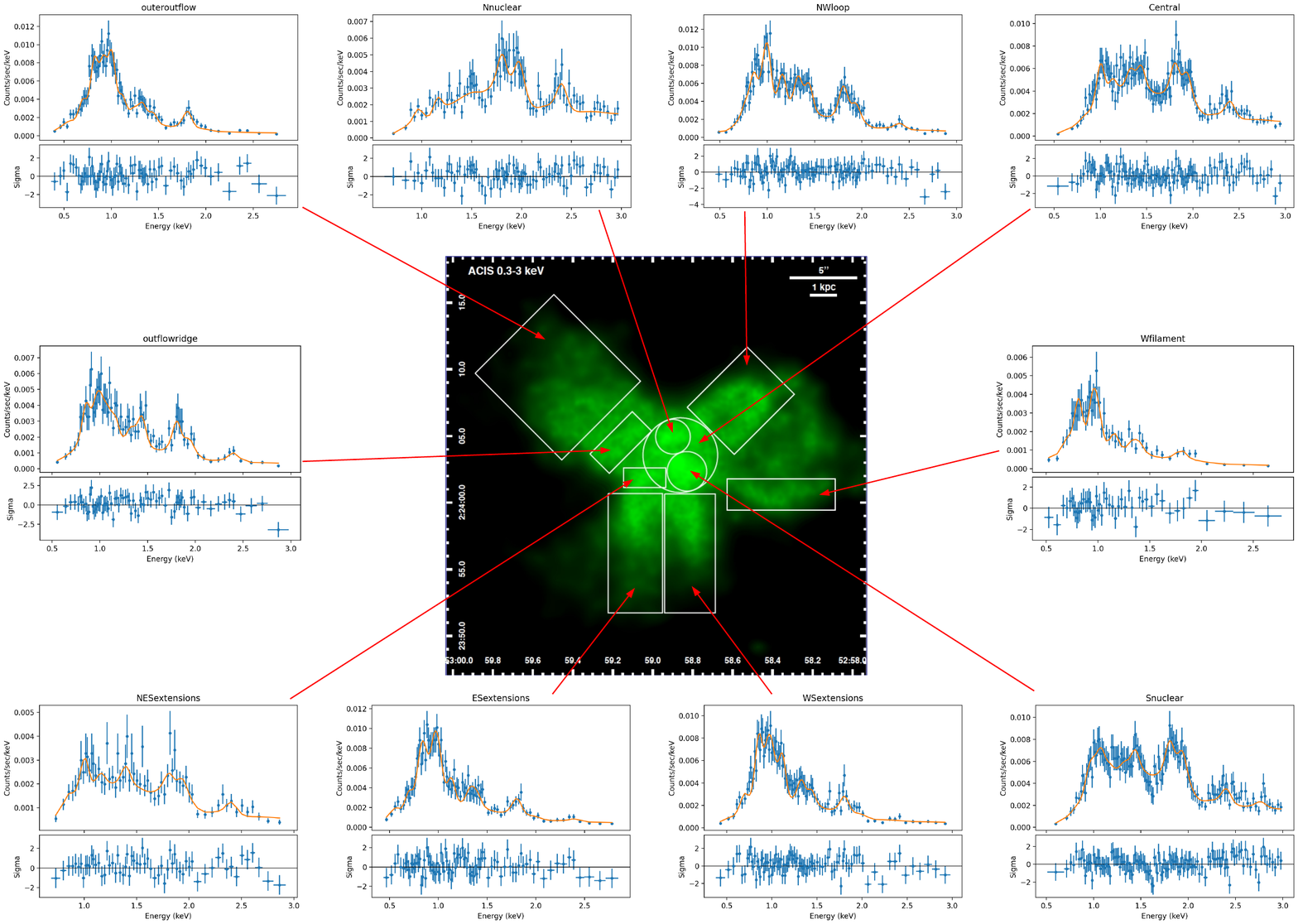}
	\caption{Best fit spectra for the phenomenological models (power-law plus Gaussian lines) for each of the extraction regions presented in Fig. \ref{fig:regions}.}\label{fig:line_regions}
\end{figure}

\begin{figure}
	\centering
	\includegraphics[scale=0.6]{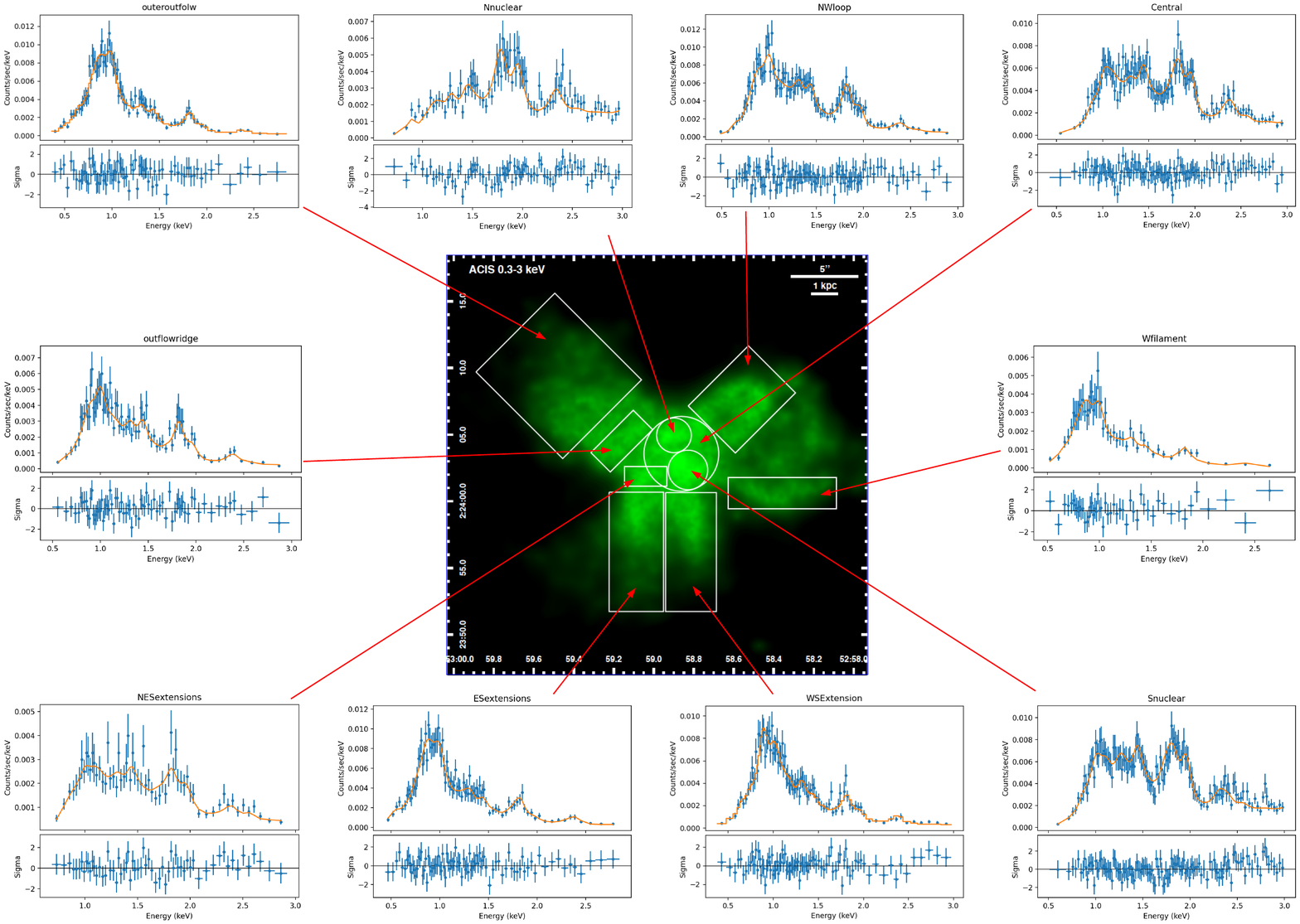}
	\caption{Best fit spectra for the physical models (thermal gas and photo-ionization components) for each of the extraction regions presented in Fig. \ref{fig:regions}.}\label{fig:phys_regions}
\end{figure}

\begin{figure}
	\centering
	\includegraphics[scale=0.5]{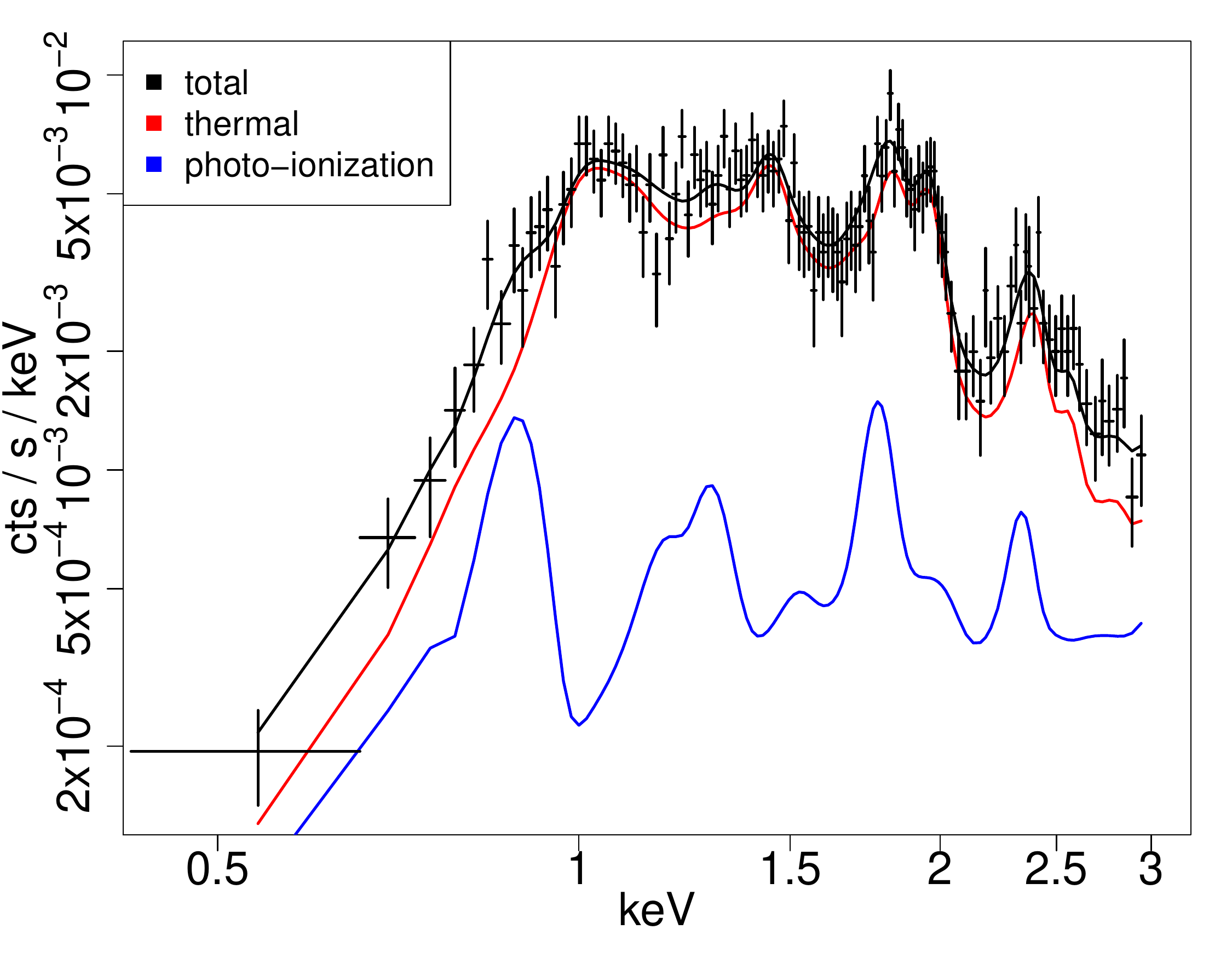}
	\caption{Decomposition of the best fit spectrum extracted in the Central region of NGC 6240 (see Fig. \ref{fig:regions}). The observed spectrum is shown with black crosses, and the best fit model is presented with a black line. The thermal and photo-ionization components are shown with red and blue lines, respectively.}\label{fig:components}
\end{figure}

\begin{figure}
	\centering
	\includegraphics[scale=0.3]{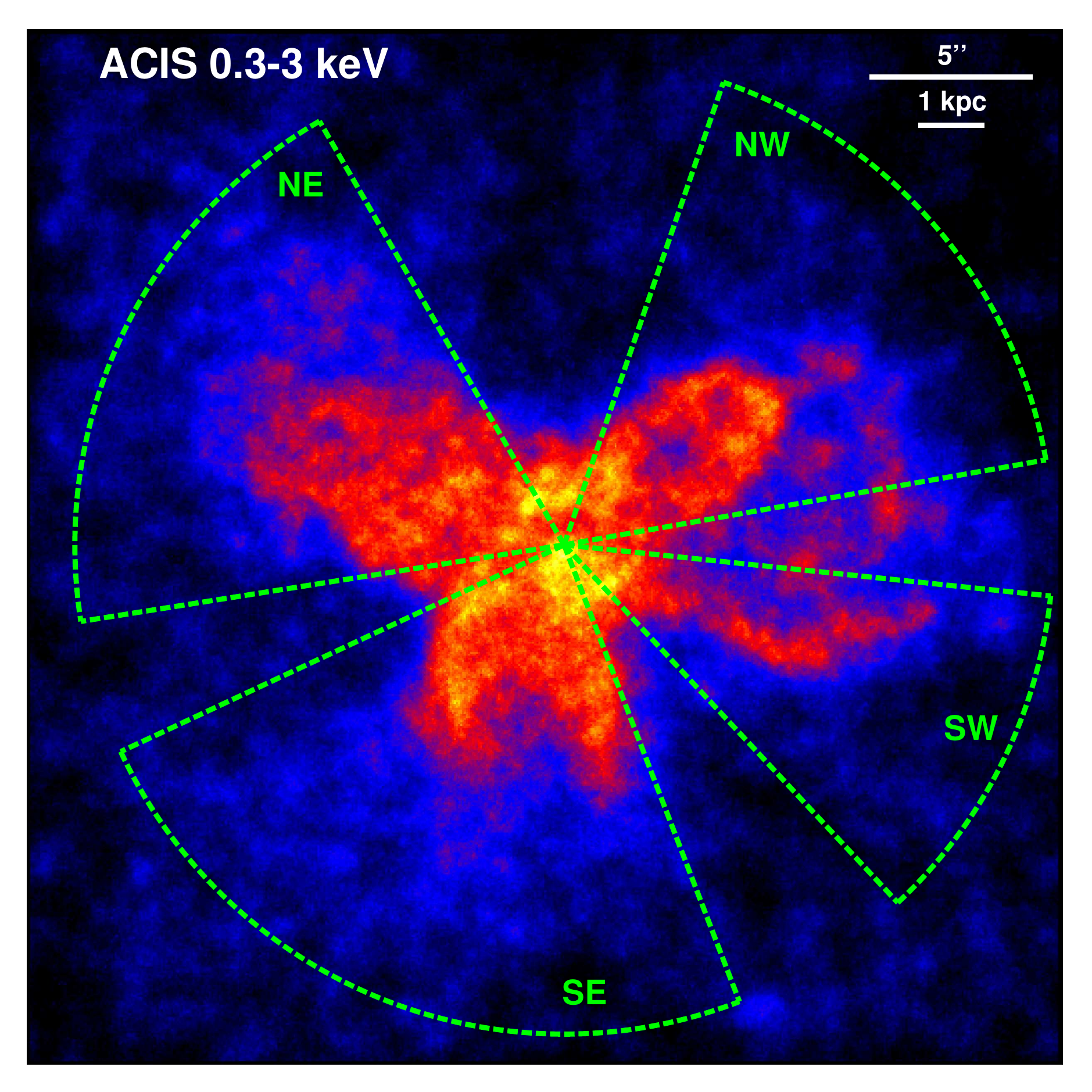}
	\includegraphics[scale=0.33]{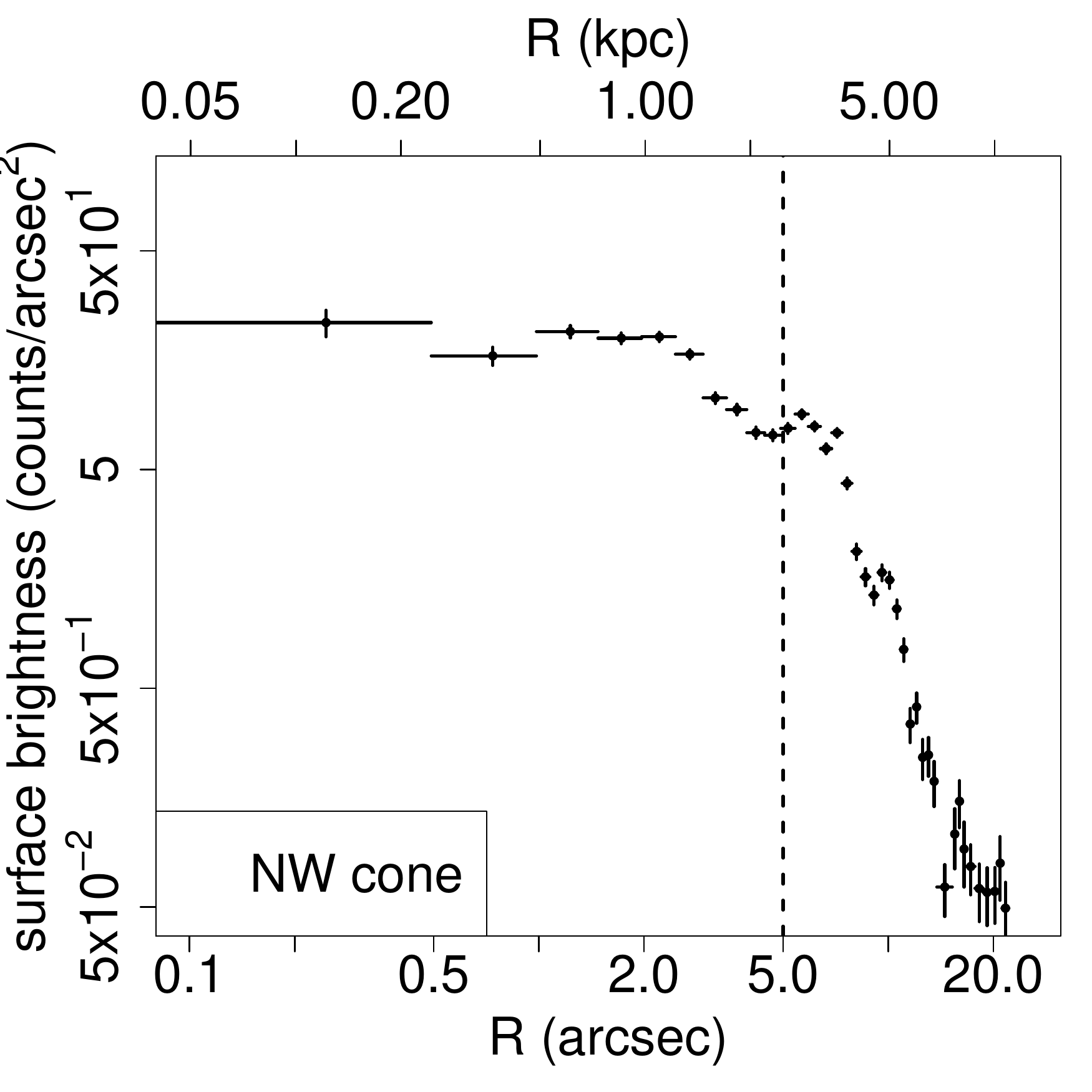}
	\includegraphics[scale=0.33]{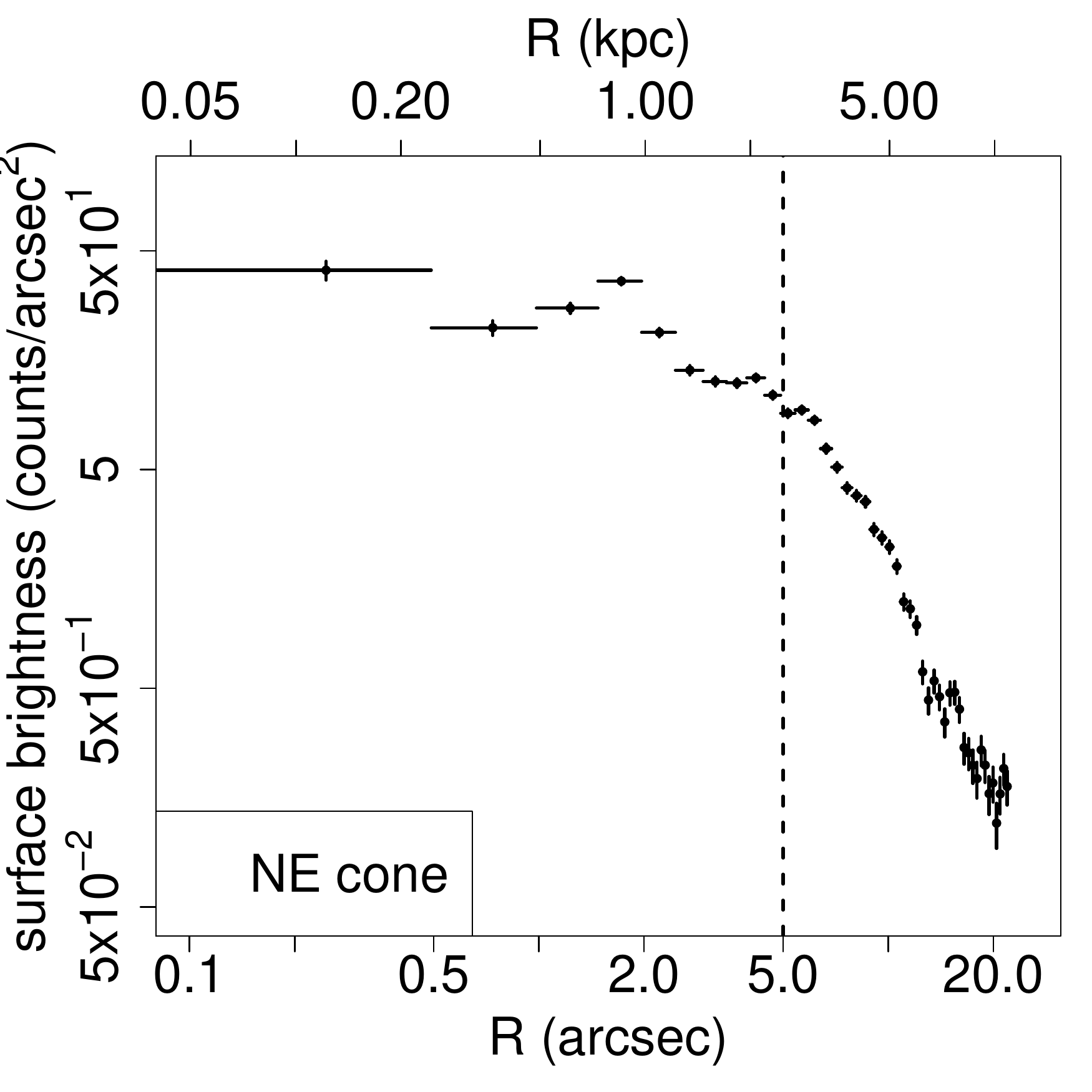}
	\includegraphics[scale=0.33]{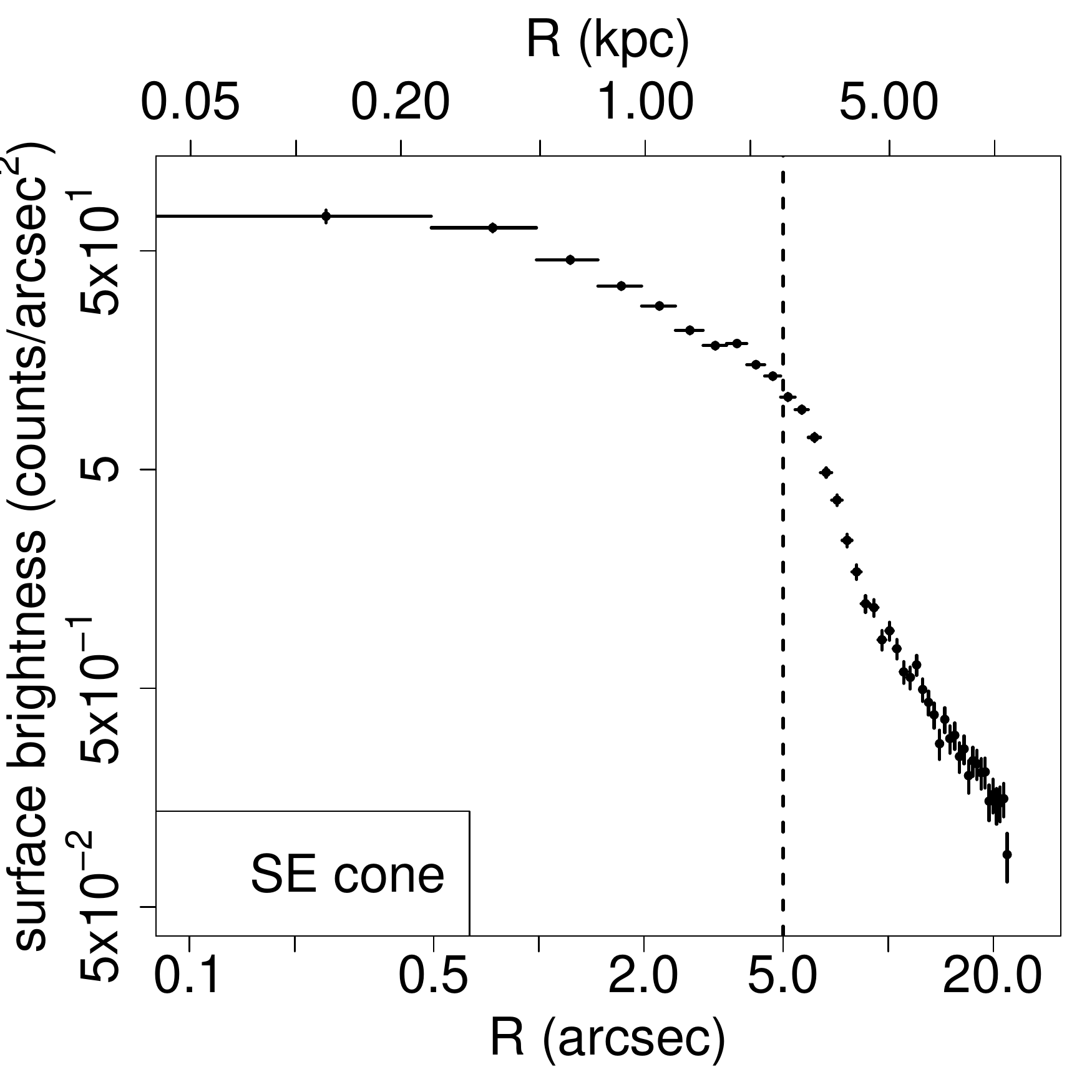}
	\includegraphics[scale=0.33]{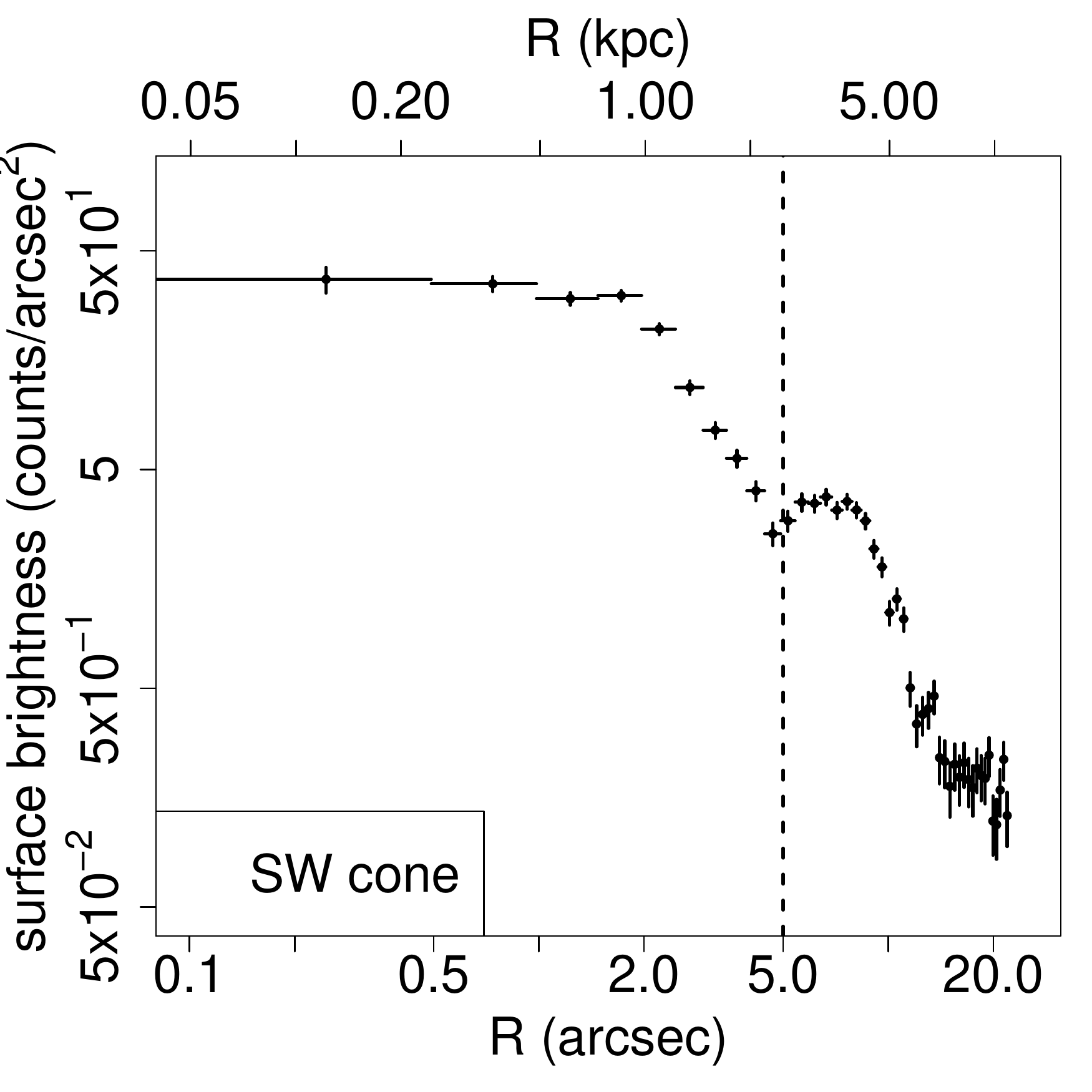}
	\includegraphics[scale=0.33]{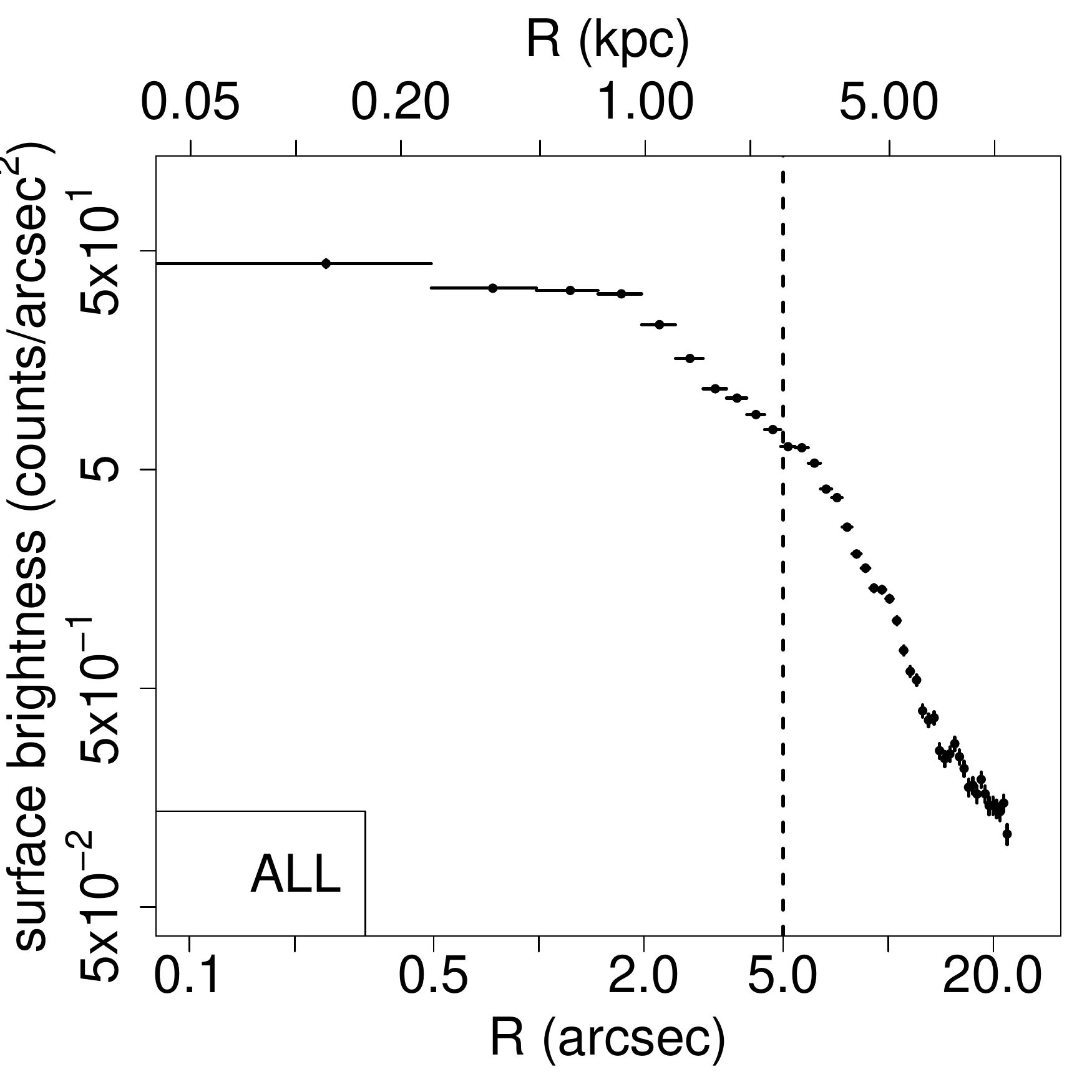}
	\caption{(Upper-left panel) Azimuthal bins used for the surface brightness profile extractions, overlaid to the EMC2 reconstructed image of the \(1/16\) sub-pixel \textit{Chandra} ACIS merged data of NGC 6240 in the \(0.3 - 3 \text{ keV}\) energy band. The surface brightness profiles extracted in the NW cone, NE cone,  SE cone and SW cone are presented in the upper-central, upper-right, lower-left and lower-central panels, respectively. In the lower-right panel is presented the profile extracted in the full \({360}^{\degree}\) sector. The vertical dashed line marks the \(5''\) radius separating the inner and outer profile regions (see main text).}\label{fig:sur_bri}
\end{figure}

\begin{figure}
	\centering
	\includegraphics[scale=0.3]{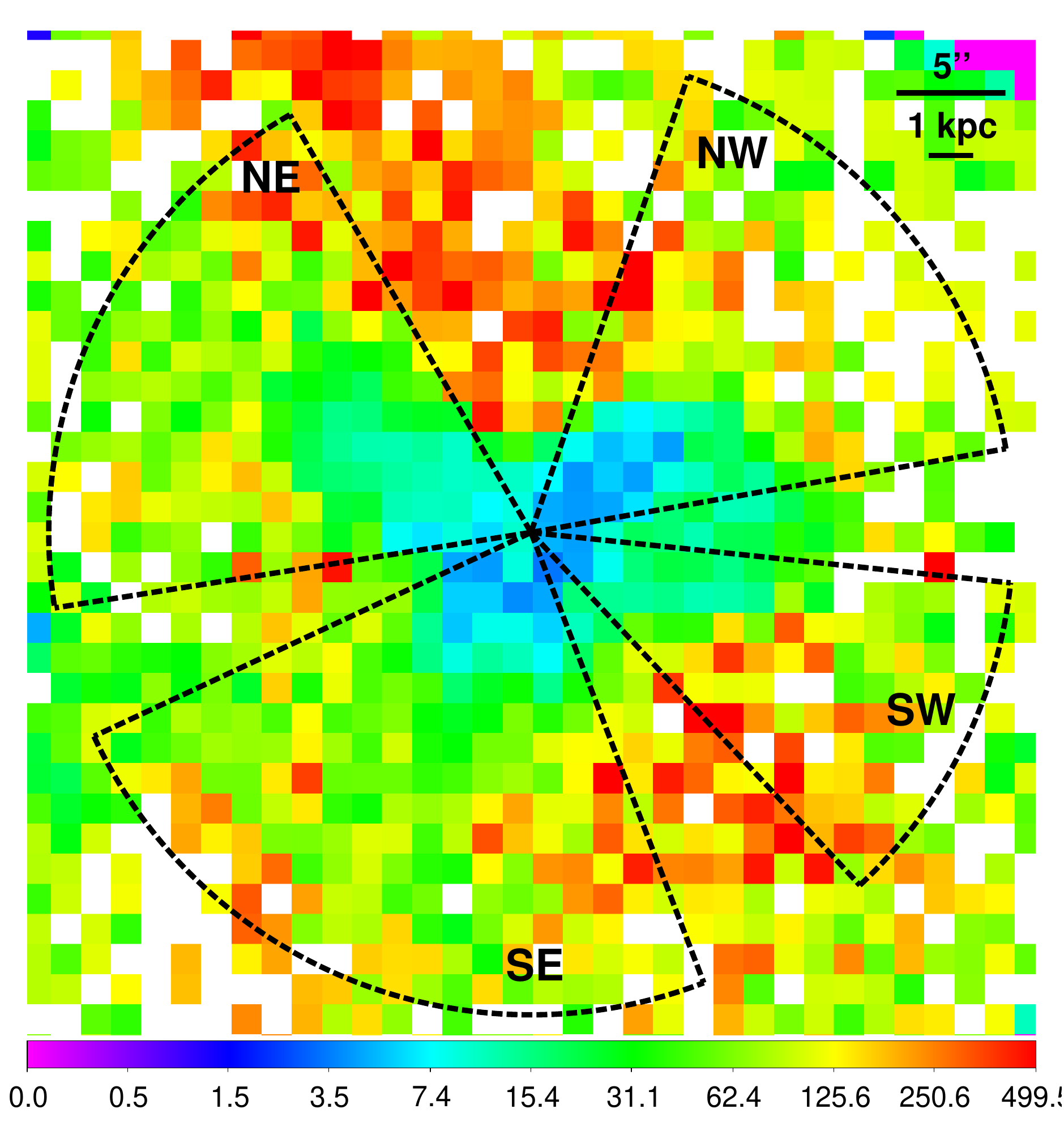}
	\includegraphics[scale=0.33]{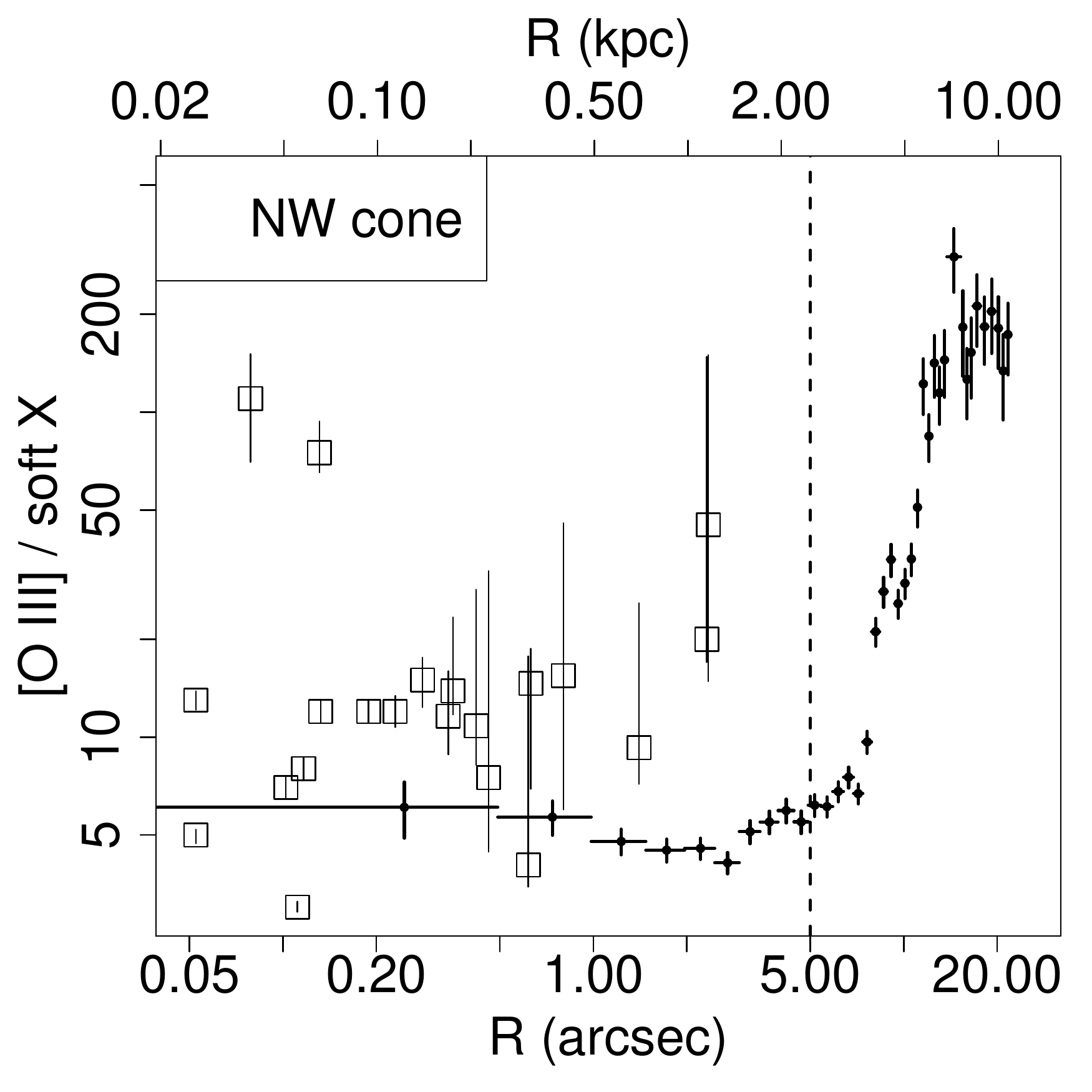}
	\includegraphics[scale=0.33]{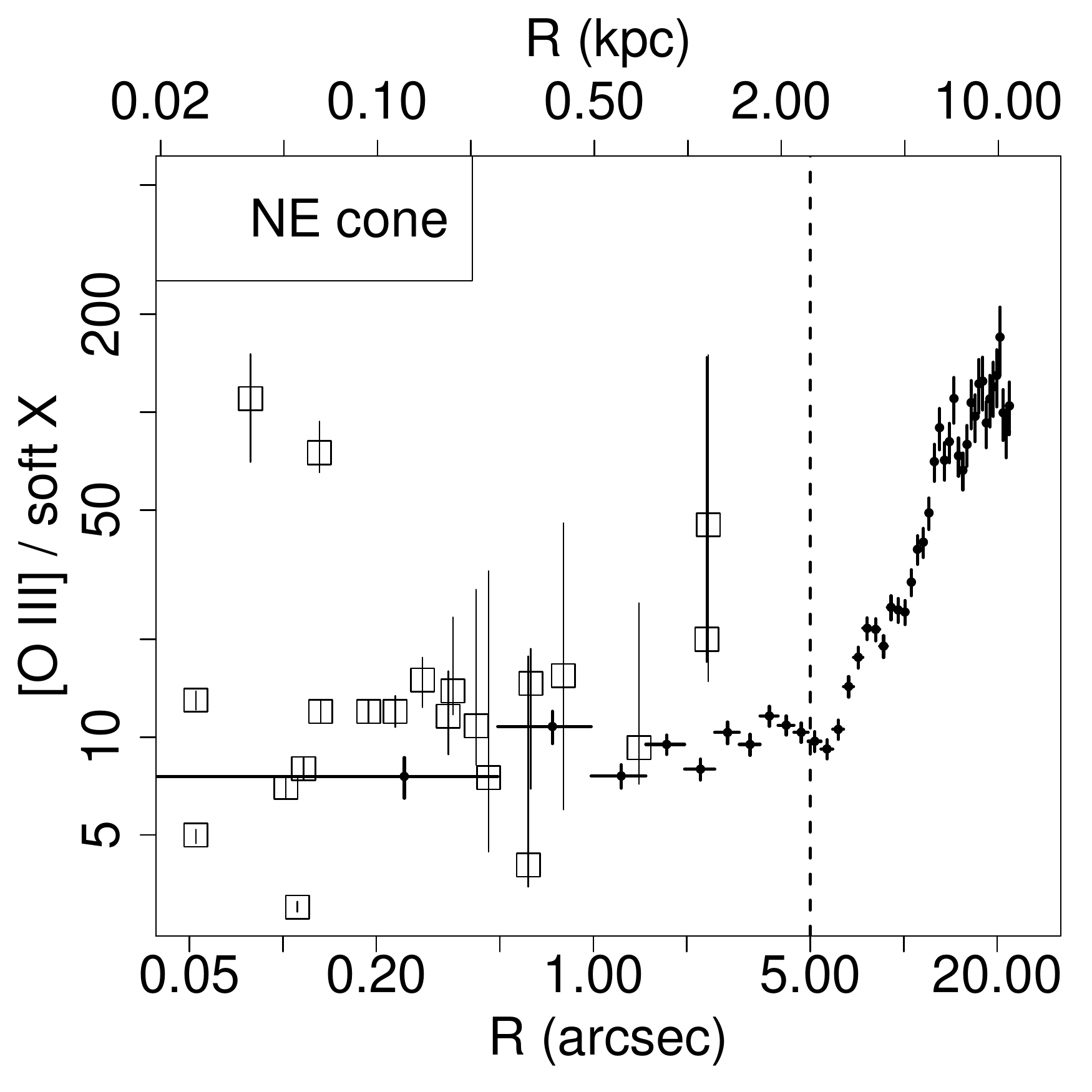}
	\includegraphics[scale=0.33]{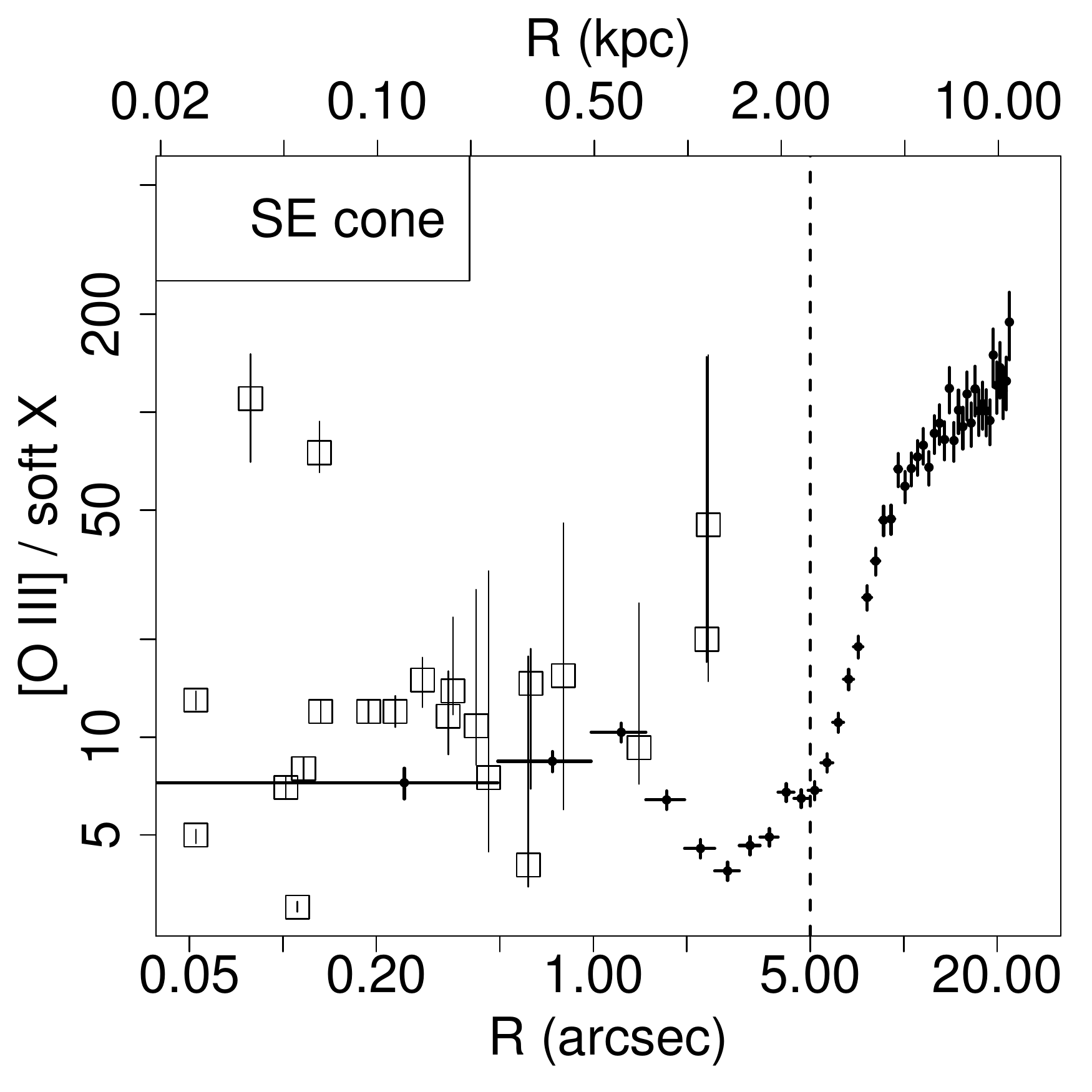}
	\includegraphics[scale=0.33]{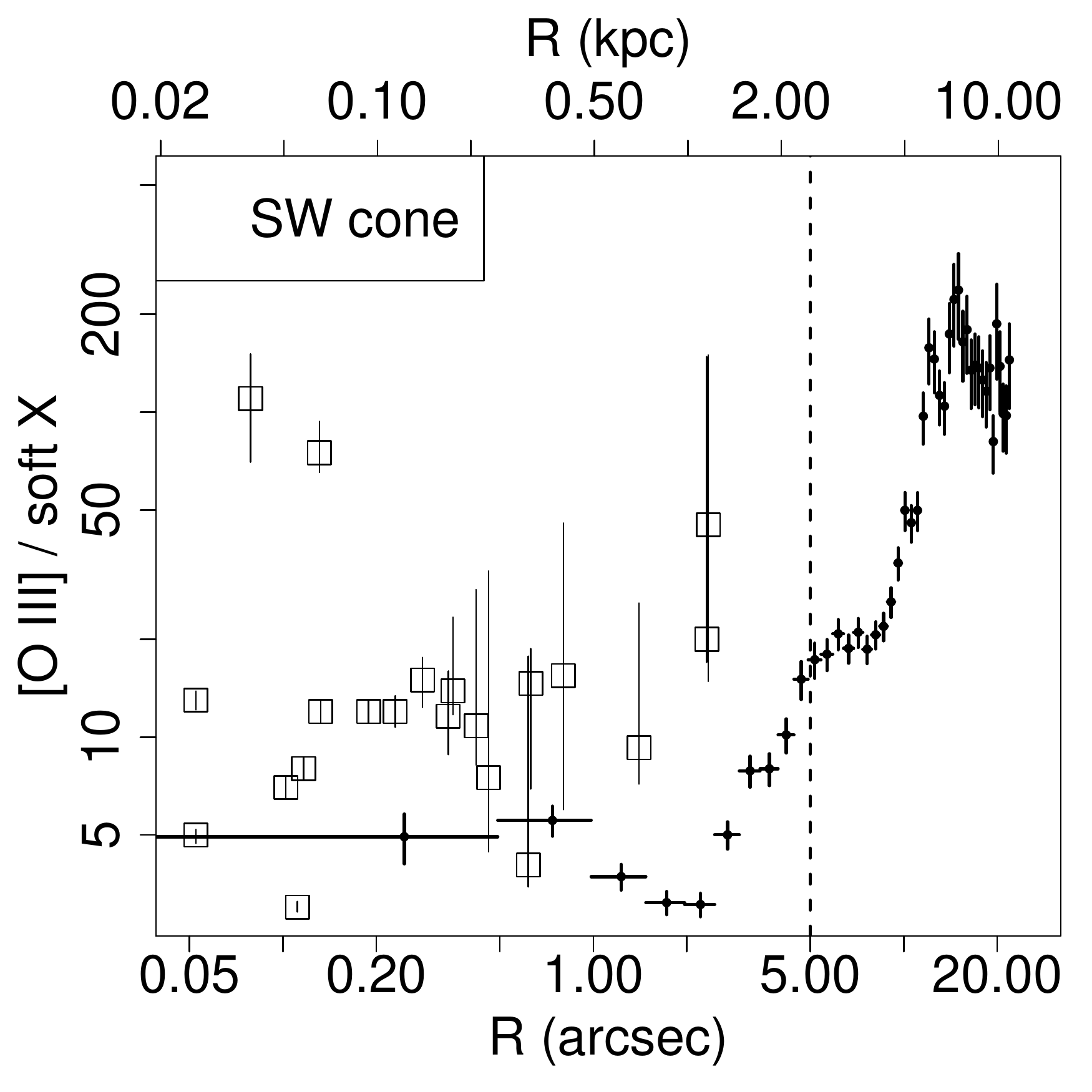}
	\includegraphics[scale=0.33]{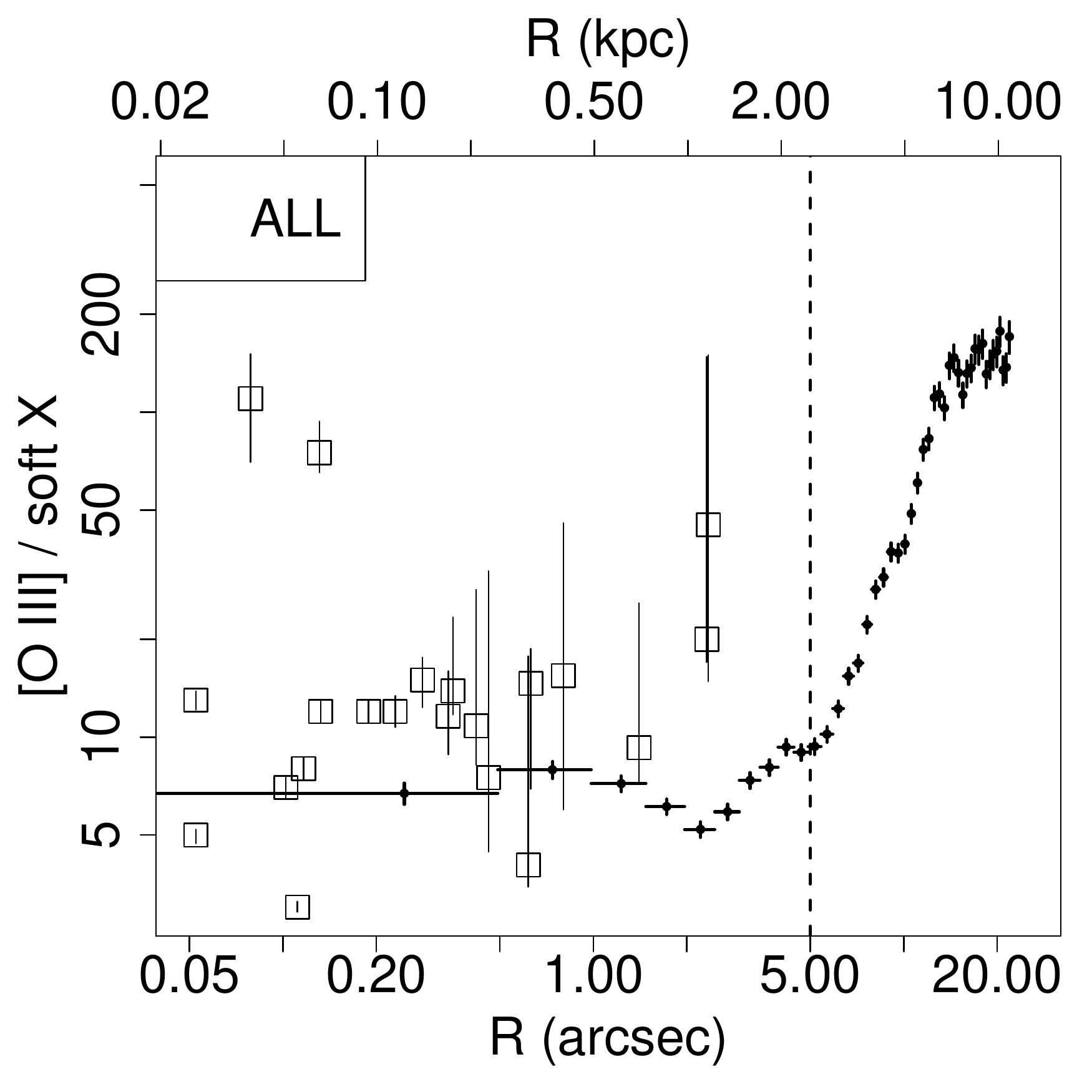}
	\caption{(Upper-left panel) Map of the [O III] to soft X-ray flux ratio, with superimposed the same azimuthal bins used for the surface brightness profile extraction (see Fig. \ref{fig:sur_bri}). The map is binned to a pixel size 20 times the HST-WFC3 with FQ508N filter image (see Fig. \ref{fig:optical-x}). The white pixels represent infinite values of the flux ratio (i.e., zero X-ray flux). The flux ratio profiles extracted in the NW cone, NE cone, SE cone and SW cone are presented in the upper-central, upper-right, lower-left and lower-central panels, respectively. In the lower-right panel is presented the profile extracted in the full \({360}^{\degree}\) sector. The vertical dashed line marks the \(5''\) radius separating the inner and outer profile regions. The squares represent the ratios measured by \citet{2011ApJ...736...62W, 2011ApJ...742...23W} in NGC 4151. }\label{fig:OIII_X}
\end{figure}

\begin{figure}
	\centering
	\includegraphics[scale=0.5]{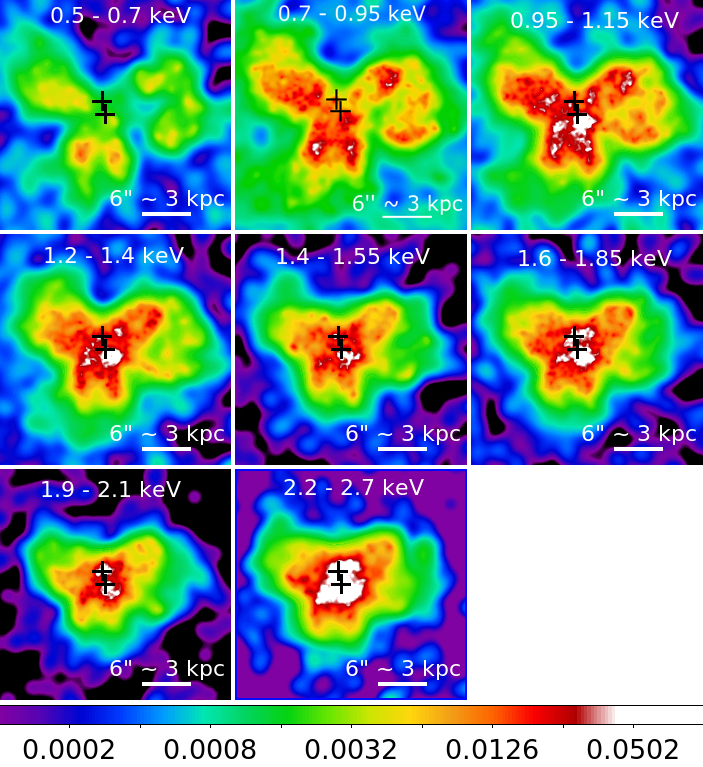}
	\caption{Narrow-band images of the \(1/16\) pixel merged data (see main text). The energy band is indicated in each panel. The data were adaptively smoothed with \(10\) counts under the Gaussian kernel; a range of kernels from \(1\) to \(30\) image pixels and \(30\) iterations were used in all cases. The color scale gives the logarithmic intensity scale in counts per image pixel, which was set to be the same for each panel, in order to high-light the larger scale lower-surface brightness features. N to the top and E to the left of each image. The crosses are the peak emission position of the two nuclear sources in the hard band image (see Fig. \ref{fig:adapsmooth_4-7}).}\label{fig:narrow-bands}
\end{figure}

\begin{figure}
	\centering
	\includegraphics[scale=1]{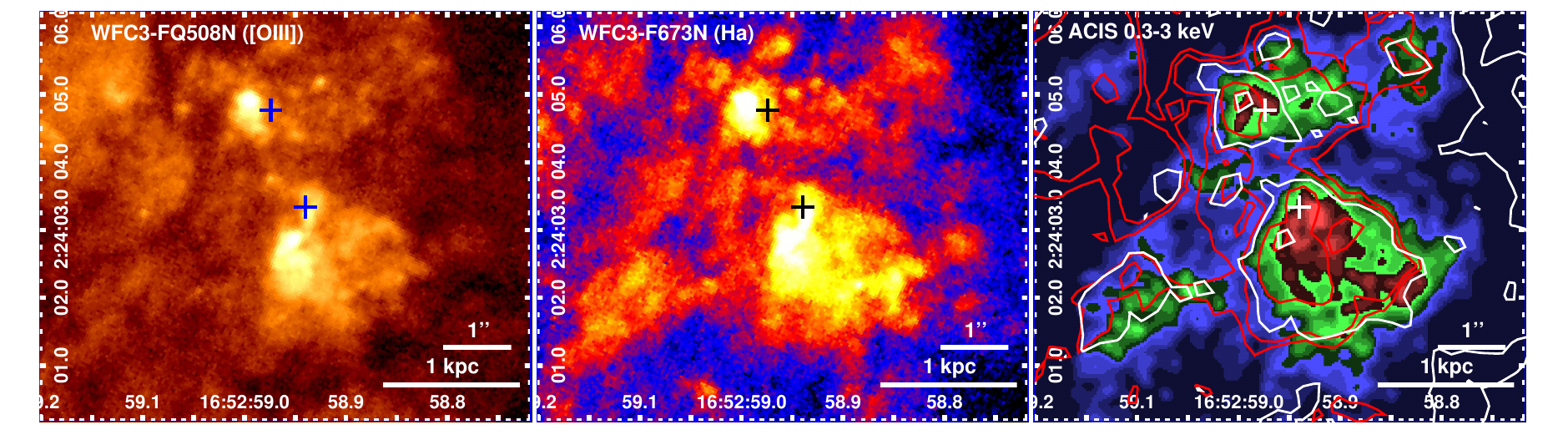}
	\caption{The central region of NGC 6240 (see Fig. \ref{fig:adapsmooth}) as imaged by HST-WFC3 with FQ508N filter (left panel), HST-WFC3 with F673N filter (center panel), and by \textit{Chandra}/ACIS-S in the \(0.3-3 \text{ keV}\) band (right panel). In the right panel we overlap to the adaptively smoothed \textit{Chandra} map the contours of [O III] and H\({\alpha}\) emissions, represented with red and white lines, respectively. The crosses are the peak emission position of the two nuclear sources in the hard band image (see Fig. \ref{fig:adapsmooth_4-7}).}\label{fig:optical-x}
\end{figure}

We have presented a detailed spectral and imaging analysis of the central \(15''\) (\(\sim 7.5 \text { kpc}\)) radius region of the double AGN merger galaxy NGC 6240 using the complete available \textit{Chandra}-ACIS data set. This {consists} of two imaging and two grating observations with combined effective ACIS-S imaging exposures of \(\sim 190 \text{ ks}\) at \(0.3-3 \text{ keV}\), \(\sim 210 \text{ ks}\) at \(3-6 \text{ keV}\), and \(\sim 360 \text{ ks}\) at higher energies. To exploit the superior \textit{Chandra}-ACIS spatial resolution, we have made use of the sub-pixel binning up to 1/16 of the native \(0.492''\) pixel size and PSF-based image  restoration techniques to separate the emission coming from the different structures observed in both {the X-ray} and optical bands. Besides the two highly obscured active galactic nuclei \citep{2003ApJ...582L..15K}, the highly disturbed central region of NGC 6240 shows different extended X-ray structures (see Fig. \ref{fig:adapsmooth}) with counterparts imaged in CO, [O III] and H\(\alpha\) line emission \citep{2013A&A...549A..51F, 2013A&A...558A..87F, 2018Natur.556..345M}. The ACIS {resolution} has been used to characterize the emission mechanisms in different spatial regions (Figs. \ref{fig:regions}, \ref{fig:line_regions}, \ref{fig:phys_regions}) and to image NGC 6240 in different energy bands (Fig. \ref{fig:narrow-bands}).

\begin{figure}
	\centering
	\includegraphics[scale=0.25]{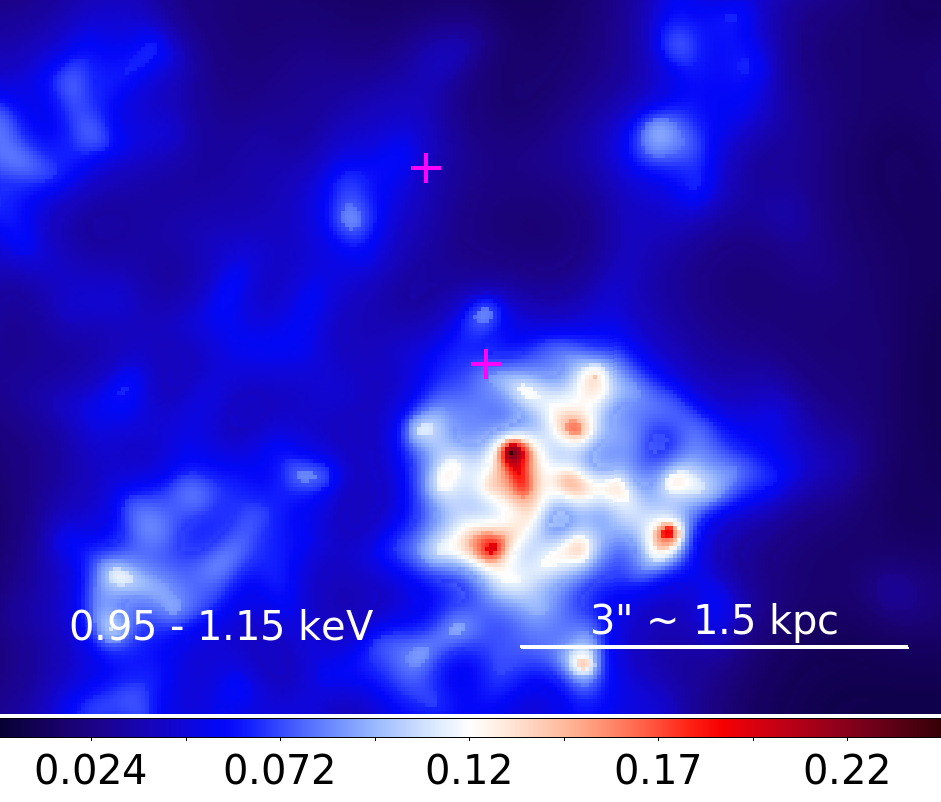}
	\includegraphics[scale=0.2506]{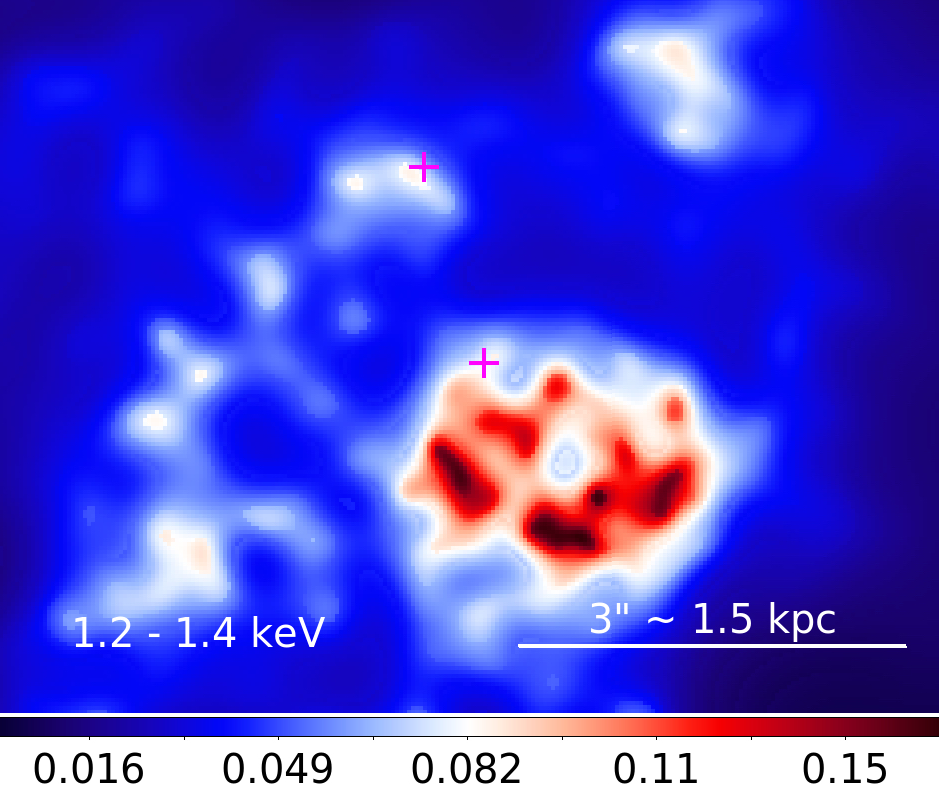}
	\includegraphics[scale=0.25]{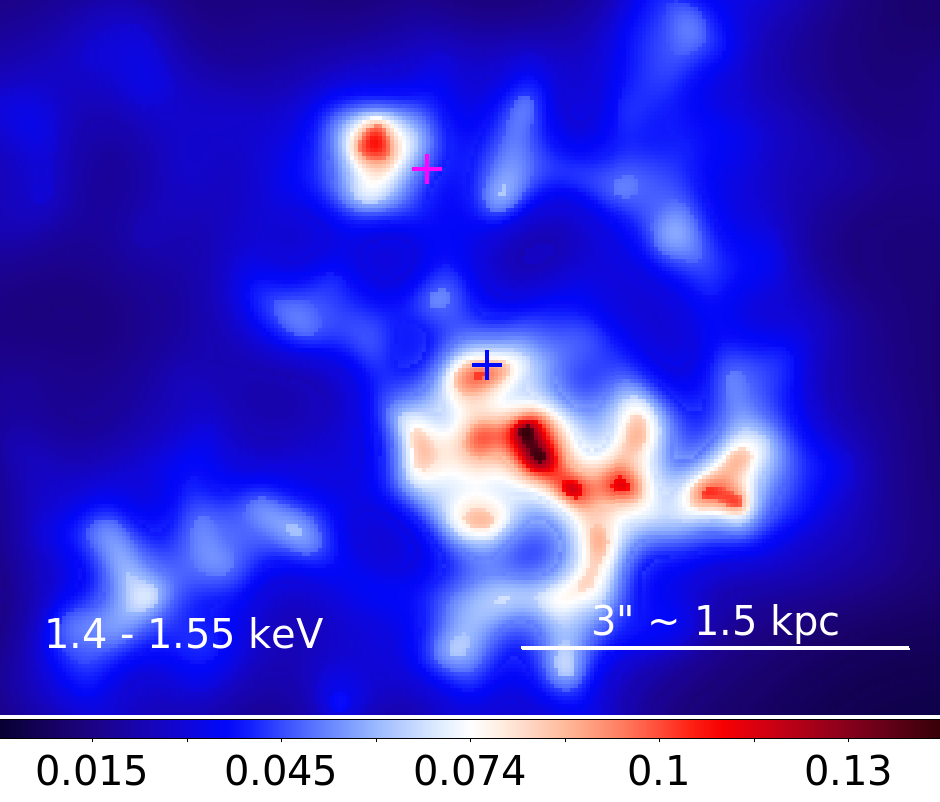}
	\includegraphics[scale=0.25]{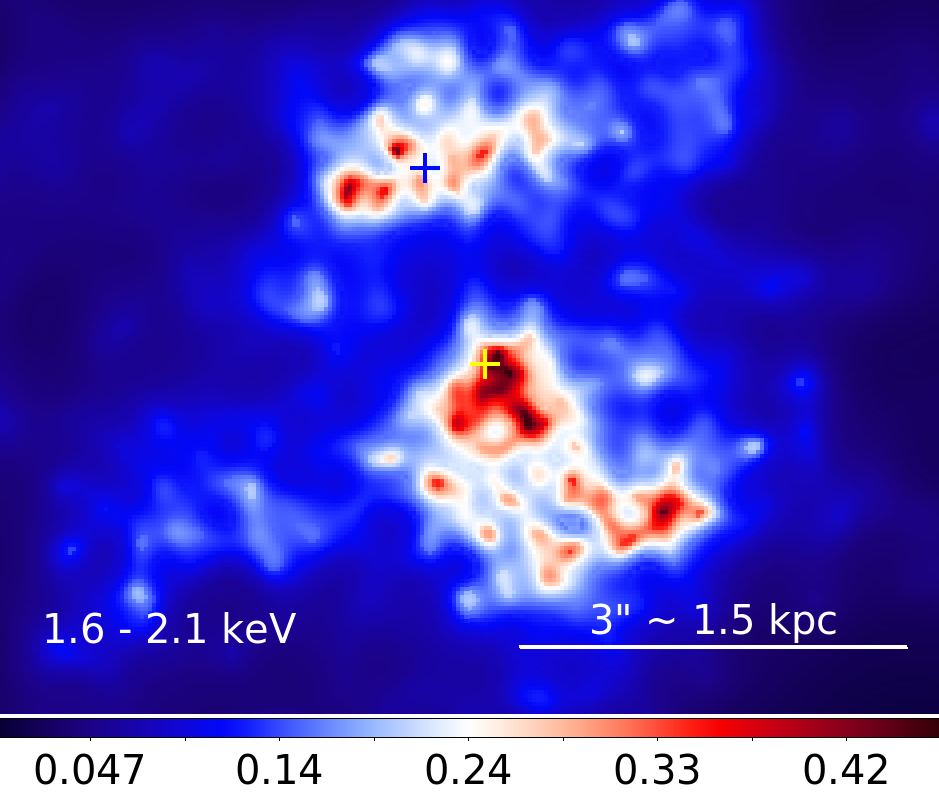}		
	\caption{Narrow-band images of the inner \(\sim 2 \text{ kpc}\) region (see main text). The energy band is indicated in each panel. From top {left}, these bands approximately correspond to Ne X, Mg XI, Mg XII and Si XIII-XIV emission lines. The data were adaptively smoothed with \(10\) counts under the Gaussian kernel; a range of kernels from \(1\) to \(30\) image pixels and \(30\) iterations were used in all cases. The color scale is linear in all {panels}, to emphasize the higher surface brightness filaments. The intensity scales are in counts per image pixel. N to the top and E to the left of each image. The crosses are the peak emission position of the two nuclear sources in the hard band image (see Fig. \ref{fig:adapsmooth_4-7}).}\label{fig:narrow-bands_zoom}
\end{figure}

\begin{figure}
	\centering
	\includegraphics[scale=0.33]{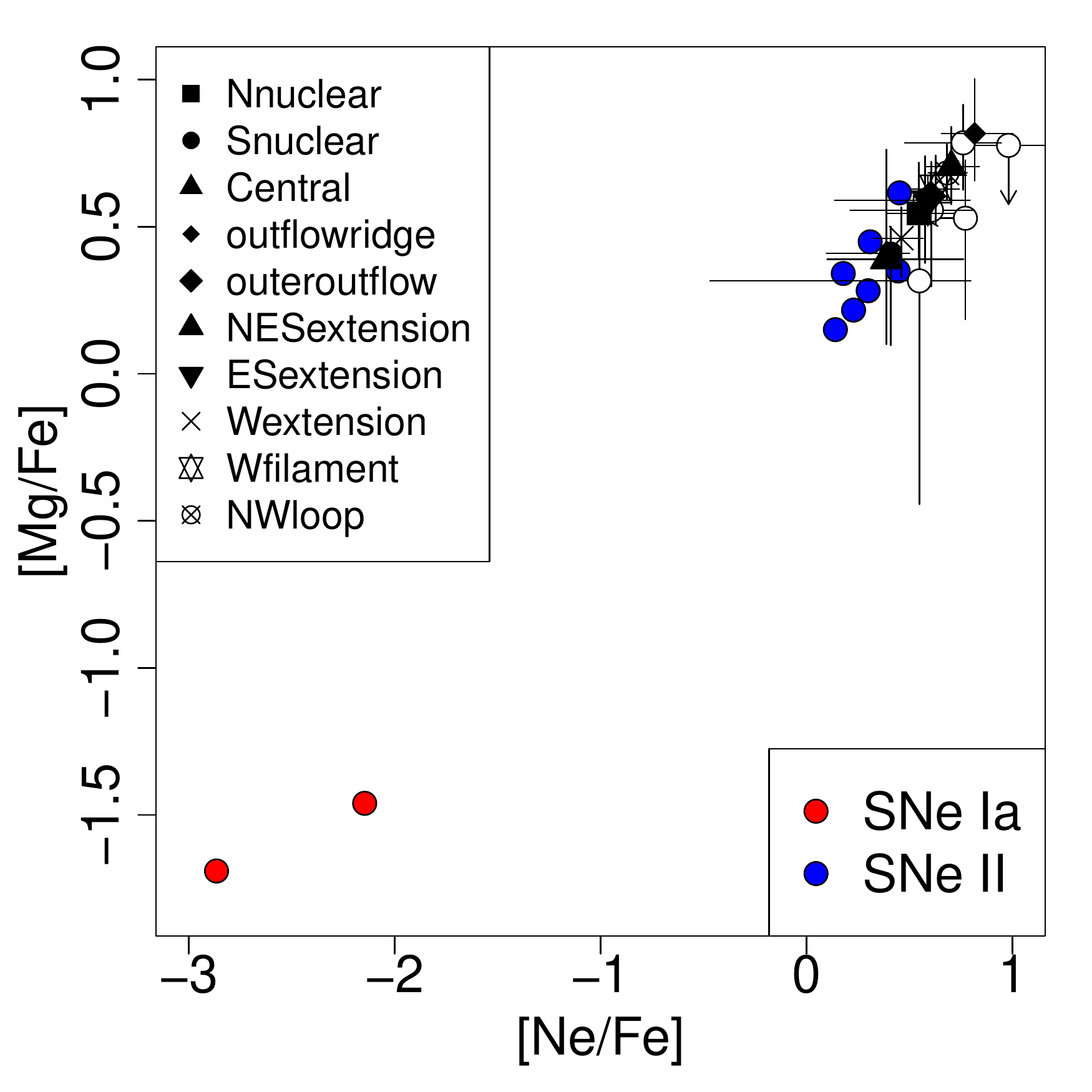}
	\includegraphics[scale=0.33]{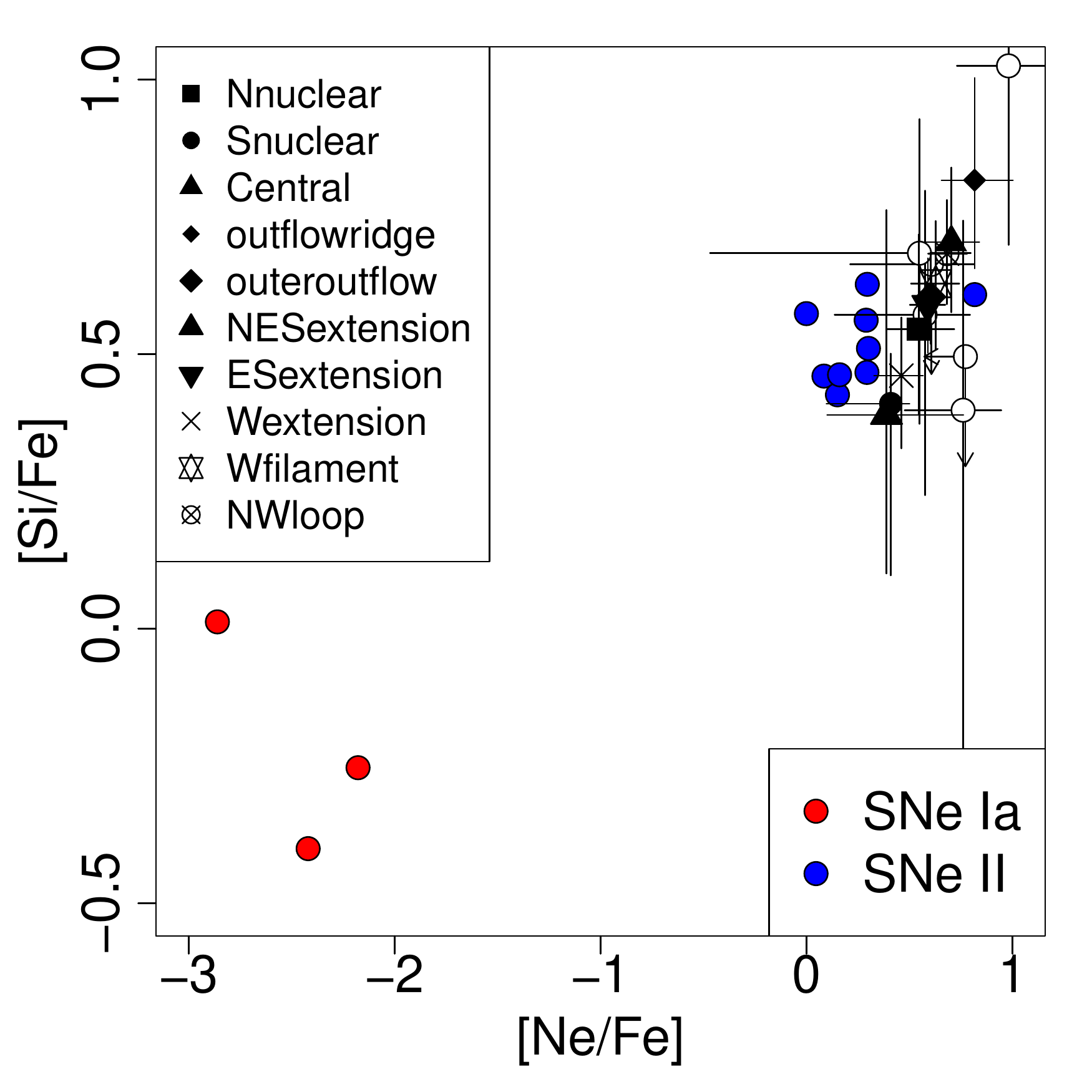}
	\includegraphics[scale=0.33]{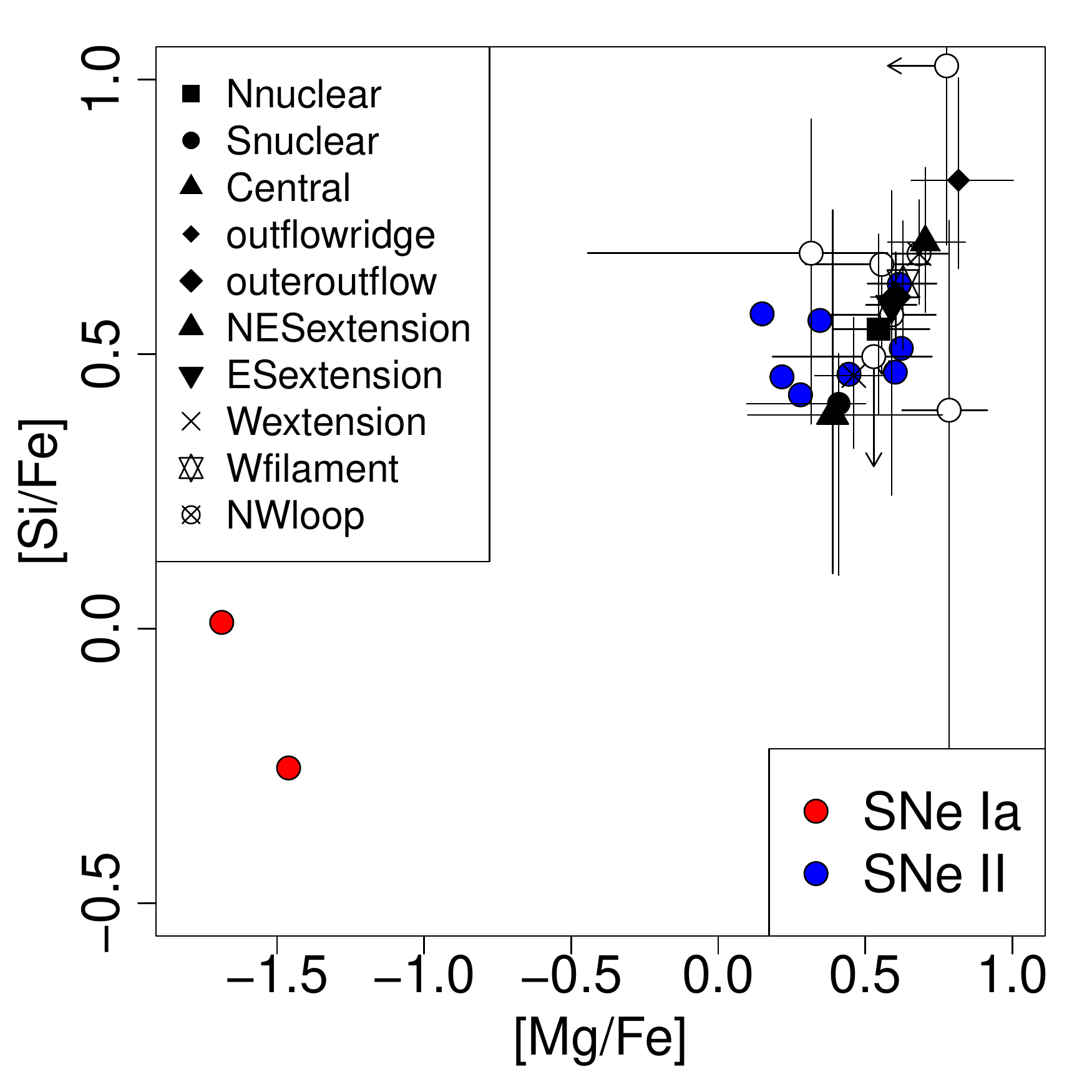}
	\caption{Ratios of the abundances of the \(\alpha\) elements to the iron abundance evaluated from the spectral fitting in the extraction regions presented in Fig. \ref{fig:regions} (black symbols). White circles represent the results from \citet{2013ApJ...765..141N}, with upper limits indicated with arrows. The expected yields {from} SNe Ia and SNe II are presented with red and blue circles, respectively \citep{2006ApJ...636..158B}.}\label{fig:abundances}
\end{figure}

The main results of our analysis of this extended emission are:
\begin{enumerate}
	\item The spectra extracted from the two nuclear regions (\(r < 3\)) are significantly harder and have larger absorption with respect to the spectra from the outer regions. The emission from the nuclear regions is dominated by Ne X and Fe XVII lines, and at higher energies by Mg XI, Si XIII and S XV, indicating contribution to the X-ray emission in the circum-nuclear region from starburst-driven winds \citep{2013A&A...558A..87F, 2014ApJ...781...55W}. The spectral analysis reveals a hot gas component with \(kT\sim 2.3 \text{ keV}\) and a mildly photo-ionized phase with \(\log{(U)}\sim 0\) in the northern nucleus, while in the southern nucleus we find a thermal gas with \(kT\sim 1.9 \text{ keV}\) and two photo-ionization components with \(\log{(U)}\sim -1\) and \(\log{(U)}\sim 1.4\). The thermal gas has a mean density of \(\sim 0.4 \text{ cm}^{-3}\). {Consistent} with the higher absorbing column estimated for the northern nucleus, its circumnuclear emission is detected at energies \(>1.6 \text{ keV}\), while for the southern nucleus this {emission} is {already visible} above \(0.95 \text{ keV}\), extending toward the WS Extension.
	\item The surface brightness profiles extracted in different directions show a {broken power-law shape}, flatter on the interior, and steeper in the outer regions. This suggests the presence of a freely expanding wind in the outer regions and some form of {hot} plasma confinement in the inner region, within \(\sim 2.5 \text{ kpc}\) (\(r < 5''\)). If the confinement is magnetic, magnetic fields of \(\sim 100\,\upmu \text{G}\) would be required, similar to those measured in the outflow of M82 \citep{2021ApJ...914...24L}.
	\item The [O III] to soft X-ray flux ratio profiles are compatible with the values measured for NGC 4151 a small radii, while for \(r > 5''\) the decreasing X-ray flux yields values of this ratio \(\sim 100\). The thermal gas at \(kT \sim 1 \text{ keV}\) {at large radii may absorb} soft X-rays from the AGN, but not the extreme ultraviolet radiation leading to a rapid increase in \(F_{\text{[O III]}}/F_{0.5-2}\) beyond \(\sim 2.5 \text{ kpc}\) where \(N_H \sim 3 \times {10}^{21} \text{ cm}^{-2}\) radially.
	\item The Outer Outflow [O III] region and the Outflow Ridge perpendicular to the outflow axis are more prominent below \(1.15 \text{ keV}\) {and are} dominated by O VII, O VIII, Ne IX, and Ne X emission lines, as observed in nearby Seyfert galaxies, and {are} related to shock excitation. Spectral analysis indicates in these regions the presence of a \(\sim 1 \text{ keV}\) thermal gas, together with two photo-ionized components with \(\log{U}\sim 0\) and \(\log{U}\sim 2\). {The arc in the SW (Fig. \ref{fig:narrow-bands_zoom}) could be connected with the other side of the bicone.}
	\item The X-ray emission from the NW Loop H\(\alpha\) region peaks between \(0.7\) and \(1.15 \text{ keV}\), but is also detected at higher energies, showing Mg XI, Si XIII and S XV lines, {that are} usually associated with starburst activity. The spectrum from this region is best fitted by a two temperature \(\sim 0.8 \text{ keV}\) and \(\sim 1.8 \text{ keV}\) thermal gas, with an additional highly photo-ionized phase with \(\log{(U)}\sim 1.3\).
	\item The ES Extension and the WS Extension regions are dominated by O VIII, Fe XX and Ne X emission. The spectral analysis indicates {the} presence of a thermal gas component with \(kT \sim 1 \text{ keV}\), and a mildly photo-ionized component with \(\log{(U)}\sim 0\), possibly connected with the same component observed in the northern nucleus. {If this feature is the southern side of a biconical outflow from the northern CT AGN, the northern side could be obscured by the dusty galactic disk (see \citealt{2018Natur.556..345M}).}
	\item The iron abundance of thermal components is found to be about solar in the nuclear regions and sub-solar in the outer regions. The ratio of \(\alpha\) elements over Fe abundances {are compatible} with SNe II yields {but not with SNe Ia yields}, confirming the importance of active star formation in NGC 6240  and giving a direct view of the enrichment of the ISM in the NGC 6240 system.
\end{enumerate}	

The emission from NGC 6240 is complex, and different physical process are at work in this source. The results of this analysis confirm the significant contribution of starburst-driven winds to the X-ray emission observed in this source, in particular in the central region and in the NW H\(\alpha\) Loop. The [O III] Outer Outflow and the Outflow Ridge regions are likely due to both photo-ionization and shock excitation, connected with the southern nucleus activity, while the southern protrusions may indicate the edges of a ionization cone connected with the northern nucleus activity. 

\begin{figure}
	\centering
	\includegraphics[scale=0.3]{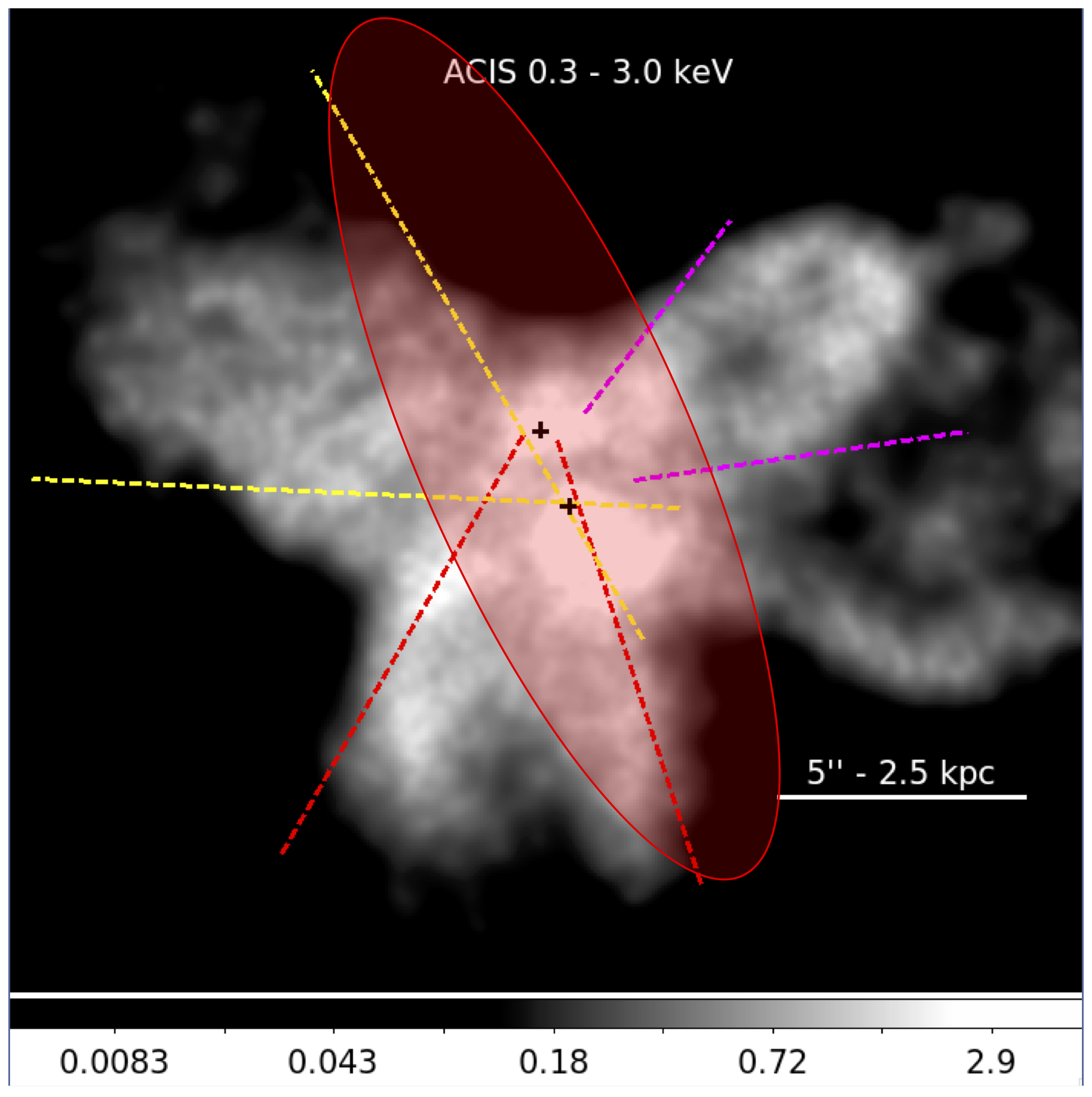}
	\caption{{Sketch of the extended X-ray emission observed in NGC 6240. The red dashed lines indicate the edges of the ionization cone connected to the northern nucleus, the yellow dashed lines indicate the edges of the ionization cone linked to the southern nucleus (possibly extending to the west), and the dashed magenta lines indicate the starburst-driven winds yielding the NW H\(\upalpha\) loop.}}\label{fig:cones}
\end{figure}

{Fig. \ref{fig:cones} sums up our interpretation of the extended X-ray emission observed in NGC 6240, with the red dashed lines delineating the edges of the ionization cone emerging from the northern nucleus, the yellow dashed lines indicating the edges of the ionization cone linked to the southern nucleus (with a putative, weak counter-cone extending to the west), and the dashed magenta lines marking the starburst-driven winds extending in the NW H\(\upalpha\) loop.}

{As} the nearest double AGN merging galaxy system, NGC 6240 {is} a unique source that provides a complex mix of different physical processes {that can be used} to study the galaxy-black hole evolution and interaction.

\begin{acknowledgments}
{We thank the anonymous referee for their useful comments and suggestions.} This work is supported by the ``Departments of Excellence 2018 - 2022" Grant awarded by the Italian Ministry of Education, University and Research (MIUR) (L. 232/2016). A.P. acknowledges financial support from the Consorzio Interuniversitario per la Fisica Spaziale (CIFS) under the agreement related to the grant MASF\_CONTR\_FIN\_18\_02. {J.W. acknowledges support from NSFC grants U1831205 and 12033004.} This work was partially supported by NASA contract NAS8-03060 (CXC). This research has made use of data obtained from the \textit{Chandra} Data Archive. This research is based on observations made with the NASA/ESA Hubble Space Telescope, obtained from the data archive at the Space Telescope Science Institute. STScI is operated by the Association of Universities for Research in Astronomy, Inc. under NASA contract NAS5-26555. This research has made use of software provided by the \textit{Chandra} X-ray Center (CXC) in the application packages CIAO, ChIPS, and Sherpa.
\end{acknowledgments}

\software{CIAO \citep{2006SPIE.6270E..1VF}, Sherpa \citep{2001SPIE.4477...76F, 2007ASPC..376..543D, 2020zndo...3944985B}, ChiPS \citep{2006ASPC..351...57G}, Cloudy \citep[c08.01][]{1998PASP..110..761F}, XSPEC \citep{1996ASPC..101...17A}.}

\newpage

\begin{sidewaystable}
\caption{Best fit results of the spectra extracted in the regions presented in Fig. \ref{fig:regions} for the phenomenological models comprising a power-law with slope frozen to \(1.8\) and red-shifted Gaussian emission lines (see Sect. \ref{sec:spectra}). For each region are presented the normalization of each line and of the power-law component, the required intrinsic absorption column density, the reduced \(\chi^2\) (with degrees of freedom indicated in parenthesis) and the \(0.3-3 \text{ keV}\) net counts (with the error indicated in parenthesis).}\label{tab:line_spectra}
\begin{center}
\resizebox{\textwidth}{!}{
\begin{tabular}{*{13}{|c}|}
\hline
\hline
\multicolumn{2}{|c|}{} & N Nuclear & S Nuclear & Central & Outflow Ridge & Outer Outflow & NES Extension & ES Extension & WS Extension & W Filament & NW Loop & All \\
\hline
Line & Rest-frame Energy (keV) & \multicolumn{11}{c|}{Line Normalization (\({10}^{-5}\text{ photons}\text{ cm}^{-2}\text{ s}^{-1}\))} \\
\hline 
 O VII triplet & \(0.569\) &  &  &  &  & \({0.20}_{-0.10}^{+0.10}\) &  & \({0.58}_{-0.35}^{+0.58}\) &  &  &  & \({4.47}_{-2.52}^{+3.08}\) \\
 Fe XVIII \(1s^2 2s^1 2p^6 \rightarrow 1s^2 2s^2 2p^4 3p^1\) & \(0.704\) & \({23.02}_{-14.74}^{+26.34}\) &  &  &  & \({0.30}_{-0.06}^{+0.06}\) &  & \({0.33}_{-0.12}^{+0.17}\) & \({0.29}_{-0.09}^{+0.10}\) & \({0.09}_{-0.04}^{+0.04}\) & & \({2.47}_{-0.77}^{+0.86}\) \\
 Fe XVII \(1s^2 2s^2 2p^5 \rightarrow 1s^2 2s^2 2p^4 3s^1\)& \(0.771\) &  &  &  &  &  &  &  &  &  & \({0.29}_{-0.11}^{+0.12}\) & \({1.72}_{-0.71}^{+0.74}\)\\
 Fe XVII \(1s^2 2s^2 2p^6 \rightarrow 1s^2 2s^2 2p^5 3d^1\) & \(0.826\) &  &  &  &  & \({0.49}_{-0.05}^{+0.05}\) &  & \({0.34}_{-0.10}^{+0.12}\) &  & \({0.24}_{-0.04}^{+0.04}\) & & \({3.41}_{-0.81}^{+0.84}\) \\
 O VIII RRC & \(0.871\) & \({2.55}_{-1.38}^{+2.04}\) &  &  & \({0.28}_{-0.06}^{+0.07}\) &  & \({0.12}_{-0.10}^{+0.13}\) & \({0.39}_{-0.10}^{+0.12}\) & \({0.58}_{-0.07}^{+0.08}\) &  & \({0.60}_{-0.09}^{+0.10}\) & \({2.05}_{-0.64}^{+0.65}\) \\
 Ne IX triplet & \(0.915\) &  &  &  &  & \({0.43}_{-0.05}^{+0.05}\) &  &  &  &  & & \({4.39}_{-0.44}^{+0.48}\) \\
 Fe XX \(1s^2 2s^2 2p^3 \rightarrow 1s^2 2s^2 2p^2 3d^1\) & \(0.965\) & \({1.97}_{-0.83}^{+1.11}\) & \({0.50}_{-0.19}^{+0.21}\) &  & \({0.16}_{-0.06}^{+0.06}\) &  &  & \({0.34}_{-0.07}^{+0.08}\) & \({0.29}_{-0.06}^{+0.07}\) & \({0.15}_{-0.03}^{+0.03}\) & \({0.22}_{-0.08}^{+0.08}\) & \\
 Ne X Ly\(\alpha\) & \(1.022\) & \({0.63}_{-0.43}^{+0.56}\) & \({0.30}_{-0.16}^{+0.16}\) & \({0.60}_{-0.10}^{+0.11}\) & \({0.17}_{-0.06}^{+0.06}\) & \({0.46}_{-0.04}^{+0.04}\) & \({0.21}_{-0.07}^{+0.08}\) & \({0.34}_{-0.06}^{+0.07}\) & \({0.27}_{-0.05}^{+0.06}\) & \({0.11}_{-0.03}^{+0.03}\) & \({0.60}_{-0.08}^{+0.08}\) & \({4.37}_{-0.30}^{+0.33}\) \\
 Fe XVIII \(1s^2 2s^2 2p^5 \rightarrow 1s^2 2s^2 2p^4 4d^1\) & \(1.095\)  &  & \({0.40}_{-0.11}^{+0.11}\) &  & \({0.12}_{-0.04}^{+0.04}\) &  &  &  &  &  & & \({0.87}_{-0.34}^{+0.34}\) \\
 Fe XXIII \(1s^2 2s^2 \rightarrow 1s^2 2s^1 3p^1\) & \(1.129\) &  &  &  &  & \({0.14}_{-0.02}^{+0.02}\) &  &  & \({0.25}_{-0.03}^{+0.03}\) &  & & \({0.77}_{-0.39}^{+0.39}\) \\
 Fe XIX \(1s^2 2s^2 2p^4 \rightarrow 1s^2 2s^2 2p^3 4d^1\) & \(1.146\) &  &  & \({0.14}_{-0.06}^{+0.06}\) &  &  &  & \({0.15}_{-0.03}^{+0.03}\) &  &  & & \\
 Fe XVII \(1s^2 2s^2 2p^6 \rightarrow 1s^2 2s^2 2p^5 7d^1\) & \(1.180\) & \({0.35}_{-0.16}^{+0.20}\) & \({0.10}_{-0.07}^{+0.07}\) &  & \({0.09}_{-0.03}^{+0.03}\) &  & \({0.05}_{-0.03}^{+0.03}\) &  &  & \({0.05}_{-0.01}^{+0.01}\) & \({0.26}_{-0.04}^{+0.04}\) & \({0.38}_{-0.23}^{+0.23}\) \\
 Fe XIX \(1s^2 2s^2 2p^4 \rightarrow 1s^2 2s^2 2p^3 5d^1\) & \(1.258\) &  &  &  &  & \({0.06}_{-0.02}^{+0.02}\) &  &  &  &  & & \({0.72}_{-0.12}^{+0.12}\) \\
 Mg XI triplet & \(1.352\) &  & \({0.06}_{-0.05}^{+0.05}\) & \({0.13}_{-0.04}^{+0.04}\) & \({0.05}_{-0.02}^{+0.02}\) & \({0.10}_{-0.02}^{+0.02}\) &  & \({0.08}_{-0.02}^{+0.02}\) & \({0.10}_{-0.02}^{+0.02}\) & \({0.02}_{-0.01}^{+0.01}\) & \({0.21}_{-0.03}^{+0.03}\) & \({0.88}_{-0.09}^{+0.09}\) \\
 Fe XXII \(1s^2 2s^2 2p^1 \rightarrow 1s^2 2s^1 2p^1 4p^1\) & \(1.423\) &  &  &  &  &  & \({0.05}_{-0.02}^{+0.02}\) & \({0.07}_{-0.02}^{+0.02}\) &  & \({0.03}_{-0.01}^{+0.01}\) & & \\
 Mg XII Ly\(\alpha\) & \(1.473\) &  & \({0.13}_{-0.04}^{+0.04}\) & \({0.14}_{-0.04}^{+0.04}\) & \({0.10}_{-0.02}^{+0.02}\) & \({0.06}_{-0.02}^{+0.02}\) &  &  & \({0.06}_{-0.02}^{+0.02}\) &  & \({0.17}_{-0.03}^{+0.03}\) & \({1.00}_{-0.08}^{+0.08}\)\\
 Si XIII triplet & \(1.839\) & \({0.14}_{-0.04}^{+0.04}\) & \({0.15}_{-0.03}^{+0.03}\) & \({0.13}_{-0.03}^{+0.03}\) & \({0.08}_{-0.02}^{+0.02}\) & \({0.08}_{-0.01}^{+0.01}\) & \({0.04}_{-0.02}^{+0.02}\) & \({0.07}_{-0.02}^{+0.02}\) & \({0.05}_{-0.02}^{+0.02}\) & \({0.02}_{-0.01}^{+0.01}\) & \({0.16}_{-0.02}^{+0.02}\) & \({1.13}_{-0.07}^{+0.07}\) \\
 Mg XII RRC & \(1.963\) &  & \({0.18}_{-0.03}^{+0.03}\) & \({0.15}_{-0.03}^{+0.03}\) & \({0.05}_{-0.01}^{+0.01}\) &  & \({0.05}_{-0.02}^{+0.02}\) &  &  &  & \({0.10}_{-0.02}^{+0.02}\) & \({0.64}_{-0.10}^{+0.10}\) \\
 Si XIV Ly\(\alpha\) & \(2.006\) & \({0.15}_{-0.03}^{+0.03}\) &  &  &  &  &  &  &  &  & & \({0.23}_{-0.09}^{+0.09}\) \\
 S XV Ly\(\alpha\) & \(2.461\) & \({0.15}_{-0.03}^{+0.03}\) & \({0.11}_{-0.03}^{+0.03}\) & \({0.11}_{-0.03}^{+0.03}\) & \({0.04}_{-0.01}^{+0.01}\) &  & \({0.04}_{-0.02}^{+0.02}\) & \({0.02}_{-0.01}^{+0.01}\) &  &  & \({0.05}_{-0.02}^{+0.02}\) & \({0.68}_{-0.06}^{+0.06}\) \\
 S XIV He\(\beta\) & \(2.884\) &  & \({0.04}_{-0.02}^{+0.02}\) &  &  &  &  &  &  &  & & \({0.27}_{-0.05}^{+0.05}\) \\
\hline
Power-law Norm. (\({10}^{-5}\text{ keV}^{-1}\text{ cm}^{-2}\text{ s}^{-1}\)) & & \({6.27}_{-0.42}^{+0.44}\) & \({6.48}_{-0.27}^{+0.27}\) & \({4.99}_{-0.21}^{+0.21}\) & \({1.24}_{-0.10}^{+0.10}\) & \({0.99}_{-0.07}^{+0.07}\) & \({2.05}_{-0.13}^{+0.14}\) & \({1.53}_{-0.09}^{+0.09}\) & \({1.72}_{-0.09}^{+0.09}\) & \({0.58}_{-0.05}^{+0.05}\) & \({2.48}_{-0.14}^{+0.14}\) & \({27.40}_{-0.53}^{+0.54}\) \\
\hline
 \(N_H\,({10}^{22}\text{ cm}^{-2})\) & & \({1.30}_{-0.11}^{+0.11}\) & \({0.54}_{-0.03}^{+0.03}\) & \({0.49}_{-0.02}^{+0.03}\) & \({0.13}_{-0.02}^{+0.03}\) & - & \({0.36}_{-0.05}^{+0.06}\) & \({0.06}_{-0.03}^{+0.04}\) & \({0.09}_{-0.02}^{+0.02}\) & - & \({0.18}_{-0.02}^{+0.02}\) & \({0.22}_{-0.01}^{+0.02}\) \\
\hline
\(\chi^2\) (d.o.f.) & & \(0.94(76)\) & \(0.84(117)\) & \(0.91(108)\) & \(1.08(62)\) & \(0.94(68)\) & \(0.94(54)\) & \(0.95(75)\) & \(0.98(79)\) & \(0.85(41)\) & \(0.93(95)\) & \(1.07(158)\) \\
\hline
Net Counts (\(0.3-3 \text{ keV}\)) & & \(2606(51)\) & \(4762(69)\) & \(3924(63)\) & \(2096(46)\) & \(2844(53)\) & \(1736(42)\) & \(3038(55)\) & \(2997(55)\) & \(1237(35)\) & \(3875(62)\) & \(35869(189)\) \\
\hline
\hline
\end{tabular}
}
\end{center}
\end{sidewaystable}

\begin{sidewaystable}
	\caption{Best fit results of the spectra extracted in the regions presented in Fig. \ref{fig:regions} for the physical models comprising thermal and photo-ionization components (see Sect. \ref{sec:spectra}). {Only best-fit models, selected on the basis of fit statistics and residual distribution, are presented here. The full list of models used in this work is presented in the Appendix. For} each region {we show} the temperature of the first (\(kT_1\)) and second (\(kT_2\)) thermal component, the iron abundance (Fe) and the abundance ratio of \(\alpha\) elements to iron (\(\alpha\)/Fe), both linked between the two thermal components, the ionization parameter of the first (\(U_1\)) and second (\(U_2\)) ionization component, the hydrogen column density of the first (\(N_{H1}\)) and second (\(N_{H2}\)) ionization component, an additional intrinsic hydrogen column density (\(N_{H}\)), and the reduced \(\chi^2\) (with degrees of freedom indicated in parenthesis). Parameters marked with an asterisk (\(^*\)) could not be constrained and {were} frozen to their best fit values.}\label{tab:line_phys}
	\begin{center}
	\resizebox{\textwidth}{!}{
		\begin{tabular}{|c|c|c|c|c|c|c|c|c|c|c|}
			\hline
			\hline
			& N Nuclear & S Nuclear & Central & Outflow Ridge & Outer Outflow & NES Extension & ES Extension & WS Extension & W Filament & NW Loop \\
			\hline
			\(kT_1 \text( keV)\)  & \({{2.32}_{-0.58}^{+1.41}}\) & \({{1.88}_{-0.21}^{+1.13}}\) & \({{1.45}_{-0.15}^{+0.11}}\) & \({{0.88}_{-0.06}^{+0.12}}\) &  \({{0.80}_{-0.04}^{+0.04}}\)  &  \({{0.84}_{-0.11}^{+0.19}}\) &  \({{0.84}_{-0.04}^{+0.07}}\)  &  \({{1.04}_{-0.06}^{+0.26}}\) &  \({{0.75}_{-0.04}^{+0.05}}\)  &  \({{0.82}_{-0.05}^{+0.11}}\) \\
			\(kT_2 \text( keV)\) & & & &  &  & \({{2.15}_{-0.29}^{+0.58}}\) &  &  & &  \({{1.82}_{-0.28}^{+0.28}}\) \\
			Fe & \({{1.22}_{-0.75}^{+4.35}}\) & \({{1.0}^*}\) & \({{0.22}_{-0.07}^{+0.11}}\) & \({{0.11}_{-0.05}^{+0.08}}\) &  \({{0.21}_{-0.07}^{+0.14}}\) &  \({{0.33}_{-0.20}^{+0.28}}\) &  \({{0.12}_{-0.03}^{+0.03}}\) &  \({{0.23}_{-0.08}^{+0.16}}\) &  \({{0.10}_{-0.03}^{+0.04}}\)  &  \({{0.34}_{-0.13}^{+0.24}}\) \\
			\(\alpha\)/Fe & \({{3.51}_{-1.05}^{+1.72}}\) & \({{2.57}_{-1.32}^{+0.60}}\) &  \({{5.06}_{-1.29}^{+1.86}}\) &  \({{6.56}_{-2.03}^{+3.52}}\) &  \({{4.02}_{-0.72}^{+0.83}}\) & \({{2.45}_{-1.19}^{+3.34}}\) &  \({{3.89}_{-0.71}^{+0.84}}\) &   \({{2.89}_{-0.76}^{+0.80}}\) &  \({{4.25}_{-1.03}^{+1.28}}\)  &  \({{4.82}_{-0.92}^{+1.22}}\) \\
			\(\log{\left({U_1}\right)}\) & \({{0.00}_{-0.10}^{+0.05}}\) & \({{-0.95}_{-0.07}^{+0.98}}\) & \({{-0.01}_{-0.19}^{+0.18}}\) & \({{0.08}_{-0.58}^{+0.33}}\) & \({{-0.50}_{-0.68}^{+0.78}}\) & & \({{-0.50}_{-0.24}^{+0.28}}\)  & \({{0.26}_{-0.16}^{+0.06}}\) &  & \({{1.25}_{-0.41}^{+0.12}}\) \\
			\(\log{\left({N_{H1}}\right)}\) & \({{22.4}^*}\) & \({{22.10}_{-0.11}^{+0.86}}\) & \({{22.5}^*}\) & \({{21.7}^*}\) & \({{22.2}^*}\) & & \({{22.9}^*}\) & \({{20.20}_{-0.30}^{+0.28}}\) & &  \({{22.1}^*}\) \\
			\(\log{\left({U_2}\right)}\) & & \({{1.43}_{-0.09}^{+0.12}}\) & & \({{2.0}^*}\) &  \({{1.9}^*}\) & &  &  &  & \\
			\(\log{\left({N_{H2}}\right)}\) & & \({{20.1}^*}\) & & \({{19.0}^*}\) &   \({{19.7}^*}\) & &  &  &  & \\
			\hline 
			\(N_H\,({10}^{22}\text{ cm}^{-2})\) & \({{1.13}_{-0.07}^{+0.10}}\) &  \({{0.69}_{-0.04}^{+0.16}}\) &  \({{0.68}_{-0.04}^{+0.05}}\) &   \({{0.38}_{-0.05}^{+0.05}}\) &  \({{0.09}_{-0.03}^{+0.03}}\) &   \({{0.68}_{-0.18}^{+0.21}}\) &  \({{0.18}_{-0.04}^{+0.03}}\) &   \({{0.40}_{-0.05}^{+0.05}}\) &  \({{0.19}_{-0.04}^{+0.05}}\)  &  \({{0.34}_{-0.03}^{+0.04}}\) \\
			\hline
			\(\chi^2\) (d.o.f.) & \({1.04(79)}\) & \({0.74(120)}\) & \({0.70(110)}\) & \({0.71(66)}\)  & \({0.68(71)}\)  & \({0.72(56)}\) & \({0.65(81)}\)  &  \({0.76(81)}\) & \({0.73(45)}\)  &  \({0.72(98)}\) \\
			\hline
			\hline
		\end{tabular}
	}
\end{center}
\end{sidewaystable}

\begin{table}
		\caption{Lines contributing to energy bands.}\label{tab:line_energy}
	\begin{center}		
			\begin{tabular}{|c|c|}
				\hline
				\hline
				Energy Band (keV) & Contributing lines \\
				\hline 
				\(0.5-0.7\) & O VII triplet, Fe XVIII \\
				\(0.7-0.95\) & Fe XVII, O VIII RRC, Ne IX\\
				\(0.95-1.15\) & Fe XX, Ne X, Fe XVIII, Fe XXIII, Fe XIX \\
				\(1.2-1.4\) & Fe XI, Mg XI triplet \\
				\(1.4-1.55\) & Fe XXII, Mg XII \\
				\(1.6-1.85\) & Si XIII triplet \\
				\(1.9-2.1\) & Mg XII RRC, Si XIV Ly\(\alpha\) \\
				\(2.2-2.7\) & Si XV Ly\(\alpha\), Si XIV He\(\beta\) \\
				\hline
				\hline
			\end{tabular}
	\end{center}
\end{table}

\clearpage

\appendix
\section{Spectral fits for physical models}

{In this appendix we present the full set of physical models - that is, comprising thermal and photo-ionization components - used to fit the spectra extracted in the regions presented in Fig. \ref{fig:regions}. All the models considered here are listed in Table \ref{tab:line_phys_long}. The models that we selected as best-fit on the basis of fit statistics and residual distribution are indicated in boldface.}

\begin{sidewaystable}
	\caption{{Full list of the best fit results of the spectra extracted in the regions presented in Fig. \ref{fig:regions} for the physical models comprising thermal and photo-ionization components (see Sect. \ref{sec:spectra}). Only models with a reduced \(\chi^2<1.2\) are shown. For each region are presented the temperature of the first (\(kT_1\)) and second (\(kT_2\)) thermal component, the iron abundance (Fe) and the abundance ratio of \(\alpha\) elements to iron (\(\alpha\)/Fe), both linked between the two thermal components, the ionization parameter of the first (\(U_1\)) and second (\(U_2\)) ionization component, the hydrogen column density of the first (\(N_{H1}\)) and second (\(N_{H2}\)) ionization component, an additional intrinsic hydrogen column density (\(N_{H}\)), and the reduced \(\chi^2\) (with degrees of freedom indicated in parenthesis). Parameters marked with an asterisk (\(^*\)) could not be constrained and were frozen to their best fit values. Parameters with two asterisks (\(^{**}\)) indicate a parameter frozen to its best fit value and whose component normalization is only constrained with an upper limit. In boldface we indicate the best-fit models, selected on the basis of fit statistics and residual distribution.}}\label{tab:line_phys_long}
	\begin{center}
		\resizebox{\textwidth}{!}{
			\begin{tabular}{|c|cccc|ccccc|cccccc|}
				\hline
				\hline
				& \multicolumn{4}{c}{N Nuclear} & \multicolumn{5}{|c}{S Nuclear} & \multicolumn{6}{|c|}{Central} \\
				\hline
				\(kT_1 \text( keV)\) & \({0.87}_{-0.05}^{+0.04}\) &  & \(\boldsymbol{{2.32}_{-0.58}^{+1.41}}\) & \({0.69}_{-0.12}^{+0.08}\) & \({1.64}_{-0.07}^{+0.06}\) & \({0.98}_{-0.06}^{+0.06}\) &  & \({1.56}_{-0.08}^{+0.07}\) & \(\boldsymbol{{1.88}_{-0.21}^{+1.13}}\) & \({1.54}_{-0.08}^{+0.08}\) & \({1.08}_{-0.10}^{+0.27}\) &  & \(\boldsymbol{{1.45}_{-0.15}^{+0.11}}\) & \({1.18}_{-0.18}^{+0.16}\) & \({1.31}_{-0.08}^{+0.18}\) \\
				\(kT_2 \text( keV)\) &  &  &  &  &  & \({6.48}_{-1.29}^{+1.64}\) &  &  &  &  & \(>2.14\) &  &  & \(<1.00\) & \\
				Fe & \({0.07}_{-0.02}^{+0.02}\) &   & \(\boldsymbol{{1.22}_{-0.75}^{+4.35}}\) & \({1.8}^*\) & \({0.10}_{-0.04}^{+0.04}\) & \({2.89}_{-2.80}^{+4.97}\) &  & \({0.45}_{-0.13}^{+0.18}\) & \(\boldsymbol{{1.0}^*}\) & \({0.08}_{-0.04}^{+0.04}\) & \({0.14}_{-0.08}^{+0.12}\) &  & \(\boldsymbol{{0.22}_{-0.07}^{+0.11}}\) & \({0.23}_{-0.09}^{+0.13}\) & \({0.63}_{-0.31}^{+0.92}\) \\
				\(\alpha\)/Fe & \({5.95}_{-1.07}^{+1.38}\) &  & \(\boldsymbol{{3.51}_{-1.05}^{+1.72}}\) & \(<0.40\) & \({9.65}_{-4.78}^{+5.08}\) & \({2.15}_{-0.78}^{+9.55}\) &  & \({3.17}_{-0.58}^{+0.76}\) & \(\boldsymbol{{2.57}_{-1.32}^{+0.60}}\) & \({11.22}_{-3.47}^{+7.69}\) & \({6.66}_{-2.68}^{+6.20}\) &  & \(\boldsymbol{{5.06}_{-1.29}^{+1.86}}\) & \({4.76}_{-1.23}^{+2.39}\) & \({2.45}_{-0.78}^{+1.43}\) \\
				\(\log{\left({U_1}\right)}\) &  & \({0.00}_{-0.12}^{+0.03}\) & \(\boldsymbol{{0.00}_{-0.10}^{+0.05}}\) & \({-0.06}_{-0.47}^{+0.12}\) &  &  & \({-0.16}_{-0.05}^{+0.07}\) & \({-0.08}_{-0.11}^{+0.10}\) & \(\boldsymbol{{-0.95}_{-0.07}^{+0.98}}\) &  &  & \({-0.19}_{-0.05}^{+0.09}\) & \(\boldsymbol{{-0.01}_{-0.19}^{+0.18}}\) & \({-0.15}_{-0.50}^{+0.49}\) & \({-0.18}_{-0.12}^{+0.21}\) \\
				\(\log{\left({N_{H1}}\right)}\) &  & \({22.3}^*\) & \(\boldsymbol{{22.4}^*}\) & \({22.29}_{-0.45}^{+0.24}\) &  &  & \({20.00}_{-0.46}^{+0.32}\) & \({22.5}^*\) & \(\boldsymbol{{22.10}_{-0.11}^{+0.86}}\) &  &  & \({19.8}^*\) & \(\boldsymbol{{22.5}^*}\) & \({22.56}_{-0.55}^{+0.65}\) & \({19.2}^*\) \\
				\(\log{\left({U_2}\right)}\) &  & \({2.0}^*\) &  & \({2.0}^*\) &  &  & \({1.77}_{-0.12}^{+0.05}\) &  & \(\boldsymbol{{1.43}_{-0.09}^{+0.12}}\) &  &  & \({1.67}_{-0.05}^{+0.05}\) &  &  & \({1.5}^*\) \\
				\(\log{\left({N_{H2}}\right)}\) &  & \({20.9}^*\) &  & \({20.9}^*\) &  &  & \({19.0}^*\) &  & \(\boldsymbol{{20.1}^*}\) &  &  & \({20.40}_{-0.56}^{+0.33}\) &  &  & \({22.3}^*\) \\
				\hline 
				\(N_H\,({10}^{22}\text{ cm}^{-2})\) & \({0.11}_{-0.02}^{+0.04}\) & \({1.19}_{-0.05}^{+0.05}\) & \(\boldsymbol{{1.13}_{-0.07}^{+0.10}}\) & \({1.68}_{-0.20}^{+0.20}\)  & \({0.60}_{-0.03}^{+0.03}\) & \({0.72}_{-0.10}^{+0.19}\) & \({1.15}_{-0.06}^{+0.09}\) & \({0.72}_{-0.04}^{+0.04}\) & \(\boldsymbol{{0.69}_{-0.04}^{+0.16}}\) & \({0.59}_{-0.03}^{+0.04}\) & \({0.64}_{-0.12}^{+0.09}\) & \({1.10}_{-0.04}^{+0.05}\) & \(\boldsymbol{{0.68}_{-0.04}^{+0.05}}\) & \({0.68}_{-0.05}^{+0.06}\) & \({1.04}_{-0.10}^{+0.08}\) \\
				\hline
				\(\chi^2\) (d.o.f.) & \(0.85(81)\) & \(1.01(82)\) & \(\boldsymbol{1.04(79)}\) & \(0.96(78)\) & \(1.18(124)\) & \(0.81(122)\) & \(0.78(123)\) & \(0.79(122)\) & \(\boldsymbol{0.74(120)}\) & \(0.84(112)\) & \(0.72(110)\) & \(0.71(111)\) & \(\boldsymbol{0.70(110)}\) & \(0.69(107)\) & \(0.69(109)\) \\
				\hline
				\hline
			\end{tabular}
		}
	\end{center}
	
	\begin{center}
		\resizebox{\textwidth}{!}{
			\begin{tabular}{|c|cccccc|cccccc|cccccc|}
				\hline
				\hline
				& \multicolumn{6}{|c}{Outflow Ridge} & \multicolumn{6}{|c|}{Outer Outflow} & \multicolumn{6}{c|}{NES Extension} \\
				\hline
				\(kT_1 \text( keV)\) & \({0.99}_{-0.05}^{+0.05}\) & \({0.86}_{-0.07}^{+0.08}\) &  & \({1.02}_{-0.08}^{+0.06}\) & \({0.86}_{-0.08}^{+0.09}\) & \(\boldsymbol{{0.88}_{-0.06}^{+0.12}}\) & \({0.85}_{-0.03}^{+0.11}\) & \({0.77}_{0.24-}^{+0.07}\) & \({0.81}_{-0.04}^{+0.04}\) & \({0.61}_{-0.21}^{+0.18}\) & \(\boldsymbol{{0.80}_{-0.04}^{+0.04}}\) & \({0.67}_{-0.09}^{+0.12}\) & \({1.78}_{-0.40}^{+0.23}\) & \(\boldsymbol{{0.84}_{-0.11}^{+0.19}}\) &  & \({1.55}_{-0.16}^{+0.14}\) & \({0.88}_{-0.14}^{+0.40}\) & \({0.72}_{-0.10}^{+0.08}\) \\
				\(kT_2 \text( keV)\) &  & \({1.56}_{-0.22}^{+0.29}\) &  &  & \({1.40}_{-0.17}^{+0.38}\) &  &  & \({1.16}_{-0.63}^{+0.26}\) &  & \({1.03}_{-0.08}^{+0.22}\) &  & \({1.14}_{-0.16}^{+0.16}\) &  & \(\boldsymbol{{2.15}_{-0.29}^{+0.58}}\) &  &  & \({2.07}_{-0.42}^{+0.60}\) &  \\
				Fe & \({0.06}_{-0.02}^{+0.03}\) & \({0.15}_{-0.05}^{+0.20}\) &  & \({0.10}_{-0.04}^{+0.09}\) & \({0.17}_{-0.04}^{+0.23}\) & \(\boldsymbol{{0.11}_{-0.05}^{+0.08}}\) & \({0.12}_{-0.02}^{+0.10}\) & \({0.21}_{-0.06}^{+0.08}\) & \({0.16}_{-0.04}^{+0.06}\) & \({0.37}_{-0.15}^{+0.32}\) & \(\boldsymbol{{0.21}_{-0.07}^{+0.14}}\) & \({0.35}_{-0.14}^{+0.30}\) & \({0.09}_{-0.09}^{+0.08}\) & \(\boldsymbol{{0.33}_{-0.20}^{+0.28}}\) &  & \({0.14}_{-0.09}^{+0.13}\) & \({0.29}_{-0.20}^{+0.30}\) & \({0.67}_{-0.37}^{+1.29}\) \\
				\(\alpha\)/Fe & \({9.91}_{-2.70}^{+4.60}\) & \({5.61}_{-1.91}^{+1.48}\) &  & \({7.39}_{-3.07}^{+3.22}\) & \({5.01}_{-1.67}^{+0.90}\) & \(\boldsymbol{{6.56}_{-2.03}^{+3.52}}\) & \({4.98}_{-0.69}^{+0.76}\) & \({3.84}_{-0.62}^{+0.74}\) & \({4.49}_{-0.74}^{+0.80}\) & \({2.80}_{-0.73}^{+0.96}\) & \(\boldsymbol{{4.02}_{-0.72}^{+0.83}}\) & \({2.99}_{-0.74}^{+0.89}\) & \({10.87}_{-4.78}^{+87.23}\) & \(\boldsymbol{{2.45}_{-1.19}^{+3.34}}\) &  & \({6.51}_{-2.65}^{+8.27}\) & \({2.98}_{-1.63}^{+5.55}\) & \({0.55}_{-0.28}^{+0.50}\) \\
				\(\log{\left({U_1}\right)}\) &  &  & \({1.17}_{-0.08}^{+0.09}\) & \({0.25}_{-0.39}^{+0.12}\) & \({0.1}^{**}\) & \(\boldsymbol{{0.08}_{-0.58}^{+0.33}}\) &  &  & \({1.9}^{*}\) & \({-1.45}_{-0.31}^{+0.40}\) & \(\boldsymbol{{-0.50}_{-0.68}^{+0.78}}\) & \({-0.6}^*\) &  &  & \({-0.15}_{-0.08}^{+0.16}\) & \({0.10}_{-0.26}^{+0.34}\) & \({-0.2}^{**}\) & \({-0.5}^{**}\) \\
				\(\log{\left({N_{H1}}\right)}\) &  &  & \({19.5}^*\) & \({21.26}_{-1.69}^{+1.17}\) & \({20.3}^{**}\) & \(\boldsymbol{{21.7}^*}\) &  &  & \({19.7}^{*}\) & \({19.0}^*\) & \(\boldsymbol{{22.2}^*}\) & \({22.2}^*\) &  &  & \({19.9}^*\) & \({22.5}^*\) & \({22.5}^{**}\) & \({20.4}^{**}\) \\
				\(\log{\left({U_2}\right)}\) &  &  & \({2.0}^*\) &  &  & \(\boldsymbol{{2.0}^*}\) &  &  &  &  & \(\boldsymbol{{1.9}^*}\) & \({1.2}^{**}\) &  &  & \({1.7}^*\) &  &  & \({1.9}^*\) \\
				\(\log{\left({N_{H2}}\right)}\) &  &  & \({21.41}_{-0.42}^{+0.31}\) &  &  & \(\boldsymbol{{19.0}^*}\) &  &  &  &  & \(\boldsymbol{{19.7}^*}\) & \({22.9}^{**}\) &  &  & \({20.6}^*\) &  &  & \({21.2}^*\) \\
				\hline 
				\(N_H\,({10}^{22}\text{ cm}^{-2})\) &  \({0.31}_{-0.04}^{+0.04}\) & \({0.31}_{-0.04}^{+0.05}\) & \({0.32}_{-0.05}^{+0.05}\) & \({0.39}_{-0.06}^{+0.13}\) & \({0.39}_{-0.10}^{+0.07}\) & \(\boldsymbol{{0.38}_{-0.05}^{+0.05}}\) & \({0.08}_{-0.05}^{+0.02}\) & \({0.06}_{-0.02}^{+0.03}\) & \({0.07}_{-0.02}^{+0.02}\) & \({0.16}_{-0.07}^{+0.06}\) & \(\boldsymbol{{0.09}_{-0.03}^{+0.03}}\) & \({0.10}_{-0.03}^{+0.04}\) & \({0.39}_{-0.04}^{+0.12}\) & \(\boldsymbol{{0.68}_{-0.18}^{+0.21}}\) & \({0.98}_{-0.06}^{+0.06}\) & \({0.52}_{-0.07}^{+0.08}\) & \({0.62}_{-0.16}^{+0.25}\) & \({1.04}_{-0.15}^{+0.16}\) \\
				\hline
				\(\chi^2\) (d.o.f.) & \(0.76(69)\) & \(0.70(67)\) & \(0.99(69)\) & \(0.74(66)\) & \(0.70(66)\) & \(\boldsymbol{0.71(66)}\) & \(0.73(74)\) & \(0.67(72)\) & \(0.69(73)\) & \(0.66(70)\) & \(\boldsymbol{0.68(71)}\) & \(0.66(70)\) & \(0.82(58)\) & \(\boldsymbol{0.72(56)}\) & \(0.76(59)\) & \(0.74(56)\) & \(0.73(55)\) & \(0.68(56)\) \\
				\hline
				\hline
			\end{tabular}
		}
	\end{center}

	\begin{center}
		\resizebox{\textwidth}{!}{
			\begin{tabular} {|c|cccccc|cccccc|ccccccc|ccccccc|}
				\hline
				\hline
				& \multicolumn{6}{|c}{ES Extension} & \multicolumn{6}{|c}{WS Extension} & \multicolumn{7}{|c}{W Filament} & \multicolumn{7}{|c|}{NW Loop} \\
				\hline
				\(kT_1 \text( keV)\) & \({0.87}_{-0.05}^{+0.04}\) &  & \(\boldsymbol{{0.84}_{-0.04}^{+0.07}}\) & \({0.81}_{-0.46}^{+0.05}\) & \({0.92}_{-0.22}^{+0.06}\) & \({0.68}_{-0.16}^{+0.17}\) & \({0.94}_{-0.04}^{+0.03}\) & \({0.86}_{-0.04}^{+0.05}\) &  & \(\boldsymbol{{1.04}_{-0.06}^{+0.26}}\) & \({0.85}_{-0.04}^{+0.08}\) & \({1.02}_{-0.05}^{+0.30}\) & \(\boldsymbol{{0.75}_{-0.04}^{+0.05}}\) & \({0.77}_{-0.06}^{+0.05}\) &  & \({0.78}_{-0.05}^{+0.06}\) & \({0.52}_{-0.16}^{+0.19}\) & \({0.76}_{-0.05}^{+0.05}\) & \({0.54}_{-0.14}^{+0.21}\) & \({0.95}_{-0.03}^{+0.03}\) & \({0.82}_{-0.05}^{+0.05}\) &  & \({1.11}_{-0.14}^{+0.07}\) & \(\boldsymbol{{0.82}_{-0.05}^{+0.11}}\) & \({1.27}_{-0.11}^{+0.16}\) & \({0.70}_{-0.16}^{+0.14}\) \\
				\(kT_2 \text( keV)\) &  &  &  & \(>0.87\) &  & \({1.13}_{-0.66}^{+0.45}\) &  & \({1.98}_{-0.28}^{+0.36}\) &  &  & \({1.49}_{-0.42}^{+0.46}\) &  &  & \({2.16}_{-0.57}^{+5.33}\) &  &  & \({0.98}_{-0.09}^{+0.23}\) &  & \({0.97}_{-0.17}^{+0.44}\) &  & \({1.80}_{-0.21}^{+0.28}\) &  &  & \(\boldsymbol{{1.82}_{-0.28}^{+0.28}}\) &  & \({1.50}_{-0.23}^{+0.38}\) \\
				Fe & \({0.07}_{-0.02}^{+0.02}\) &  & \(\boldsymbol{{0.12}_{-0.03}^{+0.03}}\) & \({0.18}_{-0.07}^{+0.09}\) & \({0.16}_{-0.05}^{+0.08}\) & \({0.19}_{-0.08}^{+0.16}\) & \({0.07}_{-0.02}^{+0.02}\) & \({0.39}_{-0.14}^{+0.21}\) &  & \(\boldsymbol{{0.23}_{-0.08}^{+0.16}}\) & \(>0.84\) & \({0.16}_{-0.05}^{+2.15}\) & \(\boldsymbol{{0.10}_{-0.03}^{+0.04}}\) & \({0.27}_{-0.14}^{+0.27}\) &  & \({0.22}_{-0.09}^{+0.41}\) & \({0.56}_{-0.31}^{+1.85}\) & \(>1.77\) & \(>0.73\) & \({0.07}_{-0.02}^{+0.02}\) & \({0.21}_{-0.06}^{+0.08}\) &  & \({0.26}_{-0.14}^{+0.14}\) & \(\boldsymbol{{0.34}_{-0.13}^{+0.24}}\) & \({0.35}_{-0.18}^{+0.64}\) & \({0.50}_{-0.26}^{+1.42}\) \\
				\(\alpha\)/Fe & \({5.80}_{-1.01}^{+1.25}\) &  & \(\boldsymbol{{3.89}_{-0.71}^{+0.84}}\) & \({3.36}_{-0.68}^{+0.87}\) & \({3.42}_{-0.82}^{+1.14}\) & \({3.01}_{-0.98}^{+1.12}\) & \({5.03}_{-0.96}^{+1.18}\) & \({2.29}_{-0.50}^{+0.60}\) &  & \(\boldsymbol{{2.89}_{-0.76}^{+0.80}}\) & \({1.49}_{-0.42}^{+0.46}\) & \({3.07}_{-0.88}^{+0.96}\) & \(\boldsymbol{{4.25}_{-1.03}^{+1.28}}\) & \({3.09}_{-0.81}^{+0.99}\) &  & \({2.46}_{-1.05}^{+1.11}\) & \({1.77}_{-0.63}^{+0.84}\) & \({1.77}_{-0.74}^{+1.02}\) & \({1.99}_{-1.05}^{+0.87}\) & \({9.52}_{-1.78}^{+2.45}\) & \({5.53}_{-0.96}^{+1.26}\) &  & \({5.76}_{-1.11}^{+1.73}\) & \(\boldsymbol{{4.82}_{-0.92}^{+1.22}}\) & \({4.90}_{-1.71}^{+2.89}\) & \({4.36}_{-1.22}^{+1.88}\) \\
				\(\log{\left({U_1}\right)}\) &  & \({1.09}_{-0.04}^{+0.05}\) & \(\boldsymbol{{-0.50}_{-0.24}^{+0.28}}\) & \({-0.35}_{-0.33}^{+0.41}\) & \({-0.53}_{-0.30}^{+0.21}\) & \({-0.38}_{-0.29}^{+0.33}\) &  &  & \({1.06}_{-0.05}^{+0.06}\) & \(\boldsymbol{{0.26}_{-0.16}^{+0.06}}\) & \({1.45}_{-0.14}^{+0.15}\) & \({-2.00}_{-0.23}^{+0.39}\) &  &  & \({0.07}_{-0.07}^{+0.11}\) & \({-0.75}_{-0.88}^{+0.37}\) & \({-1.0}^*\) & \({-0.75}_{-0.72}^{+0.63}\) & \({-3.0}^*\) &  &  & \({0.00}_{-0.13}^{+0.07}\) & \({0.33}_{-0.28}^{+0.14}\) & \(\boldsymbol{{1.25}_{-0.41}^{+0.12}}\) & \({-0.82}_{-0.42}^{+0.66}\) & \({-1.3}^*\) \\
				\(\log{\left({N_{H1}}\right)}\) &  & \({19.7}^*\) & \(\boldsymbol{{22.9}^*}\) & \({23.5}^*\) & \({22.9}^*\) & \({23.2}^*\) &  &  & \({19.8}^*\) & \(\boldsymbol{{20.20}_{-0.30}^{+0.28}}\) & \({19.9}^*\) & \({21.0}^*\) &  &  & \({20.02}_{-0.21}^{+0.31}\) & \({20.68}_{-1.46}^{+0.81}\) & \({19.0}^*\) & \({20.3}^*\) & \({21.9}^*\) &  &  & \({20.20}_{-0.26}^{+0.30}\) & \({20.5}^*\) & \(\boldsymbol{{22.1}^*}\) & \({20.70}_{-0.70}^{+0.74}\) & \({20.4}^*\) \\
				\(\log{\left({U_2}\right)}\) &  & \({2.0}^*\) &  &  & \({0.85}_{-0.25}^{+0.25}\) & \({1.2}^*\) &  &  & \({2.0}^*\) &  &  & \({0.75}_{-0.23}^{+0.12}\) &  &  & \({1.59}_{-0.07}^{+0.09}\) &  &  & \({1.76}_{-0.12}^{+0.15}\) & \({1.8}^*\) &  &  & \({1.54}_{-0.03}^{+0.04}\) &  &  & \({1.09}_{-0.07}^{+0.08}\) & \({1.31}_{-0.16}^{+0.14}\) \\
				\(\log{\left({N_{H2}}\right)}\) &  & \({22.1}^*\) &  &  & \({19.4}^*\) & \({19.4}^*\) &  &  & \({21.23}_{-0.52}^{+0.35}\) &  &  & \({20.8}^*\) &  &  & \({19.0}^*\) &  &  & \({22.3}^*\) & \({22.2}^*\) &  &  & \({20.72}_{-0.20}^{+0.18}\) &  &  & \({19.7}^*\) & \({20.50}_{-0.91}^{+0.95}\) \\
				\hline 
				\(N_H\,({10}^{22}\text{ cm}^{-2})\) & \({0.12}_{-0.02}^{+0.03}\) & \({0.12}_{-0.02}^{+0.02}\) & \(\boldsymbol{{0.18}_{-0.04}^{+0.03}}\) & \({0.17}_{-0.04}^{+0.05}\) & \({0.20}_{-0.05}^{+0.04}\) & \({0.18}_{-0.05}^{+0.04}\) & \({0.17}_{-0.02}^{+0.02}\) & \({0.13}_{-0.02}^{+0.03}\) & \({0.18}_{-0.02}^{+0.03}\) & \(\boldsymbol{{0.40}_{-0.05}^{+0.05}}\) & \({0.13}_{-0.03}^{+0.03}\) & \({0.22}_{-0.03}^{+0.05}\) & \(\boldsymbol{{0.19}_{-0.04}^{+0.05}}\) & \({0.14}_{-0.05}^{+0.06}\) & \({0.41}_{-0.04}^{+0.04}\) & \({0.29}_{-0.09}^{+0.12}\) & \({0.32}_{-0.15}^{+0.08}\) & \({0.21}_{-0.12}^{+0.11}\) & \({0.13}_{-0.06}^{+0.07}\) & \({0.34}_{-0.03}^{+0.03}\) & \({0.35}_{-0.03}^{+0.04}\) & \({0.57}_{-0.04}^{+0.04}\) & \({0.51}_{-0.24}^{+0.05}\) & \(\boldsymbol{{0.34}_{-0.03}^{+0.04}}\) & \({0.46}_{-0.06}^{+0.06}\) & \({0.41}_{-0.05}^{+0.05}\) \\
				\hline
				\(\chi^2\) (d.o.f.) & \(0.85(83)\) & \(0.99(84)\) & \(\boldsymbol{0.65(81)}\) & \(0.65(79)\) & \(0.65(79)\) & \(0.64(76)\) & \(0.93(84)\) & \(0.67(82)\) & \(0.73(84)\) & \(\boldsymbol{0.76(81)}\) & \(0.58(80)\) & \(0.79(80)\) & \(\boldsymbol{0.73(45)}\) & \(0.63(43)\) & \(1.12(44)\) & \(0.68(42)\) & \(0.60(42)\) & \(0.57(41)\) & \(0.55(41)\) & \(0.94(102)\) & \(0.76(100)\) & \(1.11(100)\) & \(0.81(100)\) & \(\boldsymbol{0.72(98)}\) & \(0.73(97)\) & \(0.72(96)\) \\
				\hline
				\hline
			\end{tabular}
		}
	\end{center}
\end{sidewaystable}

\end{document}